\documentclass[aps,prx,twocolumn,superscriptaddress]{revtex4-2}
\usepackage{graphicx}
\usepackage{amsmath,amssymb}
\usepackage{physics}
\usepackage{xcolor}
\usepackage[colorlinks=true,allcolors=blue]{hyperref}

\newcommand{\bA}{\mathbf{A}}

\newcommand{\bk}{\mathbf{k}}
\newcommand{\bq}{\mathbf{q}}
\newcommand{\bp}{\mathbf{p}}
\newcommand{\bg}{\mathbf{g}}
\newcommand{\br}{\mathbf{r}}

\newcommand{\bL}{\mathbf{L}}
\newcommand{\bK}{\mathbf{K}}

\newcommand{\bG}{\mathbf{G}}
\newcommand{\bQ}{\mathbf{Q}}

\begin{document}
\title{Theory of correlated Chern insulators in twisted bilayer graphene}
\author{Xiaoyu Wang}
\email{xiaoyuw@magnet.fsu.edu}
\affiliation{National High Magnetic Field Lab, Tallahassee, FL 32310}
\author{Oskar Vafek}
\email{vafek@magnet.fsu.edu}
\affiliation{National High Magnetic Field Lab, Tallahassee, FL 32310}
\affiliation{Department of Physics, Florida State University, Tallahassee, FL 32306}

\begin{abstract}
    Magic-angle twisted bilayer graphene is the best studied physical platform featuring moir\'e potential induced narrow bands with non-trivial topology and strong electronic correlations. Despite their significance, the Chern insulating states observed at a finite magnetic field --and extrapolating to a band filling, $s$, at zero field-- remain poorly understood. Unraveling their nature is among the most important open problems in the province of moir\'e materials. Here we present the first comprehensive study of interacting electrons in finite magnetic field while varying the electron density, twist angle and heterostrain. Within a panoply of correlated Chern phases emerging at a range of twist angles, we uncover a unified description for the ubiquitous sequence of states with the Chern number $t$ for $(s,t)=\pm (0,4), \pm(1,3),\pm(2,2)$ and $\pm(3,1)$. We also find correlated Chern insulators at unconventional sequences with $s+t\neq \pm 4$, as well as with fractional $s$, and elucidate their nature. 
\end{abstract}
\maketitle

\section{Introduction}
The twisted bilayer graphene (TBG) has been a subject of intense theoretical and experimental investigation, in no small part due to its isolated, topologicaly non-trivial, narrow bands displaying rich correlated electron physics when partially occupied \cite{Andrei2020,Balents2020}. As the twist angle between the two graphene layers is tuned toward the magic value of $\sim1.05^\circ$ \cite{Bistritzer2011}, TBG devices show a plethora of correlated phenomena including superconductivity, correlated insulating states and (quantized) anomalous Hall effect \cite{Cao2018a,Cao2018b,Tomarken2019,Park2021,Xie2019,Yankowitz2019,Sharpe2019,Nuckolls2020,Nuckolls2023,Pierce2021,Xie2021,Saito2021,Yu2022,Hoke2023,Lu2019,Das2021,Serlin2020,Stepanov2021,Wu2021,Choi2021,Choi2021b,Grover2022,Uri2020}. The non-trivial topology of the pair of narrow bands for a given valley and spin flavor is protected by the combined two-fold rotation symmetry about the out-of-plane axis $C_{2z}$ (an emergent symmetry at low twist angle) and spinless time reversal symmetry $T$ \cite{Po2019,Song2019}. The narrow band Hilbert space can thus be decomposed into a Chern $+1$ and a Chern $-1$ band \cite{Po2019,Bultinck2020,Kang2020}. One way to reveal the non-trivial topology of the narrow bands is to break $C_{2z}$ via alignment with the hexagonal boron nitride substrate (hBN) and separate the Chern bands in energy. If in addition, the valley is spontaneously polarized, thus breaking $T$, the resulting state with one electron or hole per moir\'e unit cell becomes a Chern $\pm 1$ insulator \cite{Zhang2019,Bultinck2019,TBGIV,Liu2021}. Indeed, experiments have observed anomalous Hall effect (AHE) near the filling of $3$ electrons per moir\'e unit cell \cite{Sharpe2019,Serlin2020} in hBN aligned samples. Further studies on non-aligned samples \cite{Stepanov2021,Grover2022} have also observed AHE near 1 electron per moir\'e unit cell. Theoretically, such zero-field Chern insulating states (zCIs) have been proposed to be energetically competitive near magic angle, when the Coulomb interaction exceeds the narrow bandwidth,  even without the hBN alignment \cite{Ming2020,Kang2020,Soejima2020,TBGIV}.

An external magnetic field $B$, which preserves $C_{2z}$ but breaks $T$, has been argued to be an alternative way to reveal the band topology \cite{Nuckolls2020,Das2021,Wu2021}, as evidenced by the experimental observations of correlated Chern insulating states (CCIs) with a finite Chern number $t$ and extrapolating to a band filling $s$ at $B=0$ ~\cite{Tomarken2019,Park2021,Xie2019,Nuckolls2020,Pierce2021,Xie2021,Saito2021,Yu2022,Lu2019,Das2021,Stepanov2021,Wu2021,Choi2021,Grover2022,Uri2020}. Specifically, the most prominent sequence of CCIs has
$(s,t)=(0,\pm4),\pm(1,3),\pm(2,2),\pm(3,1)$, consistent with selective population of the aforementioned $B=0$ Chern $\pm 1$ bands \cite{Nuckolls2020,Stepanov2021,Wu2021}.  These experiments also report that some CCIs are stable down to $B\sim 1$Tesla, suggesting that they originate from the zCIs \cite{Nuckolls2020,Stepanov2021}.

However, CCIs are also observed in TBG devices away from the magic angle ($\sim1.27^\circ$), where the bandwidth of the narrow bands is expected to be significantly larger than at $1.05^\circ$, without any observation of the correlated insulators at $B=0$ \cite{Yankowitz2019}. Such CCIs appear only above a critical $B$, below which they transition into nearly compressible states for a fixed $(s,t)$. Similar phenomenology has also been reported in near-magic angle devices, leading to an alternative explanation of the CCIs invoking Stoner ferromagnetism within the magnetic subbands \cite{Saito2021,Yu2022,Choi2021,Park2021}, termed Hofstadter subband ferromagnets (HSFs) \cite{Saito2021}.  As argued theoretically \cite{Zhen2019,Kwan2021,XW2023}, realistic heterostrain can also increase the bandwidth dramatically near the magic angle, likely placing many TBG devices in the intermediate coupling regime where the zCIs may not be energetically favored.

To date, the nature of these CCIs remains poorly understood. No microscopic calculation favoring {\color{black} zCI, HSF or other states} has been carried out at $B\neq 0$, nor tying them to the relevant experiments. Moreover, the interplay of the CCIs with the competing states at $B=0$ near the magic angle, such as the intervalley coherent (IVC) states \cite{Po2018,Kang2019,Bultinck2020,Vafek2021,TBGIV}, the incommensurate Kekul\'e spiral (IKS) orders  \cite{Kwan2021}, and the stripe and nematic states \cite{Kang2020,Xie2021,LiuShang2021,FangXie2023}, remains unclear.

\begin{figure*}
\centering
\includegraphics[width=\linewidth]{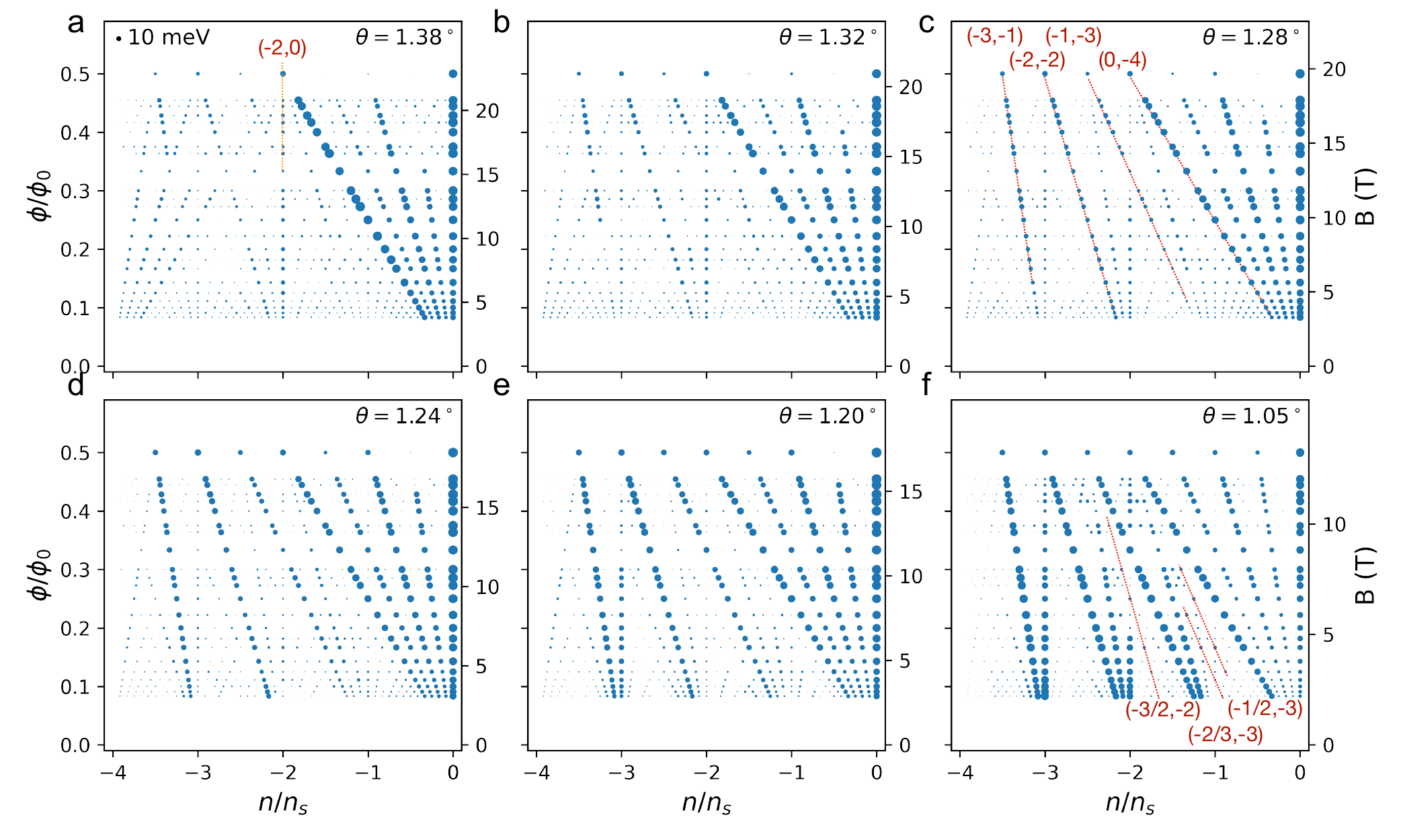}
\caption{\label{fig:StrainedPhaseDiagram} Single-electron excitation gap as a function of electron filling and magnetic flux in the presence of $0.2\%$ uniaxial heterostrain at twist angles $1.38^\circ$ (a), $1.32^\circ$ (b), $1.28^\circ$ (c), $1.24^\circ$ (d), $1.2^\circ$ (e), and $1.05^\circ$ (f). The size of the gaps are proportional to the radius of the respective solid circles, with the circle representation $10$meV shown in (a). The sequence of $(s,t)=(0,-4)$, $(-1,-3)$, $(-2,-2)$, and $(-3,-1)$ are labeled by red dashed lines in (c). Chern insulating states with fractional $s$ at $(s,t)=(-1/2,-3)$, $(-2/3,-3)$, and $(-3/2,-2)$ are labeled by dashed lines in (f). A quantum spin Hall insulating state is identified at high twist angles, and labeled by orange dashed line in (a).}
\end{figure*}

Here we report the first comprehensive study of the interacting electrons within the TBG narrow bands directly at $B\neq0$, and construct the phase diagram for a range of twist angles, $B$-fields and electron densities, with and without heterostrain. Consistent with the experimental observations, we find CCIs with $(0,\pm4)$,  $\pm(1,3)$,  $\pm(2,2)$, and $\pm(3,1)$, as shown in the Fig.~\ref{fig:StrainedPhaseDiagram} for the case with heterostrain {\color{black} and Fig.~\ref{fig:UnstrainedPhaseDiagram} for the case without heterostrain}. These figures plot the single particle excitation gap at the Fermi level obtained using the self-consistent Hartree-Fock method for each electron density and each magnetic field that we studied. In either strained or unstrained case, CCIs are found to be stablized at higher $B$ fields for twist angles as high as $1.38^\circ$ (highest twist angle studied in this work). Based on an analysis of their wavefunctions, we identify them as correlated Hofstadter ferromagnets (CHFs). Similar to HSFs, the CHFs correspond to selective population of the valley and spin flavors, but of the interaction-renormalized magnetic subbands (see Fig.~\ref{fig:120StrainedDetails}). {\color{black} Unlike HSFs however, CHFs may include -- but aren't limited to-- spin and/or valley polarized states which correspond to $B=0$ Chern insulators whose interaction renormalized bands are Landau quantized at $B\neq 0$. Although only metastable at $B=0$, such Chern insulators can be stabilized at $B\neq 0$ in the form of CHFs  as we demonstrate in the Fig.\ref{fig:HeavyLight}(d-e).} 

{\color{black} For realistic heterostrain (Fig.~\ref{fig:StrainedPhaseDiagram}), upon lowering $B$ and at a non-zero $s$ we find a phase transition into incompressible states with intervalley coherence. These states break the discrete magnetic translation symmetry but preserve the combination of the discrete magnetic translation and a $U_v(1)$ valley transformation (see Sec.~\ref{sec:iks}), and therefore are the finite $B$ analogs of the IKS states \cite{Kwan2021}, albeit carrying a non-zero Chern number $t$. Upon lowering $B$ and at larger twist angles, the IKS states transition into nearly compressible states; at lower twist angles they remain robust down to the lowest magnetic flux studied in this work. 

\begin{figure*}
\centering
\includegraphics[width=\linewidth]{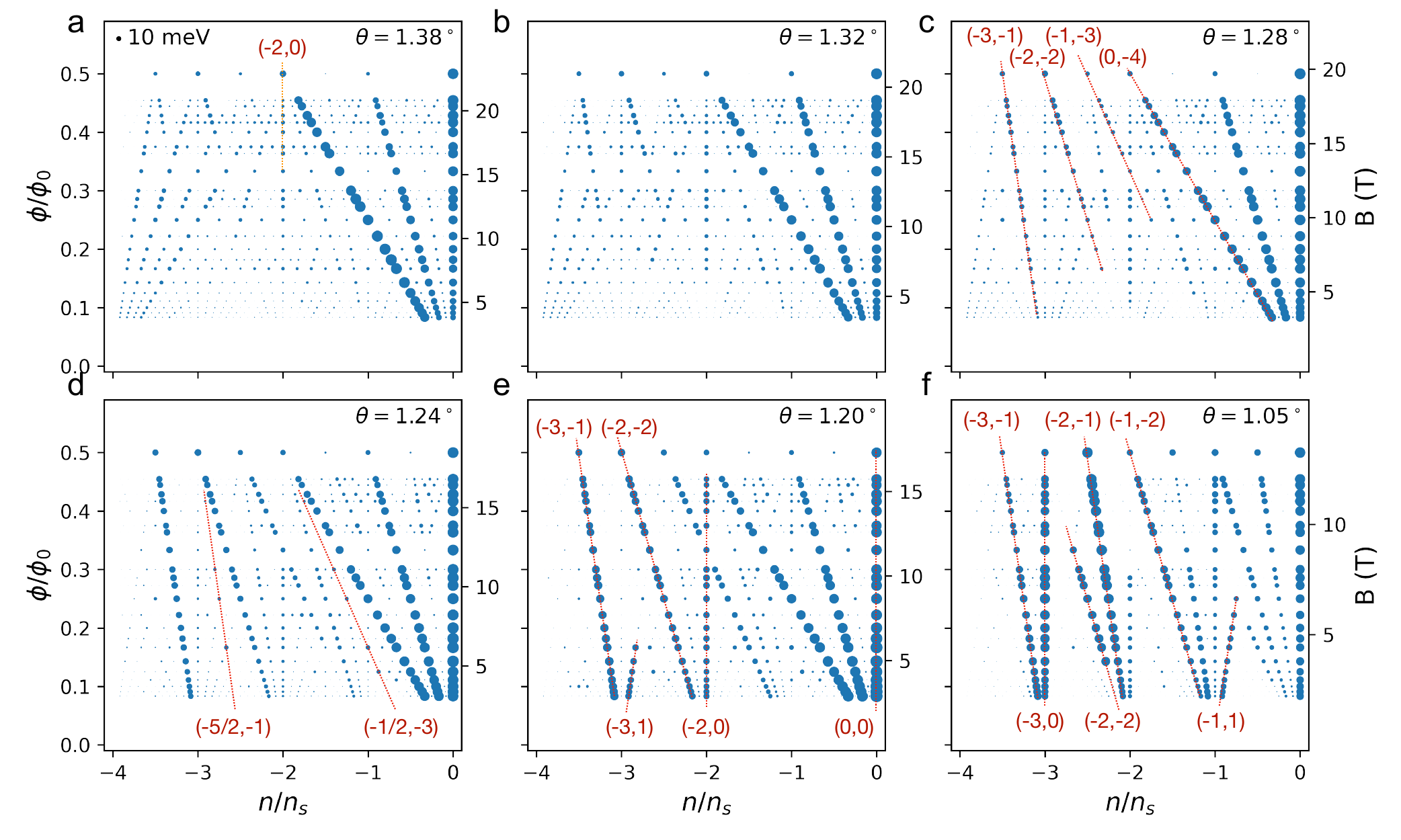}
\caption{\label{fig:UnstrainedPhaseDiagram} Single-electron excitation gap as a function of electron filling and magnetic flux in absence of heterostrain at twist angles $1.38^\circ$ (a), $1.32^\circ$ (b), $1.28^\circ$ (c), $1.24^\circ$ (d), $1.2^\circ$ (e), and $1.05^\circ$ (f). Compared to Fig.~\ref{fig:StrainedPhaseDiagram}, results at larger twist angles (upper panel) are similar, while noted differences emerge at lower twist angles. }
\end{figure*}

In the absence of heterostrain (Fig.~\ref{fig:UnstrainedPhaseDiagram}) and at larger twist angles we instead observe CHFs transitioning directly into nearly compressible states upon lowering $B$  without the intermediate IKS states. As we lower the twist angle toward $1.05^\circ$, the incompressible state at $\pm(3,1)$ extends to lower $B$ and crosses over into the finite $B$ analog of the zCI, approaching maximal sublattice polarization. We refer to this state as the strong coupling Chern insulator (sCI). Because there is no symmetry distinction between them, there is no true phase transition between CHF and the sCI, rather it is a smooth crossover as shown in the Fig.~\ref{fig:CHFSCI}(c,d). The $\pm(2, 2)$, $\pm(1, 3)$ and $(0, \pm 4)$ states also approach sCIs upon decreasing the twist angle, but they experience a first order phase transition into IVCs at low twist angles \cite{Po2018,Kang2019,Bultinck2020,Vafek2021,TBGIV}. IVCs break the $U_v(1)$ valley symmetry, but unlike the IKSs, they preserve the magnetic translation symmetry while also strongly hybridize the entirety of the narrow band Hilbert space.  The details of this transition depend sensitively on the model parameters, as shown in the Fig.~{\color{blue}S19}. 

Conversely, in the presence of heterostrain, the sCIs are absent in the Fig.~\ref{fig:StrainedPhaseDiagram}, as evidenced by the fact that the sublattice polarization remains low and far from saturating the sCI bound (Fig.~\ref{fig:CHFSCI} a,b). At the magic angle with heterostrain, we also identify Chern 0 IKS states along $(-3,0)$ and $(-2,0)$ (Fig.~\ref{fig:StrainedPhaseDiagram}f), consistent with what has been reported previously in $B=0$ Hartree-Fock calculations \cite{Kwan2021}. Unlike IKS states at a non-zero $t$, these states are less robust upon increasing $B$ and lose to nearly compressible states. Similar phenomenon has been reported experimentally in the Ref.~\cite{Saito2021,Stepanov2021,Wu2021,Das2022}.

In addition to these prominent CCIs, we also find gapped Chern states emanating from band edges and the charge neutrality point (CNP) as shown in the Figs.\ref{fig:StrainedPhaseDiagram}-\ref{fig:UnstrainedPhaseDiagram}. In the presence of heterostrain, they are quantum Hall ferromagnetic states (QHFMs) corresponding to spin-valley symmetry breaking states within a multi-flavored Landau level, with more details discussed in Figs.~{\color{blue}S6} and {\color{blue}S7}. Without heterostrain, gapped states emanating from the band edges are also identified as QHFMs (Fig.~{\color{blue}S13}). However, the gapped states emanating from the CNP assume a different character (Fig.~{\color{blue}S14}), developing intervalley coherence and a correlation gap at $B=0$ even at the largest twist angle studied. In contrast, QHFMs are field-induced symmetry breaking states, and are absent at $B=0$. At larger twist angles, a quantum spin Hall insulating state (QSH) along $(-2,0)$ is identified (see orange dashed line in Figs.~\ref{fig:StrainedPhaseDiagram}a and \ref{fig:UnstrainedPhaseDiagram}a), whose origin can be traced back to the spin-split non-interacting magnetic subbands at half flux quantum per moir\'e unit cell (Figs.~{\color{blue} S1} and {\color{blue} S11}). CCIs extrapolating to a fractional $s$ at $B=0$ can also be found at $1.05^\circ$ with heterostrain and for all twist angles in the absence of heterostrain. Further examination of their many body wavefunctions reveal that the magnetic translation symmetries are broken (Figs.~\ref{fig:FractionalCCI}, {\color{blue}S9}, {\color{blue}S16}), similar to the symmetry-broken Chern insulators (SBCIs) discussed in relation to the experiments of Ref.~\cite{Xie2021}. }

\section{Model and method}
We perform self-consistent Hartree-Fock analysis (B-SCHF) at $B\neq 0$ using the minimal continuum Hamiltonian (BM)\cite{Santos2007,Bistritzer2011}, with Coulomb interactions projected onto the narrow band Hilbert space. Here we briefly outline the formalism, additional details are in the Supplementary Information (SI). {\color{black} Our starting point is the (strained) BM Hamiltonian at rational magnetic flux ratios $\phi/\phi_0=p/q$ where $p$ and $q$ are coprime integers, $\phi$ is the magnetic flux per moir\'e unit cell and $\phi_0=h/e$ is the magnetic flux quantum. We choose a Landau gauge such that the magnetic vector potential $\bA(\br)=eB\hat{y}$, where $\hat{y}$ is defined along $\bL_2$ direction, with $\bL_{i=1,2}$ the two (strain-deformed) moir\'e unit cell vectors. The interacting Hamiltonian is invariant under discrete magnetic translation symmetries $\hat{t}_{\bL_1}(\br)=e^{-i2\pi \frac{\phi}{\phi_0} \frac{y}{|\bL_2|}} \hat{T}_{\bL_1}$ and $\hat{t}_{\bL_2}=\hat{T}_{\bL_2}$, where $\hat{T}_{\bL_{i=1,2}}$  denote discrete moir\'e translations at $B=0$. $\hat{t}_{\bL_1}$ and  $\hat{t}_{\bL_2}$ are non-commuting but satisfy $\comm{\hat{t}_{\bL_1}}{\hat{t}_{\bL_2}^q}=0$. This allows us to define a magnetic Brillouin zone described by the magnetic crystal momentum $\bk=k_1\bg_1+k_2\bg_2$, with $k_1\in[0,1)$ and $k_2 \in [0,1/q)$, and $\bg_{i=1,2}$ are moir\'e reciprocal lattice vectors (for more detailed information see SI {\color{blue}Sec.I} and {\color{blue} II}).}  We therefore first solve for the non-interacting Hofstadter spectra $\varepsilon_{\eta s r}(\bk)$ and associated eigenstates $\ket{\Psi_{\eta s r}(\bk)}$, with $\eta=\bK, \bK'$ and $s=\uparrow,\downarrow$ denoting valley and spin quantum numbers, and $r=1,\dots 2q$ is the magnetic subband index. Spin Zeeman splitting is also considered in this calculation. {\color{black}In earlier works \cite{XW2022,XW2023b}, we used the hybrid Wannier states (hWS) at $B=0$ to construct the finite $B$ Hilbert space. Although accurate and numerically efficient at low $B$, the hWS approach to TBG was shown to break down above moderate flux ratios (e.g., $\phi/\phi_0\gtrsim0.2$) \cite{XW2022}. As one of the purposes of this work is to connect the CCIs between low and high $B$, we instead solve the BM Hamiltonian by expanding in Landau level basis of each graphene layer. By scaling the upper Landau level cutoff as $\phi/\phi_0$ decreases, we ensure an accurate construction of the Hilbert space as well as the non-interacting Hofstadter spectra. While the high $B$ regime is straightforward in the LL basis, the low $B$ is computationally expensive. We were nevertherless able to reach $\phi/\phi_0=1/12$ using the LL basis, which corresponds to $\approx 2.2$Tesla at the magic angle (for all twist angle studied here we reached $B<4$Tesla). This value is therefore sufficiently low to make direct comparison with experiments. The (interacting) results shown in Figs.~\ref{fig:StrainedPhaseDiagram}-\ref{fig:UnstrainedPhaseDiagram} are for $1/12\leq \phi/\phi_0 \leq 1/2$, with the maximum value of $q$ being $12$ and $1<p<q$.}

For the model parameters studied in this work, the gap to remote Hofstadter bands does not close at the magnetic fluxes of interest. We study interaction effects by projecting the screened Coulomb interaction onto the narrow band Hilbert space. The Hamiltonian is given by: 
\begin{equation}
    H = \sum_{\eta s r, \bk} \varepsilon_{\eta s r}(\bk) d^{\dagger}_{\eta s r,\bk}d_{\eta s r,\bk} + \frac{1}{2A}\sum_{\bq} V_{\bq} \delta \hat{\rho}_{\bq} \delta \hat{\rho}_{-\bq}.
\end{equation}
Here $A$ is the total area of the system, $d_{\eta s r,\bk}$ is the electron annihilation operator, $\delta \hat{\rho}_\bq$ is the Fourier transform of the electron density operator projected onto the narrow bands, subtracting a background charge density \cite{TBGIII,Kang2020b,XW2022}. It is given by: 
\begin{equation}
\begin{split}
    \delta \hat{\rho}_{\bq} & = \sum_{\eta s, r r',\bk,\bp} \bra{\Psi_{\eta s r}(\bk)} e^{-i\bq \cdot \br } \ket{\Psi_{\eta s r'}(\bp)} \\
    & \times \left( d^{\dagger}_{\eta s r,\bk}d_{\eta s r', \bp} - \frac{1}{2}\delta_{r,r'}\delta_{\bk,\bp}\right).
\end{split}
\end{equation}
We consider a dual-gate screened Coulomb interaction of the form $V(\bq) = \frac{2\pi e^2}{\epsilon_0\epsilon_r |\bq|}\tanh\left(\frac{|\bq|\xi}{2}\right)$, with relative dielectric constant $\epsilon_r=15$ and screening length $\xi=4\sqrt{|\bL_1||\bL_2|}$. These parameters are chosen to match the overall change of chemical potential from empty to full occupation of the narrow bands in magic angle devices as extracted from the compressibility measurements  as well as STM \cite{Tomarken2019,Zondiner2020,Wong2020,Park2021,Saito2021b} (see also Fig.~{\color{blue} S2}).

{\color{black}
In the B-SCHF procedure, we minimize the total energy for a fixed particle number using many body wavefunctions expressable as product states: 
\begin{equation} \label{eq:hf_WF}
\begin{split}
    \ket{\Omega} = \prod_{n,\bk}^\prime &  \left(\sum_{s r}\alpha_{s r,\bk}^{(n)} d^{\dagger}_{\bK s r,\bk} \right. \\ 
    & \left. + \sum_{s'r'}\beta^{(n)}_{s'r',\bk+\bq_0}d^{\dagger}_{\bK' s'r',\bk+\bq_0}\right) \ket{0},
\end{split}
\end{equation}
where $\bq_0$ is an arbitrary wavevector shift between single electron states in opposite valleys, whose value is constrained by the discrete momentum mesh such that $\bk+\bq_0$ is on the same momentum mesh as $\bk$ (modulo a reciprocal lattice vector). The constrained product $\prod^\prime_{n,\bk}$ is over all the occupied states labeled by $n,\bk$. $\{\alpha^{(n)}_{s r,\bk},\beta^{(n)}_{s'r',\bk+\bq_0}\}$ are variational parameters that minimize the total energy, and they obey $\sum_{s r} |\alpha_{s r,\bk}^{(n)}|^2 + \sum_{s'r'}|\beta_{s'r',\bk+\bq_0}^{(n)}|^2=1$ for any $\{n,\bk\}$. The total energy is also optimized with respect to $\bq_0$, allowing us to probe IKS-like states (see SI {\color{blue}Sec.~III D}). An equivalent formulation of the B-SCHF procedure  (for details see SI {\color{blue}Sec.~III}) is based on the one-particle density matrix:
\begin{equation} \label{eq:den_mat}
    \hat{Q}_{\eta s r,\eta' s' r'}(\bk) \equiv \bra{\Omega} \tilde{d}^{\dagger}_{\eta s r,\bk}\tilde{d}_{\eta' s' r',\bk} \ket{\Omega},
\end{equation}
where for notational convenience we defined $\tilde{d}^{\dagger}_{\bK s r,\bk} \equiv {d}^{\dagger}_{\bK s r,\bk}$ and $\tilde{d}^{\dagger}_{\bK' s' r',\bk} \equiv {d}^{\dagger}_{\bK' s' r',\bk+\bq_0}$. Note that $\hat{Q}$ contains information about $\bq_0$. In the remaining text we discuss the numerical results based on either the one-particle density matrix or the many body wavefunction whichever is more convenient.
}

The projected Hamiltonian at $B\neq0$ is invariant under the following set of symmetries \cite{Hejazi2019,TBGIII,Jonah2022}: $C_{2z}$, valley $U_v(1)$ and spin $U_s(1)$, many-body particle-hole $P$ , magnetic translation 
symmetries generated by $\hat{t}_{\bL_1}$ and $\hat{t}_{\bL_2}$ . {\color{black} We fix the gauge such that 
\begin{align} \label{eq:mtg}
    \hat{t}_{\bL_1}d^{\dagger}_{\eta s r,\bk}\hat{t}_{\bL_1}^{-1} & =e^{-i2\pi k_1}d^{\dagger}_{\eta s r,\bk},\\
    \hat{t}_{\bL_2}d^{\dagger}_{\eta s r,\bk}\hat{t}_{\bL_2}^{-1} & =e^{-i2\pi k_2}d^{\dagger}_{\eta s r,\bk+{\phi}/{\phi_0}\bg_1}.
\end{align}
The above gauge choice is useful in identifying a magnetic translation symmetry breaking from the density matrix (see SI). 
}
In the absence of heterostrain, $C_{3z}$ and $C_{2y}T$ also leave $H$ invariant. $P$ guarantees symmetry about the charge neutrality point, and therefore we present our results for the hole filling only.

The B-SCHF calculation is carried out for a range of twist angles from $1.38^\circ$ to $1.05^\circ$, both for unstrained model as well as for realistic uniaxial heterostrain magnitude $\epsilon=0.2\%$ and orientation $\varphi=0^\circ$ (see SI and Ref.~\cite{XW2023} for the definition of the uniaxial strain orientation). We choose the Fermi velocity such that $\hbar v_F/a = 2482$meV with the graphene lattice constant $a\approx 2.46\AA$, and interlayer tunneling parameters $w_0=77$meV (intra-sublattice) and $w_1=110$meV (inter-sublattice). These parameters place the magic angle near $1.05^\circ$. The respective non-interacting Hofstadter spectra and Wannier diagrams with/without heterostrain are shown in Fig.~{\color{blue}S1} and {\color{blue}S11}.

{\color{black} At a given $B$ and electron density, obtaining the true Hartree-Fock ground state is a highly non-trivial task due to competing states of similar energy. We typically run $\sim 6$ random initializations of the single-particle density matrix, as well as several educated guesses (e.g., flavor polarized or intervalley coherent states), and report the lowest-energy state as the ground state. Optimal damping algorithm is also used to speed up the numerical convergence \cite{Bultinck2020}. This elaborate procedure turns out to be adequate (i.e. random and educated initializations converge to the same state) for establishing incompressible ground states with big excitation gaps, such as the CCIs with $(s,t)=(0,\pm4)$ $\pm(1,3)$, $\pm(2,2)$ and $\pm(3,1)$, QHFMs, and CCIs extrapolating to a fractional $s$. However it may face convergence issues for nearly compressible states, which are abundant in the phase diagrams shown in Figs.~\ref{fig:StrainedPhaseDiagram} and \ref{fig:UnstrainedPhaseDiagram}. We make no assertions regarding the nature of such nearly compressible states in this paper and mainly use them in order to highlight the contrast with the ground states with large excitation gaps or, at $B=0$, to elucidate the physics of the $B$-induced incompressible states as in the Fig.~\ref{fig:HeavyLight}(a-c).
We often find competing states close in energy, with $\lesssim 1\rm meV$ difference in the Hartree-Fock energy per moir\'e unit cell. Throughout the paper we try to adopt the philosophy of identifying these competing states \cite{Balents2020}, and comment on how small variations in model parameters (e.g., dielectric constant, kinetic terms beyond BM Hamiltonian) may tip the balance between ground states and metastable states.
}
\section{Results} 
\subsection{Phase diagram and main CCIs}
We first address the finite $B$ phase diagram for TBG subject to $0.2\%$ of heterostrain. Fig.~\ref{fig:StrainedPhaseDiagram} gives an overview of the calculated single particle excitation gap {\color{black}(i.e. the charge gap)} as a function of moir\'e unit cell filling ($n/n_s$) and magnetic flux ratio ($\phi/\phi_0$) for six twist angles $1.38^\circ$, $1.32^\circ$, $1.28^\circ$, $1.24^\circ$, $1.20^\circ$ and $1.05^\circ$. The size of the gap is proportional to the radius of the solid circle. As seen, there is a rich panoply of correlated insulating states. We start by focusing on the sequence of CCIs with $(s,t)= (0,-4),(-1,-3),(-2,-2),(-3,-1)$, which are observed for all the twist angles studied, and marked by red dashed lines in Fig.~\ref{fig:StrainedPhaseDiagram}(c). At larger twist angles,  CCIs along $(-3,-1),(-2,-2),(-1,-3)$ emerge at high $\phi/\phi_0$, and are replaced by nearly compressible states at lower $\phi/\phi_0$ via a first order phase transition. As the twist angle is lowered, they become more robust and can persist beyond the lowest flux ratio of $1/12$ studied in this work.  

\begin{figure*}
\centering
\includegraphics[width=\linewidth]{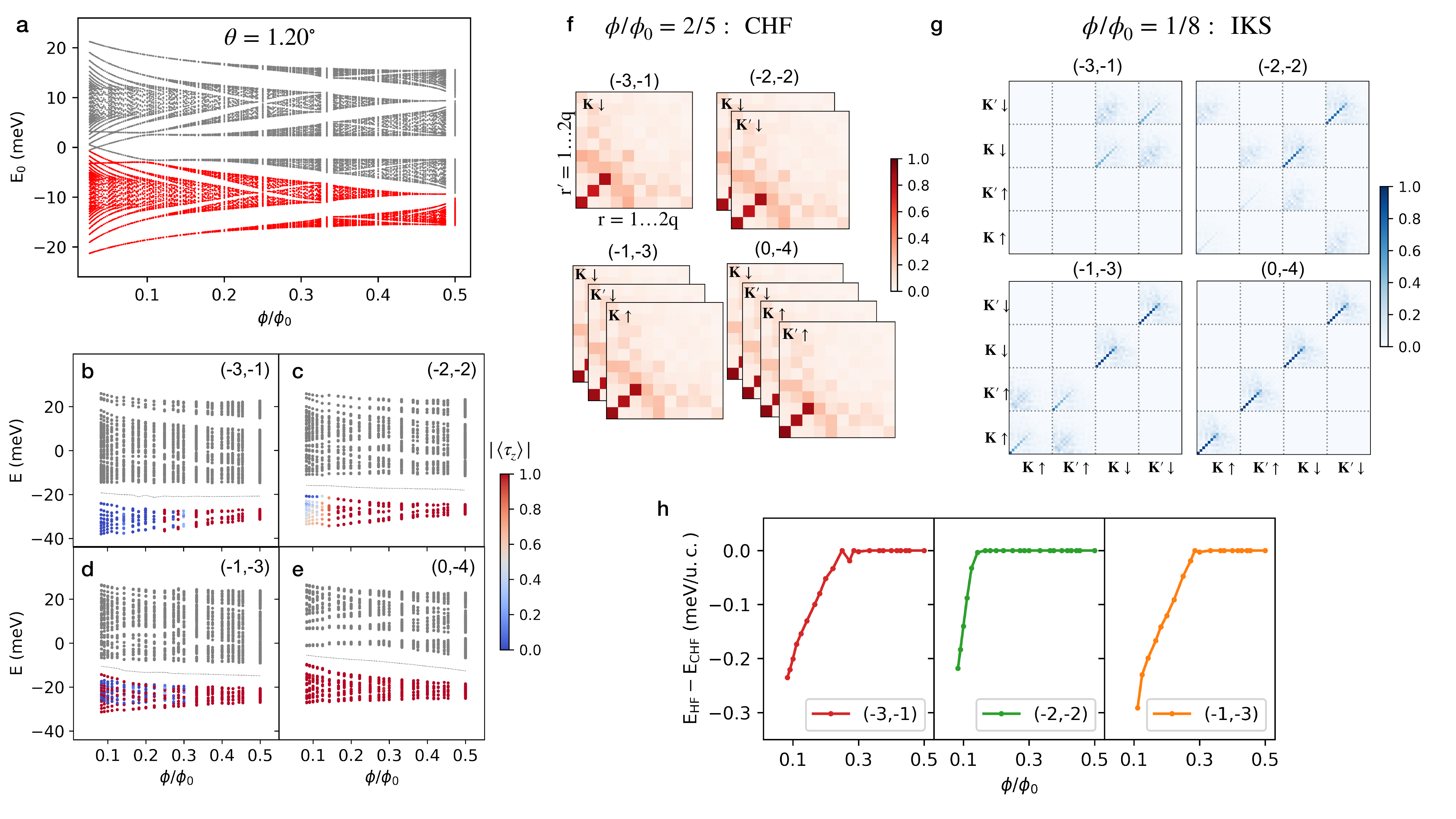}
\caption{\color{black} Compilation of B-SCHF results at $\theta=1.2^\circ$. (a) The non-interacting spectra. The group of magnetic subbands colored in red has total Chern number $-1$. (b-e) The self-consistent Hartree-Fock spectra along $(s,t)=(-3,-1)$, $(-2,-2)$, $(-1,-3)$ and $(0,-4)$. The occupied electronic states are colored by their valley polarization $|\langle \tau_z\rangle|$. $|\langle \tau_z\rangle|\rightarrow 1 (0)$ implies maximal valley polarization (maximal intervalley mixing). The alternating colors along $(-3,-1)$ near $\phi/\phi_0\approx 0.3$ are due to near energetic degeneracies ($\sim 0.05$meV) between competing states, and are not resolved within our calculation. Along $(-1,-3)$ and at $\phi/\phi_0< 0.3$, the occupied single-electronic states are maximally valley polarized in the down spin sector, and maximally intervalley mixing in the up spin sector. One-particle density matrix at higher $B$ (f) and lower $B$ (g) showing either valley and spin polarized state (CHF) or intervalley coherent state (IKS). (h) shows the Hartree-Fock energy difference per moir\'e unit cell between the ground state and the (metastable) CHF state along $(-3,-1)$, $(-2,-2)$ and $(-1,-3)$.}
\label{fig:120StrainedDetails}
\end{figure*}

{\color{black} 

To better elucidate their nature, we compile detailed results for a representative twist angle $1.20^\circ$ in Fig.~\ref{fig:120StrainedDetails}. Results for other twist angles can be found in the SI. Fig.~\ref{fig:120StrainedDetails}(a) shows the non-interacting spectra of valley $\bK$ and for one spin component (neglecting Zeeman effect). The magnetic subbands marked in red denote the Chern -1 group below the charge neutrality point. The analogous group of subbands above the charge neutrality point is related to it by particle hole symmetry, and also carries total Chern number $-1$. The remaining two subbands emanate either from the zeroth Landau levels~(zLLs) of the energetically split Dirac points ($\phi/\phi_0\gtrsim 0.1$) or the $\pm 1$ LLs ($\phi/\phi_0\lesssim 0.1$), and carry Chern number $+1$ each, such that the total Chern number of all magnetic subbands is zero. 

Fig.~\ref{fig:120StrainedDetails}(b) shows the single-particle spectra including Coulomb interactions along $(s,t)=(-3,-1)$, where the occupied states are colored according to their valley polarization $|\langle \tau_z \rangle|$, where $\tau_z$ is the Pauli matrix acting on valley degrees of freedom. $|\langle \tau_z \rangle|\rightarrow 1 (0)$ implies maximal valley polarization (maximal intervalley mixing). At $\phi/\phi_0\gtrsim 1/3$, the occupied states are maximally valley polarized, and have a large overlap onto the states marked in red in Fig.~\ref{fig:120StrainedDetails}(a).  To quantify the overlap we make use of the density matrix defined in Eq.~(\ref{eq:den_mat}), which has the spin-valley diagonal form $\mathcal{Q}_{r,r'}^{(\eta s)}(\bk)\delta_{\eta,\eta'}\delta_{s,s'}$ for a state with unbroken valley and spin symmetries. A representative $|\mathcal{Q}^{\bK\uparrow}_{r,r'}(\mathbf{0})|$ is shown in Fig.~\ref{fig:120StrainedDetails}(f). It is predominantly diagonal in the magnetic subband index, mostly occupying the lower $q-p$ magnetic subbands, i.e., the group states with total Chern number -1 marked red in Fig.~\ref{fig:120StrainedDetails}(a). For $(-2,-2)$ and $(-1,-3)$ and at higher $B$, the CCIs are also valley and spin polarized, similarly mostly populating the lower Chern -1 group of magnetic subbands for the specified valley and spin, with $|\mathcal{Q}^{(\eta s)}(\mathbf{0})|$ identical for all occupied flavors. Although these CCIs are closely related to the HSFs discussed in Ref.~\cite{Saito2021,Park2021,Choi2021}, the band structure renormalization is apparent in the non-vanishing off-diagonal matrix elements of $\mathcal{Q}^{(\eta s)}_{r,r'}(\bk)$, signifying hybridization with the higher energy subbands (marked by grey in Fig.~\ref{fig:120StrainedDetails}(a)). For this reason we refer to them as CHFs. As further demonstrated in Fig.~{\color{blue} S3}, the density matrices of the CHFs assume a much simpler structure when expressed in the eigenbasis of the valley and spin symmetric $(0,-4)$ Chern insulating state, which limits to an interaction-renormalized semimetal at $B=0$ (see Fig.~{\color{blue} S2} and also Ref.~\cite{Kwan2021}).

At lower $B$, the valley and spin polarized CCIs along $(-3,-1)$, $(-2,-2)$ and $(-1,-3)$ all transition into gapped states with strong intervalley mixing, as reflected by $|\langle \tau_z \rangle| \rightarrow 0$ for the occupied states shown in Figs.~\ref{fig:120StrainedDetails}(b-e). Along $(-3,-1)$ and $(-1,-3)$, the intervalley mixing is between the same spin species (Fig.~\ref{fig:120StrainedDetails}g), and the resulting states do not suffer from the Zeeman energy cost compared to CHFs which have the same spin polarization. However, along $(-2,-2)$ the intervalley mixing is between opposite spins (mixing between the same spins is achieved as a metastable state), leading to an extra Zeeman energy cost. This likely explains the lowered critical field for the transition between the CHF and the IKS states for $(-2,-2)$ compared to $(-3,-1)$ and $(-1,-3)$, as seen in Figs.~\ref{fig:120StrainedDetails}(b-e) by the valley polarization of occupied single-electron states. Based on a detailed analysis of the density matrix we identify these intervalley coherent CCIs as the finite $B$ analog of IKS states \cite{Kwan2021} carrying a finite Chern number. We postpone a detailed discussion of identifying IKS states to Sec.~\ref{sec:iks}. Finally in Fig.~\ref{fig:120StrainedDetails}h we show the Hartree-Fock energy difference between the ground state and the CHFs which become metastable at lower $B$. It shows that the phase transition between CHF and IKS is likely first order along $(-3,-1)$ and $(-1,-3)$, where the IKS order parameter --- qualitatively captured by the $|\hat{Q}_{\bK s r,\bK' sr'}(\bk)|$, has an abrupt onset upon lowering $B$. On the other hand, the transition along $(-2,-2)$ is most likely second order, as $|\hat{Q}_{\bK s r,\bK' sr'}(\bk)|$ gradually increases as $B$ decreases.  

The qualitative picture described above is universal across all the twist angles we have studied (see Fig.~{\color{blue}S4}). However at larger twist angles, in addition to the phase transition between CHF and IKS, upon further decreasing $B$ we observe a transition from a gapped IKS into a nearly compressible state, e.g., along $(-3,-1)$ at $1.38^\circ$ as shown in Fig.~\ref{fig:StrainedPhaseDiagram}(a). This can be attributed to the larger non-interacting bandwidth and comparatively weaker Coulomb interaction.
}

\begin{figure}[t]
\centering
\includegraphics[width=\linewidth]{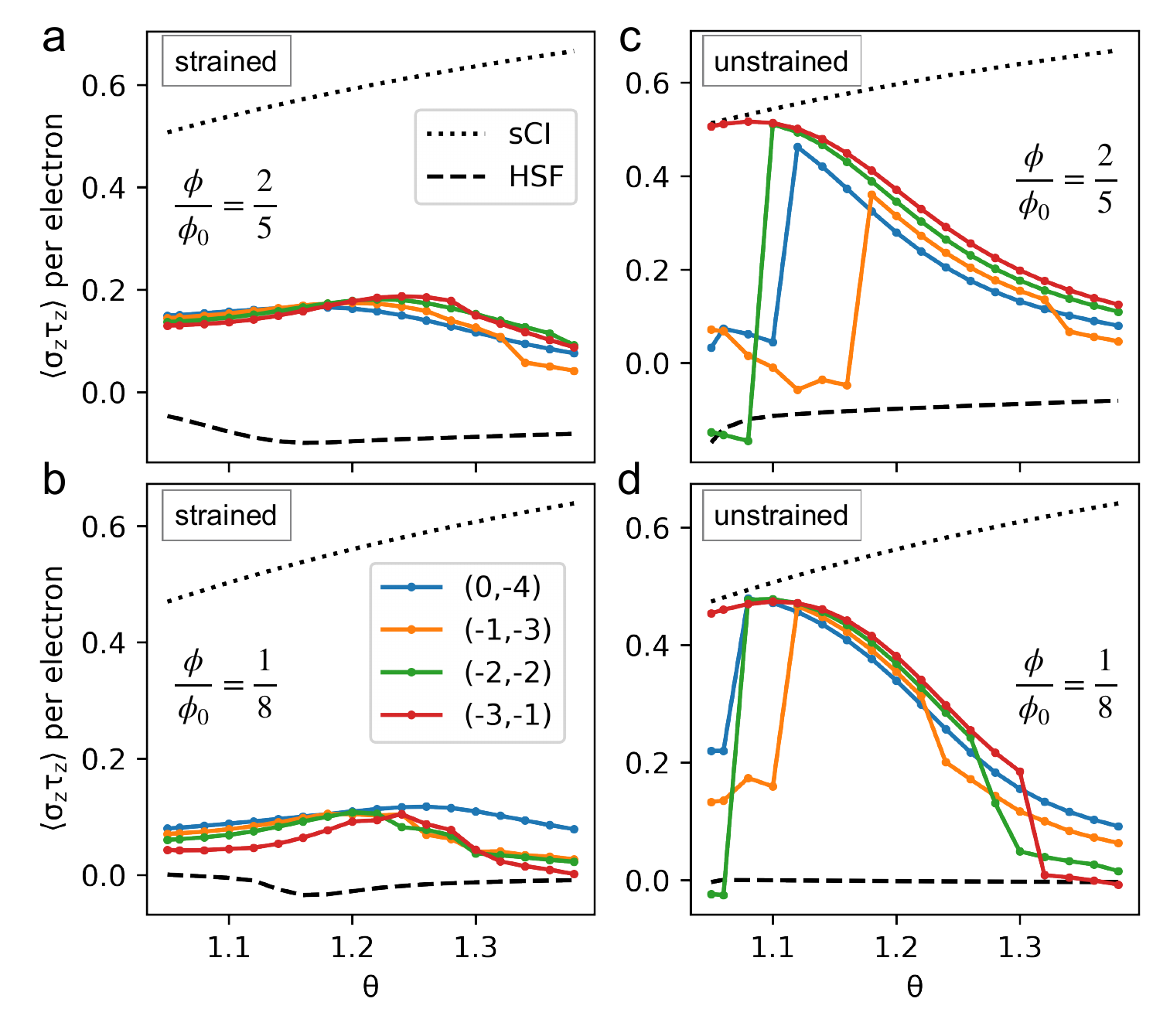}
\caption{Averaged $\sigma_z\tau_z$ per occupied single electron states at $(s,t)=(0,-4)$, $(-1,-3)$, $(-2,-2)$ and $(-3,-1)$ for (a) $\phi/\phi_0=2/5$ with heterostrain, (b) $\phi/\phi_0=1/8$ with heterostrain, (c) $\phi/\phi_0=2/5$ without heterostrain, and (d) $\phi/\phi_0=1/8$ without heterostrain. The dotted line in each panel is the value for sCI, and the dashed line corresponds to HSF.}
\label{fig:CHFSCI}
\end{figure}

{\color{black}
\subsection{IKS states} \label{sec:iks}
In previous Hartree-Fock studies at $B=0$, a notable finding is that in the presence of a small amount of uniaxial heterostrain (e.g., $0.2\%$), the ground state at integer filling fractions are the incommensurate Kekul\'e spiral ordered states \cite{Kwan2021}. Unlike the IVC states achieved when heterostrain is absent \cite{Po2018,Bultinck2020}, IKS states predominantly mix the lower energy bands (instead of the full narrow band Hilbert space) between opposite valleys of the non-interacting BM Hamiltonian, while developing a wavevector $\bQ_{\text{IKS}}$ incommensurate with the underlying moir\'e lattice. On the moir\'e scale, a modified discrete translation symmetry is preserved, i.e., discrete moir\'e translations ($\hat{T}_{\bL_j=1,2}$) followed by a $U_v(1)$ valley transformation: $e^{-\frac{i}{2}(\bQ_{\text{IKS}}\cdot\bL_j)\tau_z}\hat{T}_{\bL_j}$.

In our calculations, we find that the gapped states along $(-3,-1)$, $(-2,-2)$ and $(-1,-3)$ at lower $B$ [Fig.~\ref{fig:120StrainedDetails}(b-e,g)] also hybridize electronic states predominantly of the lower Chern $-1$ group of the non-interacting magnetic subbands, break the magnetic translation symmetries generated by $\hat{t}_{\bL_{j=1,2}}$, but preserve a modified translation $e^{-\frac{i}{2}(\bQ_{\text{IKS}}\cdot\bL_j)\tau_z}\hat{t}_{\bL_j}$. This further constrains the many body wavefunction in Eq.~(\ref{eq:hf_WF}) and the density matrix in Eq.~(\ref{eq:den_mat}), with details discussed in SI {\color{blue}Sec.III.D}. Therefore, we can identify them as the finite $B$ analog of the zero $B$ IKS states. At larger twist angles, we only find gapped IKS states with a finite Chern number (i.e. non-zero $t$),  and an absence of Chern 0 IKS states. The latter only occur close to the magic angle [e.g. Fig.~\ref{fig:StrainedPhaseDiagram}(e,f)] along $(-3,0)$ and $(-2,0)$. At higher $B$, these Chern 0 IKS states become energetically unfavorable and lose to nearly compressible states. This phenomenon has been reported in various experiments ~\cite{Saito2021,Stepanov2021,Wu2021,Das2022}.

\subsection{Robustness of the CHF-IKS phase transition against perturbations}
Given the small energy differences between the CHF and IKS states along $(-3,-1)$, $(-2,-2)$ and $(-1,-3)$ shown in Fig.~\ref{fig:120StrainedDetails}(h), it is natural to ask how robust is this phase transition to small variations in model parameters. Although this is difficult to answer definitively, it is possible to present some qualitative arguments favoring such a phase transition. 
Along $(-2,-2)$, as argued above, Zeeman coupling is expected to favor a CHF state, as it is fully spin polarized, while the IKS is not. The Zeeman energy cost of both CHF and IKS can be computed via $g_s\mu_B \tr \{\hat{Q}s_z\}$, where $g_s=2$, $\mu_B$ is the Bohr magneton, and $s_z$ is the z-component of the spin. We find that numerically switching off the Zeeman splitting increases the critical field at which the CHF-IKS transition occurs. Interestingly, the same argument does not hold along $(-3,-1)$ and $(-1,-3)$, as IKS does not suffer from extra Zeeman energy cost compared to CHF. It would appear that the relative strength of the Coulomb interaction and the non-interacting bandwidth of the relevant magnetic subbands (red colored states in Fig.~\ref{fig:120StrainedDetails}a) also plays an important role. We find numerical evidence that decreasing the Coulomb interaction (e.g., changing the relative dielectric constant from 15 to 25) makes IKS more stable and increases the critical field, conversely increasing Coulomb interaction favors a CHF state. Finally, there is also evidence that increasing the strength of the uniaxial heterostrain (e.g., from $0.2\%$ to $0.3\%$) favors IKS over CHF, and increases the critical field. 

In a recent experiment, the spin polarizations of the main CCIs near the magic angle are identified by edge conductance measurements, with $(-3,-1)$ and $(2,2)$ being spin polarized, and $(-2,-2)$ being spin unpolarized \cite{Hoke2023}. 
Our theory naturally recovers the spin polarization along $(-3,-1)$. While CHF is fully spin polarized, as mentioned, the IKS is not. Therefore, 
spin polarization along $\pm(2,2)$ could be used to differentiate between CHF and IKS states as shown in Fig.~\ref{fig:120StrainedDetails}(f,g). The experimental results therefore suggest an IKS state along $(-2,-2)$ and CHF along $(2,2)$, although to fully explain them, more careful analysis of the edge modes and modeling beyond the BM Hamiltonian ~\cite{Kang2023b} which captures the particle-hole asymmetry are necessary.
}

\subsection{Crossover to strong coupling regime}
In order to clarify the connection between the CHFs and the sCIs, we first note that the sCIs saturate the expectation value of $ \sigma_z\tau_z$ for the occupied electronic states, where $\sigma_z$ and $\tau_z$ are Pauli matrices acting in the sublattice and valley subspace respectively \cite{Bultinck2020,Kang2020,TBGIII,XW2022}.  The solid lines in
Fig.~\ref{fig:CHFSCI}(a) and (b) show calculated $\langle \sigma_z\tau_z\rangle$ per electron as a function of twist angle in the presence of heterostrain along $(0,-4)$, $(-1,-3)$, $(-2,-2)$, and $(-3,-1)$, for $\phi/\phi_0=2/5$ and $1/8$ respectively. For comparison, the upper dashed line corresponds to the sCI limit, and the lower dashed line to the HSF limit. {\color{black} This measure shows that the CHFs are quantitatively different from both the HSFs and the sCIs in the presence of heterostrain, but closer to HSFs. Intuitively, the increased $\langle\sigma_z\tau_z\rangle$ of CHFs compared to HSFs may originate in the short-ranged part of the Coulomb repulsion disfavoring two electrons sitting close to each other in real space.}

On the other hand, in the absence of heterostrain, as shown in Fig.~\ref{fig:CHFSCI}(c) and (d), the CHFs along $(-3,-1)$ smoothly cross over into the sCIs upon lowering the twist angle. For our model parameters, there is a collapse of the $\langle \sigma_z\tau_z\rangle $ along $(-2,-2)$ $(-1,-3)$ and $(0,-4)$ at lower twist angles, when the CHFs become energetically less favorable than populating the Landau quantized excitation spectra of the IVC states \cite{XW2022,Po2018,TBGIV,Keshav2023}. As further demonstrated in Fig.~{\color{blue} S19}, this transition depends sensitively on model parameters, and can be pushed toward lower $\phi/\phi_0$ (e.g., by moving toward the chiral limit, see SI and Ref.~\cite{Tarnopolsky2019}). Refs.~\cite{Stepanov2021,Nuckolls2020} report that the $(2,2)$ persist down to $\phi/\phi_0\sim 1/25$, and therefore argue that they correspond to the zCIs (more precisely, sCIs). Our quantitative calculations clearly demonstrate that such states can indeed be stablized at weak fields by small changes of the model parameters. 

{\color{black} In our earlier work \cite{XW2022} we computed the single particle excitations of IVC insulators in the strong coupling limit at $B\neq 0$, which have been demonstrated to be the ground states  at the CNP and $n/n_s=-2$. At the CNP and at low $B$ we find two-fold degenerate Landau levels (LLs) $0, \pm2,\pm 4,\dots$. At $n/n_s=\pm 2$ we find instead $0,\pm1,\pm2,\dots$. These have been corroborated by other works \cite{TBGV,Keshav2023}. Although the results in Ref.~\cite{XW2022} assumed adding a single electron or a single hole, they are expected to hold at an asymptotically low $B$ even at a small but finite density along $(0,t)$ or $(\pm2,t)$.  This is because the energy difference between competing many-body states due to addition of a small density of carriers necessary to fill the excited LLs is expected to be proportional to the number of flux quanta, while the energy difference between competing states at $B=0$ is extensive and thus proportional to the total particle number. Therefore, a finite critical $B$-field would be necessary to tip the energy balance in favor of a state such as the sCI, distinct from the one obtained by a naive rigid filling of the LLs. Closer examination of Fig.~\ref{fig:UnstrainedPhaseDiagram}(f) indeed demonstrates this. At $n/n_s=-2$, we find a prominent $(-2,-1)$ which corresponds to emptying one energetically well-separated LL from the spectra of the $(-2,0)$ IVC state, consistent with our earlier studies. Along $(-2,-2)$ and at $\phi/\phi_0<1/7$, a weaker gapped state is observed where two LLs are emptied. However at $\phi/\phi_0>1/7$ the sCI is stabilized via a first order phase transition.  At $n/n_s=0$, we similarly find a prominent $(0,-2)$ corresponding to emptying a two-fold degenerate LL of the $(0,0)$ state (with $(0,-1)$ being QHFM of $(0,-2)$). Interestingly, gapped states along $(0,-4)$ -- expected based on the results in Ref.~\cite{XW2022}-- are not observed down to $\phi/\phi_0=1/12$. Given that this corresponds to an electron filling of $n/n_s=-1/3$, we attribute the absence of $(0,-4)$ to band renormalization effects at finite electron densities not captured in Ref.~\cite{XW2022}. Fig.~\ref{fig:UnstrainedPhaseDiagram}(f) also shows gapped states emanating from $n/n_s=-3$ and $-1$ that can be characterized either as sCIs [$(-3,-1)$, $(-1,\pm 1)$] or via (de)population of the sCIs' Landau quantized excitation spectra [$(-3,0)$, $(-1,-2)$, $(-1,0)$]. We refer interested readers to details presented in Fig.~{\color{blue}S18}. We note that in the strong coupling limit there are near degeneracies between competing states. For example, along $(-1,-3)$, a Chern $-3$ sCI can be found as a metastable state, with a Hartree-Fock energy $\lesssim 0.01 \rm meV$ higher than the nearly compressible state plotted in Fig.~\ref{fig:UnstrainedPhaseDiagram}(f). }

\begin{figure*}
\centering
\includegraphics[width=\linewidth]{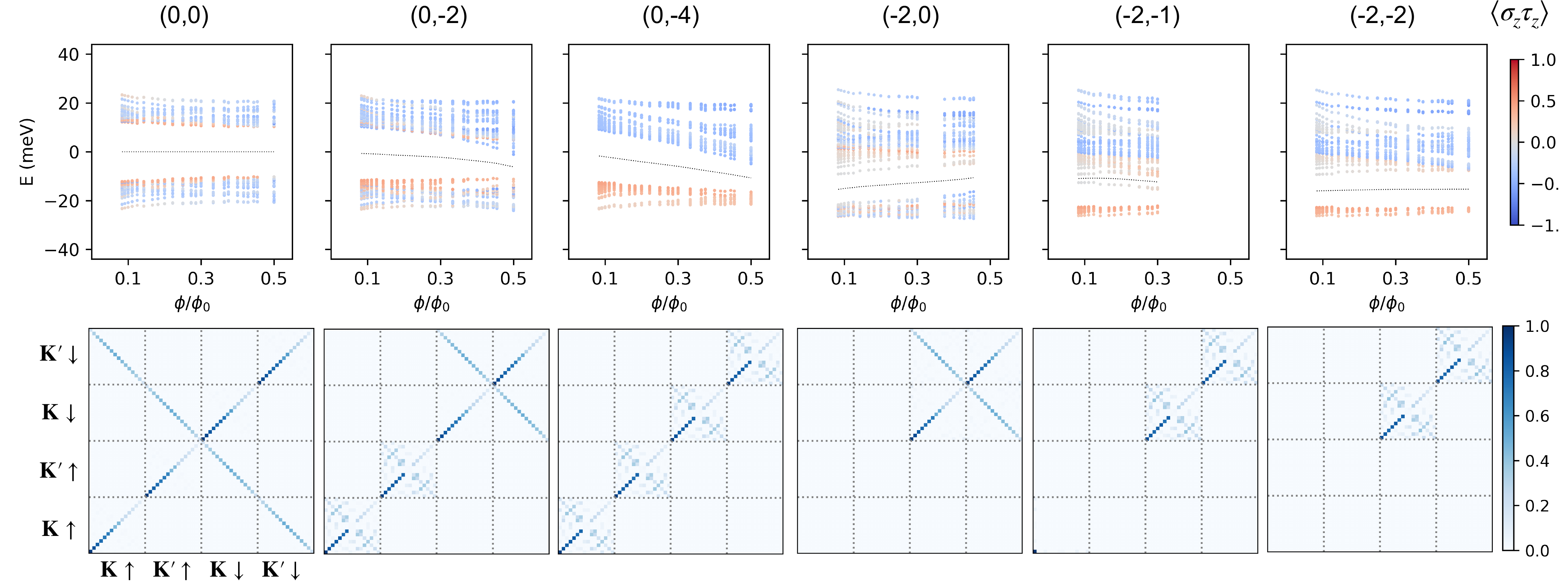}
\caption{\color{black} Hartree-Fock excitation spectra (upper panel) and representative density matrices $|\hat{Q}_{\eta sr,\eta' s'r'}(\mathbf{0})|$ calculated at $\phi/\phi_0=1/8$ (lower panel) along $(0,0)$, $(0,-2)$, $(0,-4)$, $(-2,0)$, $(-2,-1)$ and $(-2,-2)$, calculated for $1.20^\circ$ in the absence of heterostrain. Electronic states below the dashed lines in the upper panel are occupied, and the color coding represents the averaged $\sigma_z\tau_z$ of every single-electron state. In panels $(-2,0)$ and $(-2,-1)$ excitation spectra at several $\phi/\phi_0$ are omitted in the plot because they are states of different characters, either due to Landau fan crossing or the $B$-induced phase transitions as shown in Fig.~\ref{fig:UnstrainedPhaseDiagram}e. The density matrix structure for each column is as follows: Along $(0,0)$ and $(-2,0)$, the occupied electronic states form an IVC state, between opposite spin species along $(0,0)$, and between the spin $\downarrow$ species along $(0,-2)$ which is favored by Zeeman splitting. Along $(0,-4)$ and $(-2,-2)$ they are the CHF states. Along $(0,-2)$, it is a partial IVC state with IVC in the spin $\downarrow$ sector and a net Chern -2 valley polarized state in the spin $\uparrow$ sector. Along $(-2,-1)$, it is a valley/spin polarized state, populating one extra LL on top of the $(-2,-2)$ CHF state.
}
\label{fig:120UnstrainedDetails}
\end{figure*}

{\color{black}
\subsection{Main CCIs in the absence of heterostrain} \label{sec:nostrain}
Though most of the experiments so far are believed to be strongly influenced by uniaxial heterostrain, there are a few  experiments which appear to be in the ultra low heterostrain regime \cite{Lu2019,Nuckolls2023}. In addition to the main CCIs along $(0,\pm4)$, $\pm(1,3)$, $\pm(2,2)$ and $\pm(3,1)$, these experiments show that the CNP develops a gap at $B=0$ at low temperatures without any apparent hBN alignment. This is in contrast to a (gapless) semimetal found at a moderate amount of uniaxial heterostrain (e.g. $0.2\%$) \cite{Kwan2021,Nuckolls2023}. As discussed in the previous section, in the strong coupling limit, there are prominent gapped states along $(-2,-1)$, while $(-1,-3)$ and $(0,-4)$ are gapless at reasonable magnetic flux ratios. In contrast, $(-2,-1)$ does not appear to be a prominent gapped state in experiments without hBN alignment, and the main CCIs are ubiquitous. We therefore conclude that the above low heterostrain experiments \cite{Lu2019,Nuckolls2023} cannot be in the strong coupling regime with a negligible non-interacting bandwidth.

We investigate the impact of a finite non-interacting bandwidth by systematically studying the finite $B$ phase diagram for a range of twist angles from $1.38^\circ$ down to the magic angle of $1.05^\circ$, in an analogous fashion compared to studies with $0.2\%$ heterostrain. The single-electron excitation gaps at different twist angles, electron densities, and $B$ field are presented in Fig.~\ref{fig:UnstrainedPhaseDiagram}. Through separate $B=0$ Hartree-Fock calculations we demonstrate that at CNP the ground state is a gapped IVC state for all twist angles, albeit with an IVC order parameter localized to the vicinity of the $\bK$ points of the moir\'e Brillouin zone at larger twist angles. At the largest twist angle $1.38^\circ$ with the largest non-interacting bandwidth, the gap structures are very similar to that in the presence of heterostrain. Main CCIs along $(-3,-1)$, $(-2,-2)$ and $(-1,-3)$ first emerge at high $B$, and lose to nearly compressible states at lower $B$. Here the main CCIs are CHFs, without competing IVC states nearby. As the  twist angle decreases and the non-interacting bandwidth decreases, the main CCIs become stable at lower $B$. Finally at the magic angle $1.05^\circ$ where the non-interacting bandwidth is negligibly small in the BM model, main CCIs along $(-3,-1)$ remain stable, but along $(-2,-2)$ they are only stable for intermediate $B$-fields. Along $(-1,-3)$ we do not observe gapped states. Moreover, we identify a strong $(-2,-1)$ gapped state consistent with previous works \cite{XW2022,TBGV,Keshav2023}. However, gapped state along $(0,-4)$ is missing, and we attribute it to a finite $B$ and finite electron density regime where populating the rigid excitation spectra of $n/n_s=0$ IVC insulator is no longer energetically favorable.

At the twist angle $1.2^\circ$, as shown in Fig.~\ref{fig:UnstrainedPhaseDiagram}(e), the main CCIs along $(-3,-1)$, $(-2,-2)$ and $(-1,-3)$ remain robust down to the lowest magnetic flux ratio $\phi/\phi_0=1/12$ studied. Simultaneously there are strong IVC states along $(0,0)$ and $(-2,0)$. We also do not find a strong gapped state along $(-2,-1)$. Moreover, the main gapped states emanating from the CNP are along $(0,0)$, $(0,-2)$ and $(0,-4)$, consistent with the mentioned experiments. To better understand these states we present their detailed Hartree-Fock spectra and representative density matrices in Fig.~\ref{fig:120UnstrainedDetails}. At $n/n_s=0$ and along $(0,0)$, the IVC state hybridizes $\{\bK,\uparrow (\downarrow) \}$ with $\{\bK',\downarrow (\uparrow) \}$, creating bonding and antibonding states. The occupied bonding states have net Chern number 0, resulting in the gapped state along $(0,0)$. In contrast, the gapped state along $(0,-4)$ is a CHF, where the occupied group of states in each valley/spin sector contributes a Chern number $-1$. Along $(0,-2)$, we identify the gapped state as a CHF in the spin $\uparrow$ sector which accounts for the net Chern number $-2$, and an IVC in the spin $\downarrow$ sector having zero Chern number. Analogously at $n/n_s=-2$, the gapped state along $(-2,0)$ forms intervalley coherence between $\{\bK,\downarrow\}$ and $\{\bK',\downarrow\}$, whereas the gapped state along $(-2,-2)$ is a CHF. Along $(-2,-1)$, we find that the state is described by adding one Landau level worth of electrons on top of a CHF ground state along $(-2,-2)$, rather than adding one Landau level worth of holes to an IVC state along $(-2,0)$. The latter state is metastable at $1.2^\circ$ but is the Hartree-Fock ground state at $1.05^\circ$, as illustrated in Figs.~\ref{fig:UnstrainedPhaseDiagram}(e,f). Our results at twist angle $1.2^\circ$ therefore capture the most salient features of the experiments in Refs.~\cite{Lu2019,Nuckolls2023}. 

It is also interesting to note that at finite $B$, the $(0,0)$ IVC state mixes opposite spin species. Within our numerics we can find an IVC state which mixes the same spin species, which has a slightly higher energy. Nevertheless, we find that the energetic difference between these two IVC states grows as $B$ increases. For example, at $\phi/\phi_0=1/8$ the energy difference between these two states is $\sim 0.02\rm meV$ per moir\'e unit cell, but grows to $\sim0.4\rm meV$ at $\phi/\phi_0=5/11$. 
}

{\color{black}
\begin{figure*}
    \centering
    \includegraphics[width=\linewidth]{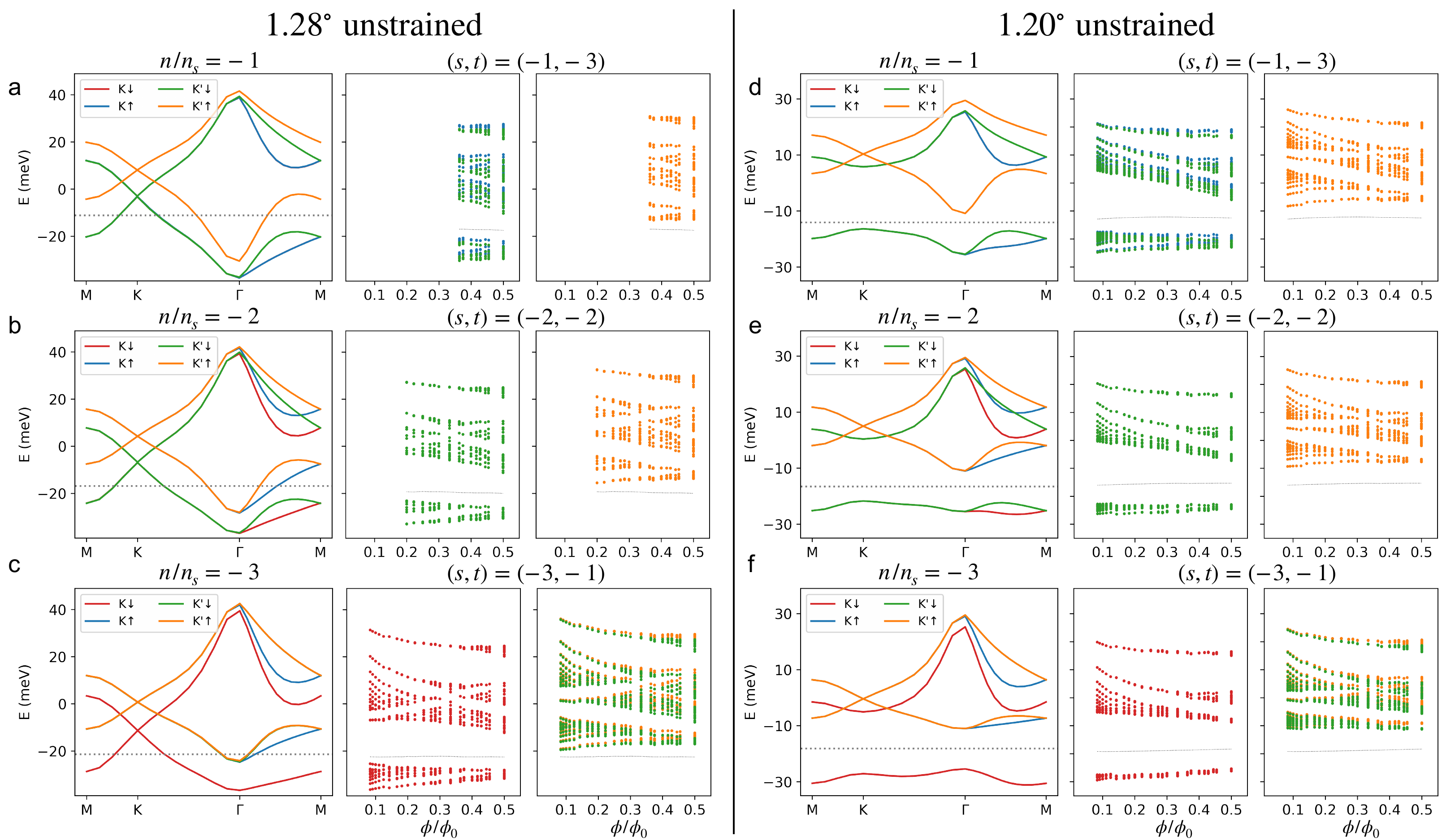}
    \caption{\label{fig:HeavyLight} Comparative study of $B=0$ Hartree-Fock spectra at integer fillings $n/n_s=-1,-2,-3$ of states that preserve valley $U_v(1)$ and spin $U_s(1)$ symmetries, and the $B\neq 0$ Hartree-Fock ground state spectra of CHFs along $(-1,-3)$, $(-2,-2)$ and $(-3,-1)$. The $B=0$ states share the same valley and spin quantum numbers with the CHFs. Panels (a-c) are at $1.20^\circ$ twist angle and panels (d-f) are at $1.28^\circ$, all in the absence of heterostrain. At $1.20^\circ$ the $B=0$ states at $n/n_s=-1,-2$ are metastable, while the state at $n/n_s=-3$ is a stable ground state. They all break the $C_{2z}T$ symmetry and are incompressible Chern states with Chern numbers $-3,-2,-1$ respectively. Although not strictly in strong coupling, they are similar to the zCIs discussed in the strong coupling limit. The spectra of CHFs smoothly extrapolate to these states as $B$ decreases even though some are metastable at $B=0$. At $1.28^\circ$, the $B=0$ states are stable ground states within our Hartree-Fock calculation, preserve $C_{2z}T$ and are compressible, but displaying sizable exchange splittings ($\sim 12$meV) reflected in the relative energy shift of the Dirac cones of different valley and spin flavors. The spectra of CHFs also extrapolate to the spectra of the $B=0$ states, albeit with some differences such as the exact locations of the zLLs of the exchange split Dirac cones. In (a) and (d), the $\{\bK,\downarrow\}$ states (red) are hidden behind $\{\bK,\uparrow\}$ (blue) at $B=0$, and hidden behind $\{\bK',\downarrow\}$ at $B\neq0$. In (b) and (e) at $B\neq0$, the spectrum of $\{\bK,\downarrow\}$ (red) is hidden behind $\{\bK',\downarrow\}$ (green), while that of $\{\bK,\uparrow\}$ (blue) behind $\{\bK',\uparrow\}$ (orange). In (c) and (f) and finite $B$,  $\{\bK,\uparrow\}$ (blue) states are hidden behind $\{\bK',\uparrow\}$ (orange).}
\end{figure*} }

{\color{black}
\subsection{Heavy-light dichotomy}
As demonstrated in both Fig.~\ref{fig:StrainedPhaseDiagram} (strained) and Fig.~\ref{fig:UnstrainedPhaseDiagram} (unstrained), a notable feature of the finite $B$ phase diagram is that the main gapped Chern states point {\it away} from the CNP as $B$ increases, e.g., along $(s,t)=\pm(1,3)$, $\pm(2,2)$ and $\pm(3,1)$. On the other hand, with a few exceptions, Chern states pointing toward the CNP with a big gap are largely missing, e.g., along $(s,t)=\pm(-1,3)$, $\pm(-2,2)$ and $\pm(-3,1)$.

As we showed in the previous section, at $1.05^\circ$, at low $B$, and in the absence of heterostrain, this can be understood from the Landau quantization \cite{XW2022,Keshav2023} of the strong coupling single-particle excitation spectra of the flavor symmetry breaking ground states at $B=0$ \cite{TBGV, Kang2021}. At integer fillings on the hole-doped side of the CNP, a doped hole (i.e. moving away from CNP) has a light mass, leading to well separated LLs (large cyclotron frequency) in relatively low $B$. In contrast, a doped electron (i.e. moving toward the CNP) has a heavy mass, and the LLs are much more densely spaced at a comparable $B$. While this explanation holds at low $B$ for $1.05^\circ$ twist angle without heterostrain, Figs.~\ref{fig:StrainedPhaseDiagram} and ~\ref{fig:UnstrainedPhaseDiagram} show that the main Landau fans point away from the CNP even at larger twist angles or when we include moderate heterostrain. Under such conditions the strong coupling limit is not reached as can be seen by the absence of the correlated insulators at $B=0$ or the presence of IKS states which do not saturate the sublattice-valley polarization (Fig.\ref{fig:CHFSCI}a,b).

In order to understand why this result persists away from the strong coupling, we start by focusing on $1.28^\circ$ in the absence of heterostrain whose phase diagram is presented in Fig.~\ref{fig:UnstrainedPhaseDiagram}(c). In Fig.~\ref{fig:HeavyLight}(a-c), the left columns show the Hartree-Fock spectra of the respective ground states at $B=0$ and $n/n_s=-1,-2,-3$. These ground states preserve the valley $U_v(1)$, spin $U_s(1)$, $C_{3z}$ and $C_{2z}T$ symmetries. As a result, the Dirac points (DPs) of any given valley/spin flavor are protected and located at the $\bK$ point of the moir\'e Brillouin zone.  Crucially however, for all three fillings there is an exchange splitting ($\sim 12$meV) between different flavors, as reflected in the relative energetic shift between the DPs of different valley/spin character. These ground states are all compressible, where the DPs are shifted above the Fermi energy, with the charge being compensated by the Fermi pockets from the exchange split bands. Moreover, there is a band renormalization effect for all flavors, reflected in the narrowing (sharpening) of the bands below (above) the DPs. The band renormalization becomes stronger as the filling is tuned to the band edge (e.g., $n/n_s=-3$). Such band flattening effect has been discussed in the literature as  a Hartree effect \cite{Choi2021b} (our calculation here also shows exchange splitting due to Fock terms). The second and third columns show the $B\neq 0$ Hartree-Fock spectra of the CHFs which share the same valley and spin quantum numbers as the $B=0$ states, and color-coded in the same manner. It is evident that spectra of the CHFs can be smoothly extrapolated to the $B=0$ Hartree-Fock dispersions, e.g. by matching the energies of the DP zLLs to DPs at $B=0$. This demonstrates that the CHFs emerge from the respective compressible ground states at $B=0$. 

Generally, the heavy-light dichotomy at high twist angles (such as $1.28^\circ$) can be understood by examining the $B=0$ dispersions. We address it using the CHF (or lack thereof) along $(-3,\pm1)$ as an example. As shown in Fig.~\ref{fig:HeavyLight}(c), the CHF is formed by populating the Chern $-1$ group of magnetic subbands below the DP zLLs of a given valley/spin flavor (in this case $\{\bK,\downarrow\}$ shown in the second column). However, Landau quantizing the $B=0$ dispersions show that CHF is not an energetically favorable state at infinitesimally weak $B$ field, as the DP LLs of the ${\bK,\downarrow}$ are buried inside the dense LLs associated with the Fermi pockets (i.e., heavier electronic states due to band flattening) of the other three flavors. Only at larger $B$, can the Chern $-1$ magnetic subband group be separated from the rest of the spectra due to the wide LL spacings of a linearly dispersing Dirac cone. This ties the finite $B$ CHF along $(-3,-1)$ directly to the Landau quantizations of the $B=0$ ground state.  Conversely, the zLLs of the DPs cannot be separated from the dense LLs, therefore the state along $(-3,1)$ is nearly compressible (unless intercepted by gapped states along another $(s,t)$). The $B=0$ and finite $B$ correspondence can also explain why the critical field for the onset of CHFs along $(-3,-1)$ occurs at a lower $B$ than $(-2,-2)$, which in turn occurs at a lower $B$ than $(-1,-3)$. As $n/n_s$ increases from $-3$ to $-1$, the flavor degeneracy of the DPs (marked by green or red crossings) increases while the flavor degeneracy of the Fermi pockets --originating from the exchange split bands-- decreases. As a result, the size of the Fermi pockets grows and the bottom of these pockets sinks deeper below the DPs. A higher $B$ is therefore necessary to separate the aforementioned Chern -1 group of magnetic subbands per flavor from the rest of the spectra. 

Next, we focus on $1.20^\circ$ in the absence of heterostrain, where interaction effects are stronger than at $1.28^\circ$, with the CHFs along $(-3,-1)$, $(-2,-2)$ and $(-1,-3)$ all persisting down to the lowest magnetic flux ratio studied in this work. The finite $B$ phase diagram is presented in Fig.~\ref{fig:UnstrainedPhaseDiagram}(e), and the comparisons to $B=0$ states are presented in Fig.~\ref{fig:HeavyLight}(d-f). These $B=0$ states are obtained by enforcing the valley $U_v(1)$ and spin $U_s(1)$ symmetries, and share the same valley/spin quantum numbers as the respective CHFs at finite $B$. While the state at $n/n_s=-3$ is the Hartree-Fock ground state, states at $n/n_s=-2,-1$ are metastable. Comparing the $B=0$ and finite $B$ spectra we conclude that the CHFs originate from the Landau quantization of these $B=0$ bands. A notable difference compared to $1.28^\circ$ is that the $C_{2z}T$ symmetry is spontaneously broken at all $n/n_s=-3,-2,-1$, making them (meta-)stable and gapped zCIs (with Chern numbers $\pm1,\pm2,\pm3$ respectively), analogous to those discussed in the strong coupling limit \cite{Bultinck2019,TBGIV}. Given that the $B=0$ states at $n/n_s=-2,-1$ are metastable, at very low $B$, these CHFs must lose to populating LLs of the excitation spectra of the respective true ground states (with IVC order, see Fig.~{\color{blue}S20}), likely via a first order phase transition. This is in contrast to higher twist angles (e.g., $1.28^\circ$) where there are no competing zCIs states nearby.

We finally address the heavy-light dichotomy for the calculations performed with $0.2\%$ of uniaxial heterostrain. Due to competing Chern 0 IKS states (gapped or compressible) which persist to higher twist angles, it is challenging to find the $B=0$ metastable states from which the finite $B$ CCIs descend, as we can in the absence of heterostrain. We therefore postpone tying the finite $B$ and $B=0$ physics to a future work. Here we instead provide an intuitive picture by examining Hartree-Fock spectra directly at finite $B$. We use CCIs (or lack thereof) along $(-3,\pm 1)$ at $1.28^\circ$ twist angle as an example. Along $(-3,-1)$, both CHF and IKS predominantly involve the lower Chern -1 group of the non-interacting magnetic subbands, separated from the zLLs and the upper Chern -1 group by a gap (see Fig.~\ref{fig:120StrainedDetails}(a) for $1.2^\circ$; at $1.28^\circ$ the structure of the magnetic subbands is qualitatively similar). The bandwidth of the lower Chern -1 group narrows upon increasing $B$, creating a favorable condition for symmetry breaking states (either CHF or IKS) driven by Coulomb interactions \cite{Saito2021}. Along $(-3,1)$, to get a similar flavor symmetry breaking state, one would involve both the lower Chern -1 group and the two zLLs. However, as $B$ increases, this Chern +1 composite of magnetic subbands has an increasing bandwidth, making it energetically more costly to break the valley/spin flavor symmetry. Conversely, it is more energetically favorable to populate two extra LLs from the nearby, exchange split, group of states on top of the $(-3,-1)$ state. Due to band flattening effect discussed above, an added electron to the bottom of the spectra of a given valley/spin flavor on the hole side of the CNP is heavy, and reflected as nearly compressible along $(-3,1)$. This is further supported numerically from Fig.~{\color{blue}S4}, where the spectra above the gap at $(-3,-1)$ (gray colored) do not have well-separated LLs. The finite $B$ perspective presented here complements the perspective of tying finite $B$ to $B=0$ (meta-)stable states, and relies on the same Hartree-Fock effects of band (magnetic subband) renormalizations and exchange splitting.

At low twist angles in the presence of heterostrain, such as $1.20^\circ$ and $1.05^\circ$ shown in Figs.~\ref{fig:StrainedPhaseDiagram}(e) and (f), the gapped IKS states along $(-3,-1)$ $(-2,-2)$ and $(-1,-3)$ remain robust down to the lowest magnetic flux ratio studied in this work, while gapped Chern 0 IKS states along $(-3,0)$ (both $(-3,0)$ and $(-2,0)$ at $1.05^\circ$) are also found. Moreover, we find that at $1.05^\circ$, the IKS wavevector of these two classes of IKS states are very different ($\approx \frac{1}{2}\bg_1$ for Chern-0 IKS and $\approx \frac{1}{q}\bg_2$ for IKS carrying a Chern number). Given the earlier discussions of competing zCI and IVC states in the absence of heterostrain and at lower twist angles, it is tempting to conjecture that the gapped IKS states along $(-3,-1)$, $(-2,-2)$ and $(-1,-3)$ may descend from $B=0$ ``topological IKS" (tIKS) states which break the $C_{2z}T$ symmetry, i.e. a distinct IKS state from those reported in the literature ~\cite{Kwan2021}. Should they exist, the tIKS state must be metastable  at $B=0$ and energetically unfavorable compared to the IKS state that preserves $C_{2z}T$. We leave more elaborate analysis of such a conjecture to future studies. 
}

\begin{figure}
    \centering
    \includegraphics[width=0.9\linewidth]{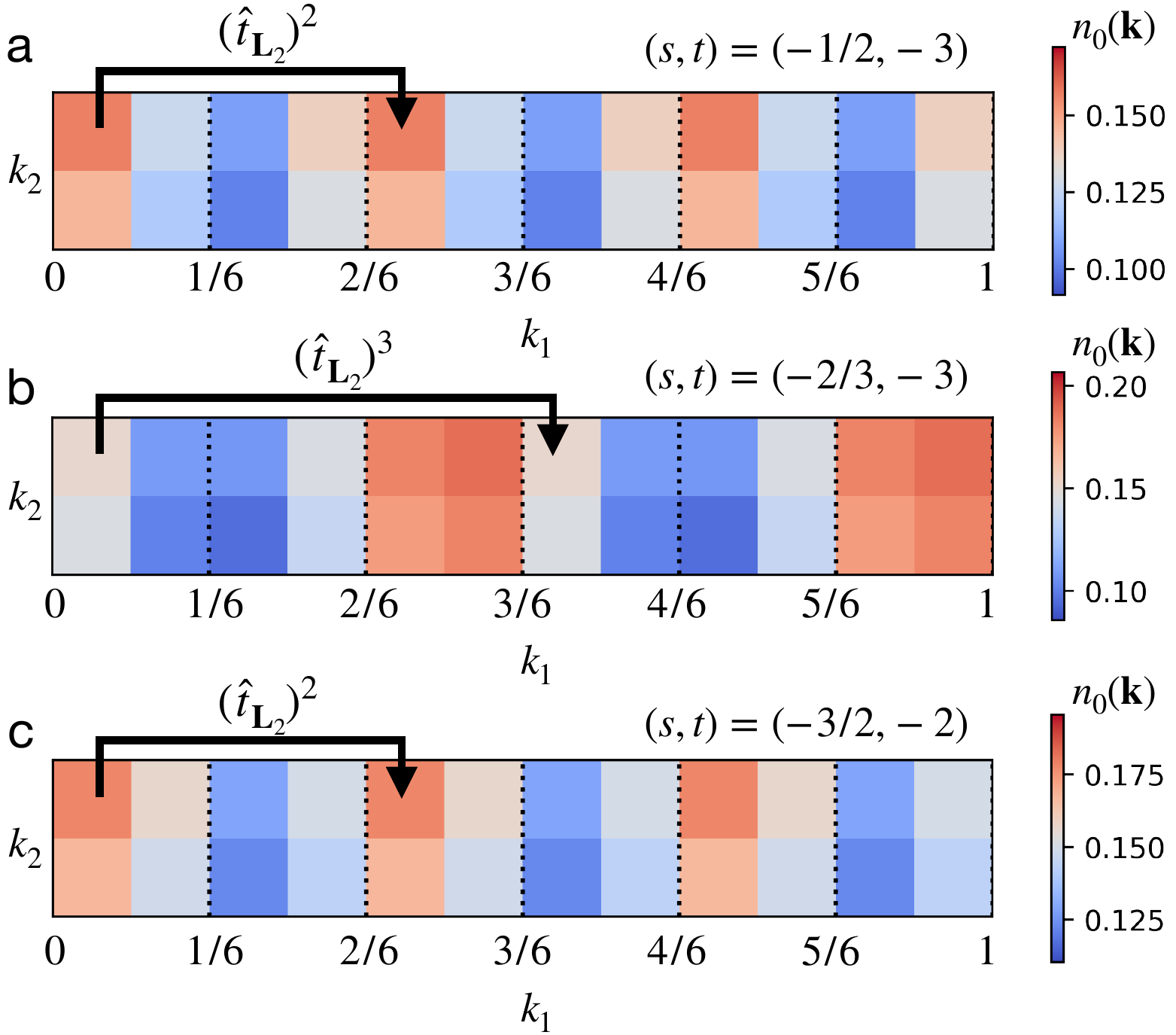}
    \caption{Occupation number $n_0(\bk)$ of the lower zLL of the non-interacting spectra in valley $\bK$ and for spin $\downarrow$. Results are obtained at $1.05^\circ$ for correlated Chern inslating states with fractional $s$, marked by the dashed lines in Fig.~\ref{fig:StrainedPhaseDiagram}(f). }
    \label{fig:FractionalCCI}
\end{figure}

\subsection{Additional CCIs}
{\color{black}
Besides the aforementioned CCIs, we also find additional correlated insulating states in the phase diagram both with and without heterostrain, as shown in Fig.~\ref{fig:StrainedPhaseDiagram} and Fig.~\ref{fig:UnstrainedPhaseDiagram} respectively.

In the presence of heterostrain, the most prominent states emanate from the CNP, and are identified as quantum Hall ferromagnetic states (QHFMs) within the zLLs of the energetically split Dirac cones \cite{QHFM_MacDonald,Kharitonov2012,Young2012} (see Fig.~{\color{blue} S7}). QHFMs emanating from the band minimum ($n/n_s=-4$) are well developed at $1.38^\circ$ but become progressively weaker as the angle decreases (see Fig.~{\color{blue} S6}). Moreover, at higher $\phi/\phi_0$ a gapped state with $(s,t)=(-2,0)$ is observed in Figs.~\ref{fig:StrainedPhaseDiagram}(a-c). We identify it as a Quantum Spin Hall (QSH) insulator due to strong spin splitting near $\phi/\phi_0=1/2$, as demonstrated in the non-interacting Hofstadter spectra in Fig.~{\color{blue} S1} and representative density matrices in Fig.~{\color{blue} S8}. At $1.05^\circ$, we also find CCIs with fractional $s$ along $(-{2}/{3},-3)$, $(-{1}/{2},-3)$, and $(-{3}/{2},-2)$, see Fig.~\ref{fig:StrainedPhaseDiagram}(f). These states break magnetic translation symmetry. We identify them as striped states with period $m$ along the $\bL_2$ direction, such that the density matrix is invariant under $(\hat{t}_{\bL_2})^m$. Their respective density matrices are shown in Fig.~{\color{blue}S9}. We use the electron occupation number of the lower zLL (see e.g. Fig.~\ref{fig:120StrainedDetails}(a)) in valley $\bK$ and for spin $\downarrow$ to illustrate the striped states. We define it as $n_0(\bk)$, and show its momentum dependence in the magnetic Brillouin zone at $\phi/\phi_0=1/6$ for  $(-{2}/{3},-3)$, $(-{1}/{2},-3)$, and $(-{3}/{2},-2)$ in Figs.~\ref{fig:FractionalCCI} (a), (b) and (c) respectively. At $(-1/2,-3)$ and $(-3/2,-2)$, the fractional part of $s$ corresponds to half-filling of a valley and spin flavor, and we identify the period of the striped state as $m=2$. At $(-2/3,-3)$, the fractional part of $s$ corresponds to two thirds filling of a flavor, and we identify the stripe period as $m=3$. While some of the fractional $s$ CCIs show intervalley coherence and others show valley/spin polarization, the energy difference between these two kinds of states are $\sim 0.05\rm meV$ per moir\'e unit cell. Within our numerical accuracy, we cannot say with certainty if either will be observed in future experimental works. Furthermore, we are also limited to probe striped states only along ${\bL_2}$ direction, but not striped states along ${\bL_1}$ or checkerboard states modulating along both directions. Nevertheless, it is clear that states with broken magnetic translation symmetries (being striped or checkerboard states) are more energetically favorable than states that preserve them (or IKS states that preserve modified magnetic translation symmetries). More careful studies of fractional $s$ states are beyond the scope of the present work, and will be left for the future. Ref.~\cite{Xie2021} observed fractional $s$ CCIs in hBN aligned TBG devices, and presented a qualitative argument based on translation symmetry breaking phases at $B=0$. Here, our results demonstrate that they can be stablized purely due to interactions and without hBN alignment.

In the absence of heterostrain and large twist angles, we identify gapped states emanating from the band minimum as QHFMs similar to the strained case (see Fig.~{\color{blue} S13}). However the most prominent gapped states emanating from the CNP are no longer QHFMs even at the largest twist angle studied. As alluded to in Sec.~\ref{sec:nostrain}, the CNP develops a small but finite correlation gap even at $1.38^\circ$, supported by separate $B=0$ Hartree-Fock studies. However the energy minimum of electron-like excitations (or maximum of hole-like excitations) remains at the hexagon corners of the moir\'e Brillouin zone. As the twist angle decreases, QHFMs emanating from the band minimum fade away similar to the cases in the presence of heterostrain, the size of the IVC gap at CNP grows and Coulomb interaction hybridizes electronic states further from the vicinity of the Dirac cones of the non-interacting band. This eventually  leads to ``inverted" excitation spectra at the magic angle $1.05^\circ$, with the energy minimum of electron-like excitations shifted to the $\Gamma$ point of the moir\'e Brillouin zone. Additionally, as in the strained case, fractional $s$ CCIs are also found at the magic angle, but persist to higher twist angles compared to strained case (see Fig.~{\color{blue} S16}). 

}

\section{Conclusion}
{\color{black} In summary, by performing a comprehensive self-consistent Hartree-Fock study of the continuum Bistritzer-MacDonald model in finite magnetic fields, we unravel the nature of the prominent correlated Chern insulators observed in a wide range of TBG experiments. For realistic heterostrain, these correlated Chern insulators are stablized at higher magnetic fields, and correspond to valley and spin polarizations of the interaction-renormalized magnetic subbands that we dub correlated Hofstadter ferromagnets (CHFs). Upon lowering magnetic field, the CHFs become energetically less favored, losing to gapped states with intervalley coherence. In the absence of heterostrain and at higher magnetic fields, the CHF crosses over to the strong coupling Chern insulating states (sCIs) as the twist angle decreases. At lower fields, competing states with intervalley coherence become more energetically favored, and the transition is marked by a collapse of the averaged sublattice polarization per occupied single-electron state. }

Our calculations also predict additional gapped correlated insulating states beyond the $(s,t)=(0,\pm4),\pm(1,3),\pm(2,2),\pm(3,1)$ sequence, notably the striped states at fractional $s$. Given that our calculations have direct access to the interaction renormalized single-electron excitation spectra at a given filling and magnetic field (see Fig.~{\color{blue} S10}), comparisons with experiments such as STM can be made to facilitate the characterization of the panoply of correlated insulating states.

\section{Acknowledgements}
X.W. and O. V. acknowledge invaluable discussions with B. Andrei Bernevig, Cyprian Lewandowski, Joe Finney, Minhao He, Jian Kang, {\color{black} Erez Berg, Nick Bultinck, Tomohiro Soejima and Tianle Wang}. X.W. acknowledges financial support from the Gordon and Betty Moore Foundation's EPiQS Initiative Grant GBMF11070, National High Magnetic Field Laboratory through NSF Grant No.~DMR-1157490 and the State of Florida. O.V. was supported by NSF Grant No.~DMR-1916958 and is partially funded by the Gordon and Betty Moore Foundation's EPiQS Initiative Grant GBMF11070. Most of the computing for this project was performed on the HPC at the Research Computing Center at the Florida State University (FSU).

\bibliography{references}

\clearpage 

\begin{widetext}

\begin{center}
  \textbf{\large Supplemental Material for ``Theory of correlated Chern insulators in twisted bilayer graphene"}
\end{center}


\setcounter{equation}{0}
\setcounter{section}{0}
\setcounter{figure}{0}
\renewcommand{\theequation}{S\arabic{equation}}
\setcounter{table}{0}
\setcounter{page}{1}

\section{TBG with heterostrain at zero magnetic field}
We consider the limit of small twist angle $\theta$ and small deformations of moir\'e superlattice due to uniaxial heterostrain, such that the moir\'e reciprocal lattice vectors are given by: $\bg_{i=1,2}=\mathcal{E}^T\bG_{i=1,2}$ \cite{Zhen2019}, where $\bG_{i}$ are reciprocal lattice vectors of the untwisted and undeformed monolayer graphene, and
\begin{equation}
    \mathcal{E} = \mathcal{T}(\theta) + \mathcal{S}(\epsilon,\varphi),
\end{equation}
where 
\begin{equation}
    \mathcal{T}(\theta) = \begin{pmatrix} 0 & -\theta\\ \theta & 0\end{pmatrix},\ \mathcal{S}(\epsilon,\varphi) = R_{\varphi}^T \begin{pmatrix} -\epsilon & 0\\ 0 & \nu\epsilon \end{pmatrix}R_{\varphi}.
\end{equation}
Here $\{\varepsilon,\varphi\}$ parameterize the uniaxial heterostrain strength and orientation, $R_\varphi$ is the two-dimensional rotation matrix, and $\nu\approx 0.16$ is the Poisson ratio. The moir\'e lattice vectors $\bL_{i=1,2}$ are uniquely defined through the relation $\bL_i\cdot \bg_j=2\pi \delta_{ij}$. Specifically,
\begin{equation}
    \bL_1 = \frac{2\pi}{(\bg_1\times\bg_2)\cdot\hat{z}}( \bg_2\times \hat{z}),\ \bL_2 = \frac{2\pi}{(\bg_1\times\bg_2)\cdot\hat{z}}( \hat{z}\times \bg_1).
\end{equation}

We consider the strained Bistritzer-MacDonald (BM) Hamiltonian discussed in our recent work \cite{XW2023}, with the continuum Hamiltonian given as:
\begin{equation}\label{eq:strained_bm}
    H_{\eta} = (\sum_{l=t,b}  H_{\eta,l}^{intra}) + H_{\eta}^{inter} ,
\end{equation}
where $\eta=\pm 1$ labels $\bK\ (\bK')$ valleys of monolayer graphene, $l=t,b$ labels the top (bottom) of the two graphene layers. The interlayer Hamiltonian is given by:
\begin{equation}
    H_{\eta}^{inter}  \approx \int \mathrm{d}^2\br \psi^\dagger_{\eta,t}(\br) \left(\sum_{j=1,2,3} T_{\eta,j} e^{-i\eta \bq_{j}\cdot\br}\right)\psi_{\eta,b}(\br) + h.c.,
\end{equation}
where $\psi_{\eta,l}(\br)\equiv (\psi_{\eta,l,A}(\br),\psi_{\eta,l,B}(\br))^T$ is a spinor in the sublattice basis for a given valley and layer. We have suppressed the spin index for simplicity. $\bq_{j=1,2,3}$ are the three nearest neighbor bonds of the reciprocal honeycomb lattice, and
\begin{equation}
    T_{\eta,j} = w_0\sigma_0+w_1\left(\cos \frac{2\pi(j-1)}{3} \sigma_x+ \eta \sin \frac{2\pi(j-1)}{3}\sigma_y\right).
\end{equation}
Here $(\sigma_0,\sigma_x,\sigma_y)$ are Pauli matrices acting on sublattice degrees of freedom. $w_0$ and $w_1$ are intra-sublattice and inter-sublattice tunneling strengths between the two layers. The ``chiral limit" discussed in Ref.~\cite{Tarnopolsky2019} corresponds to $w_0/w_1=0$. 

The intra-layer Hamiltonian is given by:
\begin{equation}
\begin{split}
     & H_{\eta,l}^{intra} =  \alpha \int \mathrm{d}^2 \br  \psi^\dagger_{\eta,l}(\br)  ( \tr[\mathcal{E}_l] \sigma_0)\psi_{\eta,l}(\br)\\
     &+  \frac{\hbar v_F}{a} \int \mathrm{d}^2 \br \psi^\dagger_{\eta,l}(\br)  \left[ (-i\nabla - \bA_{\eta,l})\cdot(\eta \sigma_x,\sigma_y) \right]\psi_{\eta,l}(\br),
\end{split}
\end{equation}
where the first term is the deformation potential that couples to the electron density. $\bA_{\eta,l}$ is the pseudovector potential that comes from changes in the inter-sublattice hopping due to deformations, and changes sign between graphene valleys. It is given as $\bA_{\eta,l} = \frac{\sqrt{3}\beta}{2a}\eta (\epsilon_{l,xx}-\epsilon_{l,yy},-2\epsilon_{l,xy})$.

We use $\alpha\approx -4.1\ \mathrm{eV}$ and $\beta \approx 3.14$. We further fix $\hbar v_F/a=2.68\mathrm{eV}$, $w_1=110\mathrm{meV}$ and $w_0/w_1=0.7$ in our calculations, and also set constants $\hbar=a=1$ in the remainder of this document. Other values of $w_0/w_1$ are also used and discussed in the manuscript as a way to stablize the strong-coupling Chern insulating states.

\section{TBG in magnetic field}
\subsection{Landau level wavefunctions of monolayer graphene} 
We begin with a brief discussion of the Landau level (LL) eigenstates of the Dirac Hamiltonian of monolayer graphene. In a finite magnetic field $B$, the Dirac Hamiltonian in valleys $\bK$ and $\bK'$ are given by (here for negative charged electrons, the minimal coupling is $\bp+e\bA$ where $e$ is positive):
\begin{align}
    \hat{H}^{\bK}_{l}(\bp + e\bA) &=  v_F \vec{\sigma}\cdot( \bp -\bK_l + e\bA),\\
    \hat{H}_l^{\bK'}(\bp+e\bA) &= - v_F \vec{\sigma}^*\cdot(\bp -\bK_l' + e\bA).
\end{align}
Here $l=1,2$ is the layer index, $\bK_l=(K_{l,x},K_{l,y})$ is the position of the Dirac cone in the reciprocal space.

We choose the Landau gauge such that the vector potential $\bA\equiv Bx \hat{y}$, where the global $x-y$ coordinate system is defined such that $\bL_2$ is along the $+\hat{y}$ direction. When TBG is subject to heterstrain, this amounts to a small angle rotation $\theta_0$ of the global coordinate system with respect to the case in absence of heterostrain. The eigenstates of the Dirac Hamiltonian are solved by going to the harmonic oscillator basis: $x=\frac{\ell}{\sqrt{2}}(a+a^\dagger)$, and $p_x=\frac{1}{i\sqrt{2}\ell}(a-a^\dagger)$, where $\ell\equiv 1/\sqrt{eB}$ is the magnetic length. 

In valley $\bK$, the particle-hole symmetric LL eigenstates are given as:
\begin{equation}
    \bra{\br} \ket{\psi_{n\gamma l}(k_2)} = e^{iK_{l,x}x}e^{i\frac{2\pi}{L_2} k_2y} \frac{1}{\sqrt{2}} \begin{pmatrix}-i\gamma e^{-i\theta_0}\phi_{n-1}(x+\tilde{k}_{2,l}\ell^2)\\ \phi_{n}(x+\tilde{k}_{2,l}\ell^2) \end{pmatrix},
\end{equation}
where  $\epsilon_{n\gamma}=\frac{v_F}{\ell}\gamma \sqrt{2n}$ is the energy of the Dirac Hamiltonian, labeled by $n=1,2,\dots$, and $\gamma=\pm 1$ corresponds to positive and negative energy solutions. We have defined $L_2\equiv |\bL_2|$. In addition, there is an anomalous zero energy state given by
\begin{equation}
   \bra{\br} \ket{\psi_{0 l}(k_2)} = e^{iK_{l,x}x}e^{i\frac{2\pi}{L_2} k_2y}  \begin{pmatrix}0\\\phi_{0}(x+\tilde{k}_{2,l}\ell^2)\end{pmatrix}
\end{equation}
which lives on the B sublattice. $\phi_n(x)$ is the eigenfunction of $a^\dagger a$, and is given to be:
\begin{equation}
    \phi_n(x) = \frac{1}{\pi^{1/4}}\frac{1}{\sqrt{2^nn!}}e^{-x^2/2\ell^2}H_n(x/\ell),
\end{equation}
where $H_n(x)$ is the Hermite polynomial. The shift in the position for a given wavevector $k_2\bg_2$ is given by:
\begin{equation}
    \tilde{k}_{2,l}\ell^2 = \left(\frac{2\pi}{L_2}k_2 - K_{l,y}\right)\ell^2.
\end{equation}

In valley $\bK'$, the Landau level wavefunctions are solved in a similar manner, and the LL eigenbasis is given by:
\begin{equation}
    \bra{\br}\ket{\psi_{n\gamma l}(k_2)} = e^{iK_{l,x}'x}e^{i\frac{2\pi}{L_2} k_2y} \frac{1}{\sqrt{2}}\begin{pmatrix}
    \phi_n(x+\tilde{k}_{2,l}\ell^2) \\ i\gamma e^{-i\theta_0} \phi_{n-1}(x+\tilde{k}_{2,l}\ell^2)
    \end{pmatrix}
\end{equation}
with eigenenergy $\epsilon_{n\gamma}=v_F\gamma\frac{\sqrt{2n}}{\ell}$, and eigenstate for the zeroth LL:
\begin{equation}
    \bra{\br}\ket{\psi_{0 l}(k_2)} =e^{iK_{l,x}'x}e^{i\frac{2\pi}{L_2} k_2y} \begin{pmatrix}
    \phi_0(x+\tilde{k}_{2,l}\ell^2) \\ 0
    \end{pmatrix}.
\end{equation}

For notational convenience we define a ket state $\ket{\Phi^{(\eta)}_{n\gamma l}(k_2)}$ such that:
\begin{equation}
    \bra{\br}\ket{\psi^{(\eta)}_{n\gamma l}(k_2)} = e^{iK_{\eta l,x}x}e^{i\frac{2\pi}{L_2}k_2y} \bra{\br}\ket{\Phi^{(\eta)}_{n\gamma l}(k_2)}.
\end{equation}
The ket state has the following real space properties:
\begin{equation}
    \bra{\br}\ket{\Phi^{(\eta)}_{n\gamma l}(k_2)} \equiv \Phi^{(\eta)}_{n\gamma l}(x+\tilde{k}_{2,l}\ell^2),
\end{equation}
where the definitions of $\Phi^{(\eta)}_{n\gamma l}(x+\tilde{k}_{2,l}\ell^2)$ are inferred from Eqs.~(S10), (S11), (S14) and (S15). 

\subsection{Magnetic translation group eigenstates}
At rational magnetic flux ratios $\phi/\phi_0=p/q$ where $\phi_0=h/e$ is the magnetic flux quantum,  $p$ and $q$ are coprime integers, the strained BM Hamiltonian is invariant under magnetic translations, $\hat{t}_{\bL_1}$ and $\hat{t}_{\bL_2}$, such that: 
\begin{equation}
    \hat{t}_{\bL_1} = e^{-i\bq_\phi \cdot (\br-\bL_1/2)}\hat{T}_{\bL_1},\ \hat{t}_{\bL_2} = \hat{T}_{\bL_2},\  \bq_\phi = \frac{2\pi}{|\bL_2|} \frac{\phi}{\phi_0}\hat{y},\ \hat{t}_{\bL_2}\hat{t}_{\bL_1} = e^{i2\pi \phi/\phi_0} \hat{t}_{\bL_1}\hat{t}_{\bL_2}.
\end{equation}
Here $\hat{T}_{\bL_{i}}\psi(\br)=\psi(\br-\bL_i)$. As a result, the eigenstates of the strained Bistritzer-MacDonald Hamiltonian are simultaneous eigenstates of the  magnetic translation group (MTG). The MTG basis states can be generated from the LL basis states discussed in the previous section:
\begin{equation}
\begin{split}
    \ket{V^{(\eta)}_{n \gamma l}(\bk)} & = \frac{1}{\sqrt{\mathcal{N}}}\sum_{s_1=-\infty}^{\infty}e^{i2\pi k_1 s_1} \hat{t}_{\bL_1}^{s_1} \ket{\psi^{(\eta)}_{n\gamma l}(k_2)} \\
    & = \frac{1}{\sqrt{\mathcal{N}}}\sum_{s_1}e^{i2\pi (k_1-k_2\frac{L_{1y}}{L_2})s_1}e^{i\frac{1}{2}s_1^2\bq_\phi\cdot \bL_1}e^{-is_1K_{\eta l,x}L_{1,x}} \ket{\psi^{(\eta)}_{n\gamma l}(k_2-s_1\frac{\phi}{\phi_0})},
\end{split}
\end{equation}
where $\mathcal{N}$ is a normalization factor, and $\bk\equiv k_1\bg_1+k_2\bg_2$. It is straightforward to check that:
\begin{align}
    \ket{V_{n\gamma l}^{(\eta)}(k_1+1,k_2)} & = \ket{V_{n\gamma l}^{(\eta)}(k_1,k_2)},\\
    \ket{V_{n\gamma l}^{(\eta)}(k_1,k_2+\frac{\phi}{\phi_0})} & = e^{iK_{l,x}L_{1,x}}e^{-i2\pi(k_1-k_2 \frac{L_{1y}}{L_2})} \ket{V_{n\gamma l}^{(\eta)}(k_1,k_2)},\\
    \hat{t}_{\bL_2}\ket{V_{n\gamma l}^{(\eta)}(k_1,k_2)} & = e^{-i2\pi k_2}\ket{V_{n\gamma l}^{(\eta)}(k_1+\frac{\phi}{\phi_0},k_2)},\\
    \bra{V_{n\gamma l}^{(\eta)}(k_1,k_2)}\ket{V_{n'\gamma'l'}^{(\eta')}(p_1,p_2)} & = \delta_{\eta,\eta'}\delta_{l,l'}\delta_{k_1,k_2}\delta_{p_1,p_2}\delta_{n,n'}\delta_{\gamma,\gamma'}.
\end{align}

Therefore, the MTG eigenstates defined in $(k_1,k_2)\in [0,1)\times[0,\frac{\phi}{\phi_0})$ form a complete and orthornomal basis set in a finite magnetic field. This is referred to as the magnetic Brillouin zone. In order to find the eigenstates of the narrow bands for strained BM Hamiltonian in $B$, we also need to compute the matrix elements of the interlayer tunneling terms in the above MTG basis states. From now on we work with rational flux ratio $\phi/\phi_0=p/q$, where $p$ and $q$ are coprime. 

\subsection{Matrix elements for a generic operator $A_{\bq}\equiv \hat{A}e^{-i\bq\cdot\br}$}
We first work out the matrix elements for a generic operator in the MTG eigenstates, $A_{\bq}=\hat{A}e^{-i\bq\cdot\br}$, where $\hat{A}$ is a space-independent matrix in the layer, sublattice, and valley basis. For example, the Fourier transform of the electron density operator,  $\hat{\rho}_{\bq}\equiv \hat{\rho}e^{-i\bq\cdot\br}$, has $\hat{\rho}=\hat{I}$, where $\hat{I}$ is the identity matrix. We denote $\bq=q_1\bg_1+q_2\bg_2$ and $\bp=p_1\bg_1+p_2\bg_2$. We therefore have:
\begin{equation}
\begin{split}
& \bra{V^{(\eta)}_{n\gamma l}(\bk)}A_{\bq}\ket{V^{(\eta')}_{n'\gamma'l'}(\bp)} \\
= & \frac{1}{\mathcal{N}}\sum_{s_1s_1'} e^{-i2\pi k_1 s_1}e^{i2\pi p_1 s_1'} \bra{\psi_{n\gamma l}^{(\eta)}(k_2)}\hat{t}_{\bL_1}^{-s_1}\hat{A}e^{-i\bq\cdot \br} \hat{t}_{\bL_1}^{s_1'}\ket{\psi^{(\eta')}_{n'\gamma'l'}(p_2)} \\
= &\frac{1}{\mathcal{N}} \sum_{s_1s_1'} e^{-i2\pi (k_1+q_1) s_1}e^{i2\pi p_1 s_1'}  \bra{\psi_{n\gamma l}^{(\eta)}(k_2)}\hat{A}e^{-i\bq\cdot \br} \hat{t}_{\bL_1}^{s_1'-s_1}\ket{\psi^{(\eta')}_{n'\gamma'l'}(p_2)} \\
= & \delta_{[p_1-k_1-q_1]_1,0}\sum_{s}e^{i2\pi p_1 s}  \bra{\psi_{n\gamma l}^{(\eta)}(k_2)}\hat{A}e^{-i\bq\cdot \br} \hat{t}_{\bL_1}^{s}\ket{\psi^{(\eta')}_{n'\gamma'l'}(p_2)} \\
= & \delta_{[p_1-k_1-q_1]_1,0}\sum_{s}e^{i2\pi (p_1-p_2\frac{L_{1y}}{L_2}) s} e^{\frac{i}{2}s^2\bq_\phi\cdot\bL_1}e^{-isK_{\eta'l',x}L_{1,x}}  \bra{\psi_{n\gamma l}^{(\eta)}(k_2)}\hat{A}e^{-i\bq\cdot \br} \ket{\psi^{(\eta')}_{n'\gamma'l'}(p_2-\frac{sp}{q})}\\
= & \delta_{[p_1-k_1-q_1]_1,0}\sum_{s}\delta_{p_2-sp/q,k_2+q_2}e^{i2\pi (p_1-p_2\frac{L_{1y}}{L_2}) s} e^{\frac{i}{2}s^2\bq_\phi\cdot\bL_1}e^{-isK_{\eta'l',x}L_{1,x}}  \bra{\Phi_{n\gamma l}^{(\eta)}(k_2)}\hat{A}e^{-i\tilde{q}_{x}x} \ket{\Phi^{(\eta')}_{n'\gamma'l'}(k_2+q_2)}\\
\equiv & \delta_{[p_1-k_1-q_1]_1,0}\sum_{s}\delta_{p_2-sp/q,k_2+q_2}e^{i2\pi (p_1-p_2\frac{L_{1y}}{L_2}) s} e^{\frac{i}{2}s^2\bq_\phi\cdot\bL_1}e^{-isK_{\eta'l',x}L_{1,x}} \mathcal{A}^{\eta\eta'}_{n\gamma l,n'\gamma'l'}(k_2,k_2+q_2).
\end{split}
\end{equation}
The notation $[b]_a$ represents $b$ modulo $a$, with $a > 0$. We have defined: 
\begin{equation}
    \tilde{q}_{x} \equiv q_x + K_{\eta l,x} - K_{\eta'l',x}.
\end{equation}

We proceed to work out the 1D integration by noting that:
\begin{equation}
    \bra{x}\ket{\Phi^{(\eta)}_{n\gamma l}(k_2)} = \Phi^{(\eta)}_{n\gamma l}(x+\tilde{k}_{2,\eta l}\ell^2) = e^{i p_x \tilde{k}_{2,\eta l}\ell^2} \Phi^{(\eta)}_{n\gamma l}(x) \equiv \hat{T}_x(\tilde{k}_{2,\eta l}\ell^2)\Phi^{(\eta)}_{n\gamma l}(x),
\end{equation}
and as a result:
\begin{equation}
\begin{split}
    \mathcal{A}^{\eta\eta'}_{n\gamma l,n'\gamma'l'}(k_2,k_2+q_2) & = \bra{\Phi_{n\gamma l}^{(\eta)}(0)}\hat{T}_x(-\tilde{k}_{2,\eta l}\ell^2)\hat{A}e^{-i\tilde{q}_{x}x} \hat{T}_x(\widetilde{(k+q)}_{2,\eta'l'}\ell^2)\ket{\Phi^{(\eta')}_{n'\gamma'l'}(0)}\\
    & = e^{i\tilde{q}_x\tilde{k}_{2,\eta l}\ell^2}\bra{\Phi_{n\gamma l}^{(\eta)}(0)}\hat{A}e^{-i\tilde{q}_{x}x}e^{i\tilde{q}_y\ell^2p_x}\ket{\Phi^{(\eta')}_{n'\gamma'l'}(0)}\\
    & = e^{i\tilde{q}_x\tilde{k}_{2,\eta l}\ell^2} e^{\frac{i}{2}\tilde{q}_x\tilde{q}_y\ell^2}\bra{\Phi_{n\gamma l}^{(\eta)}(0)}\hat{A}e^{c_-a+c_+a^\dagger}\ket{\Phi^{(\eta')}_{n'\gamma'l'}(0)}\\
    & \equiv e^{i\tilde{q}_x\tilde{k}_{2,\eta l}\ell^2} e^{\frac{i}{2}\tilde{q}_x\tilde{q}_y\ell^2} \hat{\mathcal{A}}_{n\gamma l,n'\gamma' l'}^{\eta\eta'},
    \end{split}
\end{equation}
where $\hat{\mathcal{A}}^{\eta\eta'}_{n\gamma l,n'\gamma'l'}$ can be expressed in terms of associated Laguerre polynomials \cite{XW2022}, and:
\begin{equation}
    \tilde{q}_y \equiv q_y + K_{\eta l,y} - K_{\eta'l',y}, \ c_{\pm} = -i \frac{\ell}{\sqrt{2}}(\tilde{q}_x\mp i\tilde{q}_y).
\end{equation}

Next we consider  the implications of the $\delta$ function constraints. Firstly we have:
\begin{equation}
    q_1 = p_1 - k_1 + r,\ r\in \mathbb{Z}. 
\end{equation}
Secondly we have:
\begin{equation}
    q_2 = p_2 - k_2 - \frac{sp}{q}, s\in \mathbb{Z}.
\end{equation}
This shows that above matrix elements are non-vanishing only if: 
\begin{equation}
    \bq \equiv \bp - \bk + r\bg_1 - \frac{sp}{q}\bg_2 ,\ r,s\in \mathbb{Z}.
\end{equation}

Finally, we have the following expression for the matrix elements: 
\begin{align*}
    & \bra{V^{(\eta)}_{n\gamma l}(\bk)}A_{\bq}\ket{V^{(\eta')}_{n'\gamma'l'}(\bp)} \\
= &  \delta_{[p_1-k_1-q_1]_1,0}\sum_{s}\delta_{p_2-sp/q,k_2+q_2}e^{i2\pi (p_1-p_2\frac{L_{1y}}{L_2}) s} e^{\frac{i}{2}s^2\bq_\phi\cdot\bL_1}e^{-isK_{\eta'l',x}L_{1,x}} e^{i\tilde{q}_x\tilde{k}_{2,\eta l}\ell^2} e^{\frac{i}{2}\tilde{q}_x\tilde{q}_y\ell^2} \hat{\mathcal{A}}^{\eta\eta'}_{n\gamma l,n'\gamma'l'}.
\end{align*}
with $\tilde{\bq}\equiv \bq + \bK_{\eta l}-\bK_{\eta'l'}$.

\subsection{Matrix elements of strained BM Hamiltonian}
\subsubsection{Interlayer terms} 
The interlayer terms are a special case of the generic $A_\bq$ discussed in the previous section. 
For bottom to top layer tunneling we have (see Eq.~(S24)):
\begin{equation}
\begin{split}
    & \bra{V^{(\eta)}_{n\gamma b}(\bk)}T_{\eta,j}e^{-i\eta \bq_{j}\cdot\br}\ket{V^{(\eta')}_{n'\gamma't}(\bp)} \\
= &  \delta_{\eta\eta'}\delta_{[p_1-k_1-q_1]_1,0}\sum_{s}\delta_{p_2-sp/q,k_2+q_2}e^{i2\pi (p_1-p_2\frac{L_{1y}}{L_2}) s} e^{i\frac{s^2}{2}\bq_\phi\cdot\bL_1}e^{-isK_{\eta t,x}L_{1,x}} e^{i\tilde{q}_x\tilde{k}_{2,\eta l}\ell^2} e^{\frac{i}{2}\tilde{q}_x\tilde{q}_y\ell^2} \hat{\mathcal{T}}^{(\eta,j)}_{n\gamma b,n'\gamma't},
\end{split}
\end{equation}
where $\bq_j = \mathbf{0},\mathbf{g}_2,\mathbf{g}_1+\mathbf{g}_2$. For notational convenience we have also defined $\bq_{j}\equiv q_1 \bg_1+q_2\bg_2$ where $q_{1,2}$ are integers (with implicit dependence on $j$). Plugging these into the $\delta$-function constraints we have: 
\begin{equation}
    p_1=k_1,\ p_2-k_2-\frac{sp}{q} \in \mathbb{Z}.
\end{equation}
We can split the momentum $p_2,k_2$ into strips of $[0,1/q),\dots[(p-1)/q,p/q)$, by redefining: 
\begin{equation}
    p_2 \rightarrow p_2 + r_2'/q,\ k_2 \rightarrow k_2 + r_2/q,\ k_2,p_2\in[0,1/q).
\end{equation}
It is clear then that $(k_1,k_2)\in[0,1)\times[0,1/q)$ are good quantum numbers under moire periodic potential, and that: 
\begin{equation}
    r_2'=r_2+qq_2+sp \Rightarrow r_2=[r_2+qq_2]_p,\ s= -(r_2+qq_2)/ p.
\end{equation}

\subsubsection{Narrow band eigenstates}
From above analysis, it is clear that the narrow band eigenstates in the Landau gauge are labeled by $(k_1,k_2)\in[0,1)\times[0,1/q)$, and an additional $2q$ quantum numbers per valley and spin. Generally we can denote them as: 
\begin{equation}
    \ket{\Psi^{(\eta)}_a(\bk)} \equiv \sum_{n\gamma l }\sum_{r_2=0}^{p-1}U_{n\gamma lr_2,a}^{(\eta)}(\bk) \ket{V^{(\eta)}_{n\gamma l}(k_1,k_2+r_2/q)},\ a=1,\dots,2q.
\end{equation}

\subsubsection{Degeneracy of magnetic subbands generated by $\hat{t}_{\bL_2}$}
The non-interacting Hamiltonian $H_\text{non-int}$ in valley $\eta$ acting on the narrow band eigenstates is: 
\begin{equation}
    H_{\text{non-int}}\ket{\Psi^{(\eta)}_a(\bk)} = \varepsilon^{(\eta)}_{a}(\bk) \ket{\Psi^{(\eta)}_a(\bk)}.
\end{equation}
We apply the magnetic translation $\hat{t}_{\bL_2}$ to these energy eigenstates. Note that they are eigenstates of $\hat{t}_{\bL_1}$ and $\hat{t}^q_{\bL_2}$; the reason why $\hat{t}_{\bL_2}$ is applied $q$ times is due to non-commuting nature of magnetic translation operators $\hat{t}_{\bL_1}$ and $\hat{t}_{\bL_2}$. We see that:
\begin{equation}
\begin{split}
     \hat{t}_{\bL_2}\ket{\Psi^{(\eta)}_a(\bk)} & = \sum_{n\gamma l}\sum_{r_2=0}^{p-1} U^{(\eta)}_{n\gamma lr_2,a}(\bk) \hat{t}_{\bL_2}\ket{V_{n\gamma l}^{(\eta)}(k_1,k_2+\frac{r_2}{q})}\\
     & = \sum_{n\gamma l}\sum_{r_2=0}^{p-1} U^{(\eta)}_{n\gamma lr_2,a}(\bk) e^{-i2\pi (k_2+r_2/q)} \ket{V_{n\gamma l}^{(\eta)}(k_1+p/q,k_2+\frac{r_2}{q})}.
\end{split}
\end{equation}
Furthermore since $[H_\text{non-int},\hat{t}_{\bL_2}]=0$, we have:
\begin{equation}
    H_\text{non-int} \hat{t}_{\bL_2}\ket{\Psi^{(\eta)}_{a}(\bk)} = \hat{t}_{\bL_2} H_\text{non-int} \ket{\Psi^{(\eta)}_{a}(\bk)} = \epsilon^{(\eta)}_{a}(\bk) \hat{t}_{\bL_2}\ket{\Psi^{(\eta)}_{a}(\bk)},
\end{equation}
namely that $\hat{t}_{\bL_2}\ket{\Psi^{(\eta)}_{a}(\bk)}$ is an energy eigenstate at $([k_1+p/q]_1,k_2)$. 

This means that the tower of states are periodic with respect to $k_1\rightarrow k_1 + 1/q$, and that all the distinct energy solutions can be found in the domain of $(k_1,k_2)\in[0,1/q)\times[0,1/q)$. Furthermore, the $\hat{t}_{\bL_2}$ translation gives us one way of gauge fixing between $k_1$ values in different strips of $[0,1/q)$. Specifically, we can define:
\begin{equation}
    \ket{\Psi^{(\eta)}_{a}([k_1+p/q]_1,k_2)} \equiv  e^{i2\pi k_2}\hat{t}_{\bL_2}\ket{V_{a}(k_1,k_2)},
\end{equation}
as the gauge-fixed eigenstate wavefunction at $([k_1+p/q]_1,k_2)$. This amounts to the following definition: 
\begin{equation}
    U^{(\eta)}_{n\gamma lr,a}([k_1+r_1p/q]_1,k_2) = e^{-i2\pi r_1 (r_2/q)}U^{(\eta)}_{n\gamma lr,a}(k_1,k_2) ,\ r_1=0,\dots q-1. 
\end{equation}


\subsection{Matrix elements of $A_{\bq}=\hat{A}e^{-i\bq\cdot\br}$ with respect to narrow band eigenstates}
This can be computed as follows: 
\begin{equation}
\begin{split}
    & \bra{\Psi^{(\eta)}_{a}(\bk)}A_{\bq} \ket{\Psi^{(\eta')}_{b}(\bp)}\\
= & \sum U^{(\eta)*}_{n\gamma l r_2,a}(\bk)U^{(\eta')}_{n'\gamma' l' r_2',b}(\bp)\bra{V^{(\eta)}_{n\gamma l}(k_1,k_2+r_2/q)}A_{\bq}\ket{V^{(\eta')}_{n'\gamma' l'}(p_1,p_2+r_2'/q)}\\
= & \sum U^{(\eta)*}_{n\gamma l r_2,a}(\bk)U^{(\eta')}_{n'\gamma' l' r_2',b}(\bp) \delta_{[p_1-k_1-q_1]_1,0}\\
& \times \sum_{s}\delta_{p_2-sp/q,k_2+q_2}e^{i2\pi (p_1-p_2\frac{L_{1y}}{L_2}) s} e^{i\frac{s^2}{2}\bq_\phi\cdot\bL_1}e^{-isK_{\eta'l',x}L_{1,x}} e^{i\tilde{q}_x\tilde{k}_{2,\eta l}\ell^2} e^{\frac{i}{2}\tilde{q}_x\tilde{q}_y\ell^2} \hat{\mathcal{A}}^{\eta\eta'}_{n\gamma l,n'\gamma'l'}.
\end{split}
\end{equation}
Note that here the $\delta$-constraint depends on $r_2,r_2'$, and matrix elements are only non-vanishing for: 
\begin{equation}
    \bq = \bp - \bk + \frac{r_2'-r_2}{q}\bg_2 + r\bg_1 - \frac{sp}{q}\bg_2.
\end{equation}

Computing above bruteforce is costly. However we could make use of $\hat{t}_{\bL_2}$ to reduce the amount of calculations by a factor of $q^2/(2q-1)$. This is seen by noting that:
\begin{equation}
\begin{split}
    & \bra{V^{(\eta)}_{n\gamma l}([k_1+r_1p/q]_1,k_2)} A_\bq \ket{V^{(\eta)}_{n\gamma l}([p_1+r_1p/q]_1,p_2)}\\
    = & e^{i2\pi r_1(p_2-k_2)}\bra{V^{(\eta)}_{n\gamma l}(\bk)} \hat{t}_{\bL_2}^{-r_1} A_\bq \hat{t}_{\bL_2}^{r_1}\ket{V^{(\eta)}_{n'\gamma' l'}(\bp)} \\
    = &  e^{i2\pi r_1(p_2-k_2-q_2)} \bra{V^{(\eta)}_{n\gamma l}(\bk)}  A_\bq\ket{V^{(\eta)}_{{n'\gamma' l'}}(\bp)} \\
    = & e^{-i2\pi r_1 n/q} \bra{V^{(\eta)}_{n\gamma l}(\bk)}  A_\bq\ket{V^{(\eta)}_{n'\gamma' l'}(\bp)},\ n\in \mathbb{Z}.
\end{split}
\end{equation}
In the last step we made use of the constraint $q_2 = p_2-k_2+n/q$. Therefore, rather than calculating $q^2$ blocks of matrices corresponding to differences in $k_1,p_1$, we only need to compute $2q-1$ blocks, and generate all remaining ones by multiplying a global phase factor. 

\section{Self-consistent Hartree-Fock method in finite magnetic field} 
We proceed to discuss the Hartree-Fock procedure in a finite magnetic field (B-SCHF). We consider the strained BM Hamiltonian in the presence of screened Coulomb interactions, and project it onto the finite field Hilbert space. The Hamiltonian is given by: 
\begin{equation}  \label{eq:intHamB}
    H = \sum_{\eta s a, \bk} \varepsilon^{(\eta)}_{sa}(\bk) d^{\dagger}_{\eta s a,\bk}d_{\eta s a,\bk} + \frac{1}{2A}\sum_{\bq} V_{\bq} \delta\hat{\rho}_{\bq} \delta \hat{\rho}_{-\bq},
\end{equation}
where $A$ is the area of the system, $\varepsilon^{(\eta)}_{sa}(\bk)$ is the Hofstadter spectra including spin Zeeman splitting, and $V_{\bq}$ is the Fourier transform of the screened Coulomb interaction. The projected density operator is given by: 
\begin{equation}
    \delta \hat{\rho}_{\bq} = \sum_{\eta s} \bra{\Psi^{(\eta)}_{a}(\bk)}\hat{\rho} e^{-i\bq \cdot \br } \ket{\Psi^{(\eta)}_{b}(\bp)} \left( d^{\dagger}_{\eta s a,\bk}d_{\eta s b, \bp} - \frac{1}{2}\delta_{a,b}\delta_{\bk,\bp}\right)
\end{equation}
where $\hat{\rho}=\hat{I}$. We emphasize again that the matrix elements are non-zero only for wavevectors satisfying: 
\begin{equation}
    \bq = \bp - \bk + \frac{r_2'-r_2}{q} \bg_2 + r \bg_1 - \frac{sp}{q} \bg_2.
\end{equation}
Here on we introduce the matrix notation for the structure factor: 
\begin{equation}
    \hat{\Lambda}^{\eta s}_{\bq}(a\bk,b\bp) \equiv \bra{\Psi^{(\eta)}_{a}(\bk)}\hat{\rho} e^{-i\bq \cdot \br } \ket{\Psi^{(\eta)}_{b}(\bp)}.
\end{equation}
where we use $s$ to label spin. Pay attention that in prior contexts we have used $s$ to denote an integer.

As a result, the Coulomb piece of the Hamiltonian can be written as: 
\begin{equation}
    H_U = \frac{1}{2A}\sum_{\bq} V_{\bq} \hat{\Lambda}_{\bq}^{\eta s}(a\bk,b\bp) \hat{\Lambda}_{\bq}^{\eta' s'*}(\beta \bp',\alpha \bk') \left( d^{\dagger}_{\eta s a,\bk}d_{\eta s b, \bp} - \frac{1}{2}\delta_{a,b}\delta_{\bk,\bp}\delta_{\bq,\bG}\right) \left( d^{\dagger}_{\eta' s' \alpha,\bk'}d_{\eta' s' \beta, \bp'} - \frac{1}{2}\delta_{\alpha,\beta}\delta_{\bk',\bp'}\delta_{\bq,\bG}\right),
\end{equation}
where $\bG=m\bg_1+n\bg_2$ is the moir\'e reciprocal lattice vector. We note first on the constraint of the $\delta$-functions on the allowed values of momentum. Specifically we have:
\begin{equation}
    \bq = \bp - \bk + m \bg_1 + \frac{n}{q}\bg_2 = \bk'-\bp' + m'\bg_1+\frac{n'}{q}\bg_2.
\end{equation}
Since $p_1-k_1\in(-1,1)$ and $p_2-k_2\in(-1/q,1/q)$, the following constraint applies: 
\begin{equation}
    \bp-\bk = \bk'-\bp',\ m=m',\ n=n'.
\end{equation}

\subsection{Product state and one-particle density matrix}
At a given filling, we consider a ground state $\ket{\Omega}$ to be a product state, given by partial fillings of the narrow band states $\ket{\Psi^{\eta s}_a(\bk)}$. $\ket{\Omega}$ can always be  expressed in terms of a product of Bogoliubov quasiparticle creation operators acting on vacuum,
\begin{equation}
    \ket{\Omega} = \prod_{i,\bk}' \gamma_{i,\bk}^\dagger \ket{0},
\end{equation}
where number of $\gamma_{i,\bk}^{\dagger}$ appearing in the product is constrained by the filling. Generally the Bogoliubov quasiparticles are related by the original fermions $d_{\eta s a,\bk}$ via a unitary transformation: 
\begin{equation}
    d_{\eta s a,\bk} = \sum_{i} U_{\eta s a,i}(\bk) \gamma_{i,\bk},\ U(\bk)U^\dagger(\bk) = I.
\end{equation}
The Hartree-Fock procedure is to find the optimal set of unitary rotations $\{U_{\eta s a,i}(\bk)\}$  which minimizes the total energy. 

Alternatively we can define a one-particle density matrix (note here we used a different definition compared to the main text, and for clairity we also changed the notation from $\hat{Q}$ in the main text to $\hat{P}$ here):
\begin{equation} \label{eq:den_mat_SM}
    \hat{P}^{\eta s,\eta's'}_{a,b}(\bk) \equiv \bra{\Omega}d^{\dagger}_{\eta s a,\bk}d_{\eta' s' b,\bk} \ket{\Omega} - \frac{1}{2}\delta_{\eta,\eta'}\delta_{s,s'}\delta_{a,b}.
\end{equation}
It is related to $\{U_{\eta s a,i}(\bk)\}$ via: 
\begin{equation}
     \hat{P}^{\eta s,\eta's'}_{a,b}(\bk) = \sum_{i,\bk}' U^{*}_{\eta s a,i}(\bk) U^T_{i,\eta' s' b}(\bk) - \frac{1}{2}\delta_{\eta,\eta'}\delta_{s,s'}\delta_{a,b}.
\end{equation}

Note that $\hat{t}_{\bL_2}$ takes $\bk$ to $\bk+\phi/\phi_0 \bg_1$, and as a result, a difference of the density matrix at $\hat{P}(\bk)$ and $\hat{P}(\bk+\phi/\phi_0\bg_1)$ allows us to probe magnetic translation symmetry breaking of $\hat{t}_{\bL_2}$.

\subsection{Mean field energy}
Here we calculate the mean field energy $E_{\Omega}$ with respect to the density matrix $\hat{P}$ defined above. For simplicity and generality we derive $E_\Omega$ with the following Hamiltonian instead of the notation used in Eq.~\ref{eq:intHamB}: 
\begin{equation}
    \hat{H} = \sum_{\alpha,\beta} \hat{T}_{\alpha,\beta} d^{\dagger}_\alpha d_{\beta} + \frac{1}{2A}\sum_{\bq} V_{\bq} \Lambda^{\bq}_{a,b}\Lambda^{-\bq}_{\alpha,\beta} \left(d^\dagger_a d_b -\frac{1}{2}\delta_{a,b}\right)\left(d^\dagger_\alpha d_\beta -\frac{1}{2}\delta_{\alpha,\beta}\right).
\end{equation}

Firstly the contribution to the energy from the kinetic term is given as: 
\begin{equation}
    E_{\Omega}^{(1)} = \bra{\Omega} \sum_{\alpha,\beta} \hat{T}_{\alpha,\beta} d^{\dagger}_\alpha d_\beta\ket{\Omega} = \sum_{\alpha,\beta } \hat{T}_{\alpha,\beta}\left(\hat{P}_{\alpha,\beta}+\frac{1}{2}\delta_{\alpha,\beta}\right) \equiv \tr{\hat{T}(\hat{P}^T+\frac{1}{2}I)}.
\end{equation}

Next we address the Coulomb piece $E_{\Omega}^{(2)}$. We define $A(\Omega)\equiv \bra{\Omega} \left(d^\dagger_a d_b -\frac{1}{2}\delta_{a,b}\right)\left(d^\dagger_\alpha d_\beta -\frac{1}{2}\delta_{\alpha,\beta}\right)\ket{\Omega}$, and: 
\begin{equation}
    \begin{split}
        A(\Omega) = & \frac{1}{4}\delta_{a,b}\delta_{\alpha,\beta} - \frac{1}{2}\left[\delta_{a,b}\left(\hat{P}_{\alpha,\beta}+\frac{1}{2}\delta_{\alpha,\beta}\right) + \delta_{\alpha,\beta}\left( \hat{P}_{a,b}+\frac{1}{2}\delta_{a,b}\right)\right] + \delta_{\alpha,b} \left(  \hat{P}_{a,\beta}+\frac{1}{2}\delta_{a,\beta}\right)  + \bra{\Omega} d^\dagger_a d^\dagger_\alpha d_\beta d_b\ket{\Omega}.
    \end{split}
\end{equation}
The last term can be evaluate as follows: 
\begin{equation}
\begin{split}
     \bra{\Omega} d^\dagger_a d^\dagger_\alpha d_\beta d_b\ket{\Omega} & = U^*_{a,i}U^*_{\alpha,j}U_{\beta,k}U_{b,l} \bra{\Omega} \gamma^\dagger_i \gamma^\dagger_j \gamma_k \gamma_l \ket{\Omega} \\
        & = U^*_{a,i}U^*_{\alpha,j}U_{\beta,k}U_{b,l} \left( \delta_{jk}\delta_{il}-\delta_{jl}\delta_{ik}\right) \\
        & = U^*_{a,i}U^*_{\alpha,j}U_{\beta,j}U_{b,i} - U^*_{a,i}U^*_{\alpha,j}U_{\beta,i}U_{b,j} \\
        & = \left(\hat{P}_{a,b}+\frac{1}{2}\delta_{a,b}\right)\left(\hat{P}_{\alpha,\beta}+\frac{1}{2}\delta_{\alpha,\beta}\right) - \left(\hat{P}_{a,\beta}+\frac{1}{2}\delta_{a,\beta}\right)\left(\hat{P}_{\alpha,b}+\frac{1}{2}\delta_{\alpha,b}\right). 
\end{split}
\end{equation}
Collecting all the terms we get:
\begin{equation}
    \begin{split}
    E_{\Omega}^{(2)} & =\frac{1}{2A}\sum_{\bq}V_{\bq} \Lambda^{\bq}_{a,b}\Lambda^{-\bq}_{\alpha,\beta}  \left( \hat{P}_{a,b}\hat{P}_{\alpha,\beta} - \hat{P}_{a,\beta}\hat{P}_{\alpha,b} + \frac{1}{4}\delta_{\alpha,b}\delta_{a,\beta}\right) \\
    & = \frac{1}{2A}\sum_{\bq}V_{\bq} \tr{\Lambda^\bq\hat{P}^T}\tr{\Lambda^{-\bq}\hat{P}^T} - \frac{1}{2A}\sum_{\bq}V_{\bq}  \tr{\Lambda^{\bq}\hat{P}^T\Lambda^{-\bq}\hat{P}^T} + \frac{1}{8A}\sum_{\bq} V_{\bq} \sum_{a,b}|\Lambda^{\bq}_{a,b}|^2.
\end{split}
\end{equation}

Therefore, the total mean field energy is given as: 
\begin{equation} \label{eq:mf_energy}
    E_{\Omega} = \tr{\hat{T}\hat{P}^T} + \frac{1}{2A}\sum_{\bq}V_{\bq} \tr{\Lambda^\bq\hat{P}^T}\tr{\Lambda^{-\bq}\hat{P}^T} - \frac{1}{2A}\sum_{\bq}V_{\bq}  \tr{\Lambda^{\bq}\hat{P}^T\Lambda^{-\bq}\hat{P}^T} + c_\Omega,
\end{equation}
where $c_{\Omega}$ is a constant number independent of the density matrix. 
\subsection{Energy optimization and Hartree-Fock Hamiltonian}
Eq.~\ref{eq:mf_energy} needs to be minimized subject to the constraint that $\left(\hat{P}+\frac{1}{2}I\right)=\left(\hat{P}+\frac{1}{2}I\right)^2$ is a projector onto the narrow band Hilbert space. This is equivalent to minimizing the following ``free energy" \cite{Kwan2021},
\begin{equation}
    F(\hat{P},\hat{X}) = E_\Omega -  \tr{\hat{X}\left[\left(\hat{P}^{T}+\frac{1}{2}I\right)^2-\left(\hat{P}^{T}+\frac{1}{2}I\right)\right]}.
\end{equation}
Here $\{\hat{X}\}$ are the Lagrange multipliers. If $\hat{P}_c$ is the desired density matrix, then for any small perturbation $\hat{P}_1$ satisfying $\tr \{\hat{P}_1\}=0$, the coefficient to linear order term in $\hat{P}_1$ must vanish. We obtain: 
\begin{equation}\label{eq:tmp1}
\begin{split}
    &F(\hat{P}_c+\hat{P}_1,\hat{X}) - F(\hat{P}_c,\hat{X}) \\
    \approx & \tr{\hat{T}\hat{P}_1^T} + \frac{1}{A}\sum_{\bq} \tr{V_{\bq} \tr{\Lambda^{-\bq}\hat{P}_c^T} \Lambda^\bq\hat{P}_1^T} - \frac{1}{A}\sum_{\bq}V_{\bq}  \tr{\Lambda^{\bq}\hat{P}_c^T\Lambda^{-\bq}\hat{P}_1^T} - \tr{(\hat{X}\hat{P}_c^{T}+\hat{P}_c^{T}\hat{X})\hat{P}_1^T},\\
    \equiv & \tr{\left\{ \hat{H}_{MF}(\hat{P}_c)-\left[\hat{X}\left(\hat{P}_c^{T}+\frac{1}{2}I\right)+\left(\hat{P}_c^{T}+\frac{1}{2}I\right)\hat{X}-\hat{X}\right]\right\}P_1^T},
\end{split}
\end{equation}
where on the last line we defined the Hartree Fock mean field Hamiltonian as: 
\begin{equation} \label{eq:hf_Hamiltonian}
    \hat{H}_{MF}(\hat{P}_c) \equiv \hat{T} + \frac{1}{A}\sum_{\bq} V_{\bq} \tr{\Lambda^{-\bq}\hat{P}_c^T} \Lambda^\bq - \frac{1}{A}\sum_{\bq}V_{\bq}  \Lambda^{\bq}\hat{P}_c^T\Lambda^{-\bq}.
\end{equation}
Requiring Eq.~\ref{eq:tmp1} to vanish leads to the following mean field equation: 
\begin{equation}
    \hat{H}_{MF}(\hat{P}_c)-\left[\hat{X}\left(\hat{P}_c^{T}+\frac{1}{2}I\right)+\left(\hat{P}_c^{T}+\frac{1}{2}I\right)\hat{X}-\hat{X}\right]= 0. 
\end{equation}
Observing the mean field equation, we first see that a solution must satisfy
\begin{equation}
    \hat{H}_{MF}(\hat{P}_c)\left(\hat{P}_c^{T}+\frac{1}{2}I\right) = \left(\hat{P}_c^{T}+\frac{1}{2}I\right) \hat{H}_{MF}(\hat{P}_c) = \left(\hat{P}_c^{T}+\frac{1}{2}I\right)\hat{X}\left(\hat{P}_c^{T}+\frac{1}{2}I\right). 
\end{equation}
This is equivalent to the following commutation relations: 
\begin{equation} \label{eq:self_consistency}
\comm{\hat{H}_{MF}(\hat{P}_c)}{\hat{P}_c}=0,\ \comm{\hat{X}}{\hat{P}_c}=0.
\end{equation}

Finally, with respect to the optimized $\hat{P}_c$, the total energy is given by: 
\begin{equation}
    E_\Omega(\hat{P}_c) = \tr{\hat{T}\hat{P}_c^T + \frac{1}{2}\left( \hat{H}_{MF}(\hat{P}_c)-\hat{T}\right)\hat{P}_c^T} + c_\Omega.
\end{equation}

Our B-SCHF results are obtained by iteratively solving the self-consistency equations (Eq.~\ref{eq:self_consistency}). In practice, we run both random initializations and educated guesses for the density matrix $\hat{P}$, such as a strong coupling Chern insulating state, flavor polarized state, intervalley coherent state, etc. Typically, every point on the phase diagram in Fig.~1 and Fig.~2 of the main text is a result of  $\sim 20$ initializations of the density matrix.

For each magnetic flux ratio $\phi/\phi_0$, we study momentum space mesh consisting of $n_q q$ points along $\bg_1$ direction in the range of $[0,1)$, and $n_q$ points along $\bg_2$ direction in the range of $[0,1/q)$, where $n_q$ is chosen as the maximum integer such that $n_q q\le 14$.  Our calculations presented in the main text correspond to 23 magnetic flux ratios between $1/12$ and $1/2$, corresponding to all $\{p,q\}$ satisfying $q\le 12$ and $p<q$. 

\subsection{Probing IKS like states}
In our B-SCHF calculations, the density matrix defined in Eq.~\ref{eq:den_mat_SM} can probe states with intervalley coherence, where $\hat{Q}^{\bK s,\bK' s'}_{a,b}(\bk)$ becomes nonzero. Numerically, we also look for the finite $B$ analog of the incommensurate Kekul\'e spiral ordered states (IKS) discussed in Ref.~\cite{Kwan2021}. In zero magnetic field, IKS has intervalley coherence, and breaks the moir\'e translation symmetry $\hat{T}_{\bL_i},\ i=1,2$. However the state preserves the discrete moir\'e translation symmetry followed by  U(1) valley gauge transformations: 
\begin{equation}
    e^{-i \frac{1}{2}\bQ_\text{IKS}\cdot \bL_i \hat{\tau}_z}\hat{T}_{\bL_i},
\end{equation}
where $\bQ_\text{IKS}$ is the IKS wavevector, $\hat{\tau}_z$ is the Pauli matrix acting on the valley space ($\hat{\tau}_z\ket{\bK}=\ket{\bK},\ \hat{\tau}_z\ket{\bK'}=-\ket{\bK'}$). In momentum space, this corresponds to a non-vanishing density matrix element: $\langle d^{\dagger}_{\bK, \alpha \bk} d_{\bK' \beta \bk+\bq_0} \rangle \neq 0$ only if $\bq_0\equiv \bQ_\text{IKS}$. 

In finite $B$, the discrete moir\'e translations are replaced by the non-commuting magnetic translations $\hat{t}_{\bL_i}$. We seek analogs of the zero field IKS states by looking for intervalley coherent states which preserves the magnetic translation symmetry followed by  U(1) valley gauge transformations: 
\begin{equation}
    \tilde{t}_{\bL_i} \equiv e^{-i\frac{1}{2}\bQ_\text{IKS}\cdot \bL_i \hat{\tau}_z} \hat{t}_{\bL_i}.
\end{equation}

The many body wavefunction for such  state can be generally written as follows: 
\begin{equation}
    \ket{\Psi_\text{IKS}} = \prod_{n,\bk}'  \prod_{r=0,\dots q-1} \left(\sum_{s a}\alpha^{(n)}_{s a,\bk}d^{\dagger}_{\bK s a,\bk+r\frac{\phi}{\phi_0}\bg_1} + \sum_{s'b}\beta^{(n)}_{s' b,\bk}d^{\dagger}_{\bK' s' b,\bk+\bq_0+r\frac{\phi}{\phi_0}\bg_1} \right) \ket{0}.
\end{equation}
Here, $n$ is the band index of the Bogoliubov quasiparticles, $\bk=k_1\bg_1+k_2\bg_2$, with $k_1,k_2\in[0,1/q)$. The constrained ($'$) product is over all occupied Bogoliubov quasiparticle states. We have split the magnetic Brillouin zone in to $q$ strips along the $\bg_1$ axis, and label each strip by $r=0,\dots q-1$. $\bq_0 = q_{0,1}\bg_1+q_{0,2}\bg_2$ is the relative wavevector between electrons in opposite valleys. $\{\alpha,\beta\}$ are coefficients that we shall constrain below to make above an IKS state respecting the $\tilde{t}_{\bL_i}$ symmetries. 

Note that from the gauge fixing procedure discussed previously, the following relation applies,
\begin{equation}
    \hat{t}_{\bL_1} d^{\dagger}_{\eta s a,\bk} \hat{t}_{\bL_1}^{-1} = e^{-i 2\pi k_1} d^{\dagger}_{\eta s a,\bk},\ \hat{t}_{\bL_2} d^{\dagger}_{\eta s a,\bk} \hat{t}_{\bL_2}^{-1} = e^{-i 2\pi k_2} d^{\dagger}_{\eta s a,\bk+\frac{\phi}{\phi_0}\bg_1}.
\end{equation}
Note here that $\hat{t}_{\bL_2}$ boosts the magnetic momentum $\bk$ by the magnetic flux ratio $\phi/\phi_0$ along the $\bg_1$ axis.  

We begin by noting that $\ket{\Psi_\text{IKS}}$ is an eigenstate of $e^{-i\frac{1}{2}\bq_0\cdot \bL_1 \hat{\tau}_z}\hat{t}_{\bL_1}$, as: 
\begin{equation}
\begin{split}
    & e^{-i\frac{1}{2}\bq_0\cdot \bL_1 \hat{\tau}_z}\hat{t}_{\bL_1} \ket{\Psi_\text{IKS}} \\
    = &  \prod_{n,\bk,r}' \left( e^{-i\frac{1}{2}\bq_0\cdot\bL_1 }e^{-i2\pi (k_1+r\frac{\phi}{\phi_0})}\sum_{s a}\alpha^{(n)}_{s a,\bk}d^{\dagger}_{\bK s a,\bk+r\frac{\phi}{\phi_0}\bg_1} + e^{i\frac{1}{2}\bq_0\cdot\bL_1 }e^{-i2\pi (k_1+q_{0,1}+r\frac{\phi}{\phi_0})} \sum_{s'b}\beta^{(n)}_{s' b,\bk}d^{\dagger}_{\bK' s' b,\bk+\bq_0+r\frac{\phi}{\phi_0}\bg_1} \right) \ket{0} \\
    = & \left( \prod_{n,\bk,r}' e^{-i 2\pi (k_1+\frac{q_{0,1}}{2}+r\frac{\phi}{\phi_0})}\right) \ket{\Psi_\text{IKS}}.
\end{split}
\end{equation}

Note also that: 
\begin{equation}
\begin{split}
    & e^{-i\frac{1}{2}\bq_0\cdot \bL_2 \hat{\tau}_z}\hat{t}_{\bL_2} \ket{\Psi_\text{IKS}} \\
    = &  \prod_{n,\bk,r}' \left( e^{-i\frac{1}{2}\bq_0\cdot\bL_2 }e^{-i2\pi k_2}\sum_{s a}\alpha^{(n)}_{s a,\bk}d^{\dagger}_{\bK s a,\bk+r\frac{\phi}{\phi_0}\bg_1} + e^{i\frac{1}{2}\bq_0\cdot\bL_2 }e^{-i2\pi (k_2+q_{0,2})} \sum_{s'b}\beta^{(n)}_{s' b,\bk}d^{\dagger}_{\bK' s' b,\bk+\bq_0+(r+1)\frac{\phi}{\phi_0}\bg_1} \right) \ket{0} \\
    = & \left( \prod_{n,\bk,r}' e^{-i 2\pi (k_2+\frac{q_{0,2}}{2})}\right) \prod_{n,\bk,r}' \left( \sum_{s a}\alpha^{(n)}_{s a,\bk}d^{\dagger}_{\bK s a,\bk+r\frac{\phi}{\phi_0}\bg_1} + \sum_{s'b}\beta^{(n)}_{s' b,\bk}d^{\dagger}_{\bK' s' b,\bk+\bq_0+(r+1)\frac{\phi}{\phi_0}\bg_1} \right) \ket{0}.
\end{split}
\end{equation}
Note that by constraining: 
\begin{equation}
\begin{split}
    \alpha_{sa,\bk+(r-1)\frac{\phi}{\phi_0}}^{(n)} & = \alpha_{sa,\bk+r\frac{\phi}{\phi_0}}^{(n)} e^{i \theta_{n\bk r}}, \\
    \beta_{s'b,\bk+(r-1)\frac{\phi}{\phi_0}}^{(n)} & = \beta_{s'b,\bk+r\frac{\phi}{\phi_0}}^{(n)} e^{i (\theta_{n\bk r}-\varphi)}, \\
    \alpha_{sa,\bk+(r-1)\frac{\phi}{\phi_0}}^{(n)*}\beta_{s'b,\bk+(r-1)\frac{\phi}{\phi_0}}^{(n)} & =\alpha_{sa,\bk+r\frac{\phi}{\phi_0}}^{(n)*}\beta_{s'b,\bk+r\frac{\phi}{\phi_0}}^{(n)} e^{-i\varphi},
\end{split}
\end{equation}
where $\theta_{n\bk}$ is independent on the spin and magnetic subband quantum numbers, and $\varphi$ is independent of all quantum numbers, we have: 
\begin{equation}
e^{-i\frac{1}{2}\varphi \hat{\tau}_z}e^{-i\frac{1}{2}\bq_0\cdot \bL_2 \hat{\tau}_z}\hat{t}_{\bL_2} \ket{\Psi_\text{IKS}}
    =  \left( \prod_{n,\bk,r}' e^{-i 2\pi (k_2+\frac{q_{0,2}}{2})-i\frac{\varphi}{2}}e^{-i\theta_{n\bk r}}\right) \ket{\Psi_{\text{IKS}}}.
\end{equation}

To summarize, for a state given by: 
\begin{equation}
    \ket{\Psi_\text{IKS}} = \prod_{n,\bk}'  \prod_{r=0,\dots q-1} \left(\sum_{s a}\alpha^{(n)}_{s a,\bk}d^{\dagger}_{\bK s a,\bk+r\frac{\phi}{\phi_0}\bg_1} + \sum_{s'b}\beta^{(n)}_{s' b,\bk}d^{\dagger}_{\bK' s' b,\bk+\bq_0+r\frac{\phi}{\phi_0}\bg_1} \right) \ket{0}
\end{equation}
and satisfying the constraints: 
\begin{equation}
\alpha_{sa,\bk+(r-1)\frac{\phi}{\phi_0}}^{(n)} = \alpha_{sa,\bk+r\frac{\phi}{\phi_0}}^{(n)} e^{i \theta_{n\bk r}}, \ 
    \beta_{s'b,\bk+(r-1)\frac{\phi}{\phi_0}}^{(n)} = \beta_{s'b,\bk+r\frac{\phi}{\phi_0}}^{(n)} e^{i (\theta_{n\bk r}-\varphi)},
\end{equation}
it is a finite $B$ analog of the zero field IKS state, with intervalley coherence and preserves magnetic translations followed by a U(1) valley gauge transformation: 
\begin{equation}
    \tilde{t}_{\bL_i} \equiv e^{-i\frac{1}{2}\bQ_\text{IKS}\cdot \bL_i \hat{\tau}_z} \hat{t}_{\bL_i}.
\end{equation}

The IKS wavevector is given by: 
\begin{equation}
    \bQ_{\text{IKS}} = \bq_0 + \frac{\varphi}{2\pi} \bg_2.
\end{equation}
Note due to the cyclic nature of the state under $\hat{t}_{\bL_2}$, $\frac{\varphi}{2\pi} = n \frac{\phi}{\phi_0}$ where $n\in \mathbb{Z}$. 

IKS-like states can be studied using the density matrix as defined in Eq.~\ref{eq:den_mat_SM}. On the technical side, for any $\bq_0$, we can displace the momentum space mesh of valley $\bK'$ from valley $\bK$ by $\bq_0$, before carrying out the (magnetic) translationally invariant B-SCHF calculations. This is how we probed these states in the paper, by optimizing the ground state energy with respect to $\bq_0$.   

We look at the constraints imposed on the density matrix (Eq.~\ref{eq:den_mat_SM}) by an IKS ground state. Without confusion, we simplify the notations by redefining fermionic operators in valley $\bK'$ as $d_{\bK's b,\bk+\bq_0} \equiv d_{\bK' s b, \bk}$ in case $\bq_0\neq 0$, and keep the definition of fermionic operators the same in valley $\bK$. We get:  
\begin{equation}
\begin{split}
     & \hat{P}^{\eta s,\eta' s'}_{a,b}(\bk) + \frac{1}{2}\delta_{a,b}\delta_{\eta,\eta'}\delta_{s,s'}\\
     = & \bra{\Psi_{\text{IKS}}} d^{\dagger}_{\eta s a,\bk}d_{\eta' s' b,\bk}\ket{\Psi_{\text{IKS}}} \\
    = &\bra{\Psi_{\text{IKS}}} \tilde{t}_{\bL_2}^{-1}\tilde{t}_{\bL_2}d^{\dagger}_{\eta s a,\bk}d_{\eta' s' b,\bk} \tilde{t}_{\bL_2}^{-1}\tilde{t}_{\bL_2}\ket{\Psi_{\text{IKS}}} \\
     = & e^{i \pi (\eta'-\eta)({Q_{\text{IKS},2}}-q_{0,2})} \left[ \hat{P}^{\eta s,\eta' s'}_{a,b}\left(\bk+ \frac{\phi}{\phi_0}\bg_1\right) + \frac{1}{2}\delta_{a,b}\delta_{\eta,\eta'}\delta_{s,s'} \right]\\
     = & e^{i \frac{\varphi}{2} (\eta'-\eta)} \hat{P}^{\eta s,\eta' s'}_{a,b}\left(\bk+ \frac{\phi}{\phi_0}\bg_1\right) + \frac{1}{2}\delta_{a,b}\delta_{\eta,\eta'}\delta_{s,s'},
\end{split}
\end{equation}
where $\eta,\eta' = +(-) 1$ denoting valley $\bK$ ($\bK'$). This property of the density matrix for an IKS-like state can be used to reduce the calculation of the Hartree-Fock energy (Eq.~\ref{eq:mf_energy}) by a factor of $q$. 

\subsection{Stripe states}
Unlike IKS states, stripe states break the magnetic translation symmetries, and lead to density/spin modulations on the moir\'e scale. In our code, we can probe stripe states commensurate with the enlarged unit cell. Specifically in the Landau gauge, the unit cell is chosen to be 1 moir\'e unit cell along $\bL_1$ direction and $q$ moir\'e unit cells along $\bL_2$ direction. This allows us to probe stripe states with periods divisible by $q$. For example, at $\phi/\phi_0=1/8$, we could probe period 2, 4 stripes along $\bL_2$, but not period 3, 6 and so on. We could not probe checkerboard states, nor stripe states with modulations along $\bL_1$. This is important to keep in mind when examining the phase diagrams in the paper.

\clearpage
\section{Extended Figures}
\setcounter{figure}{0}
\renewcommand{\thefigure}{S\arabic{figure}}

\subsection{Extended results in the presence of heterostrain}
Here we present B-SCHF results obtained for various twist angles in the presence of heterostrain. 

\begin{figure}[h]
\centering 
\includegraphics[width=\linewidth]{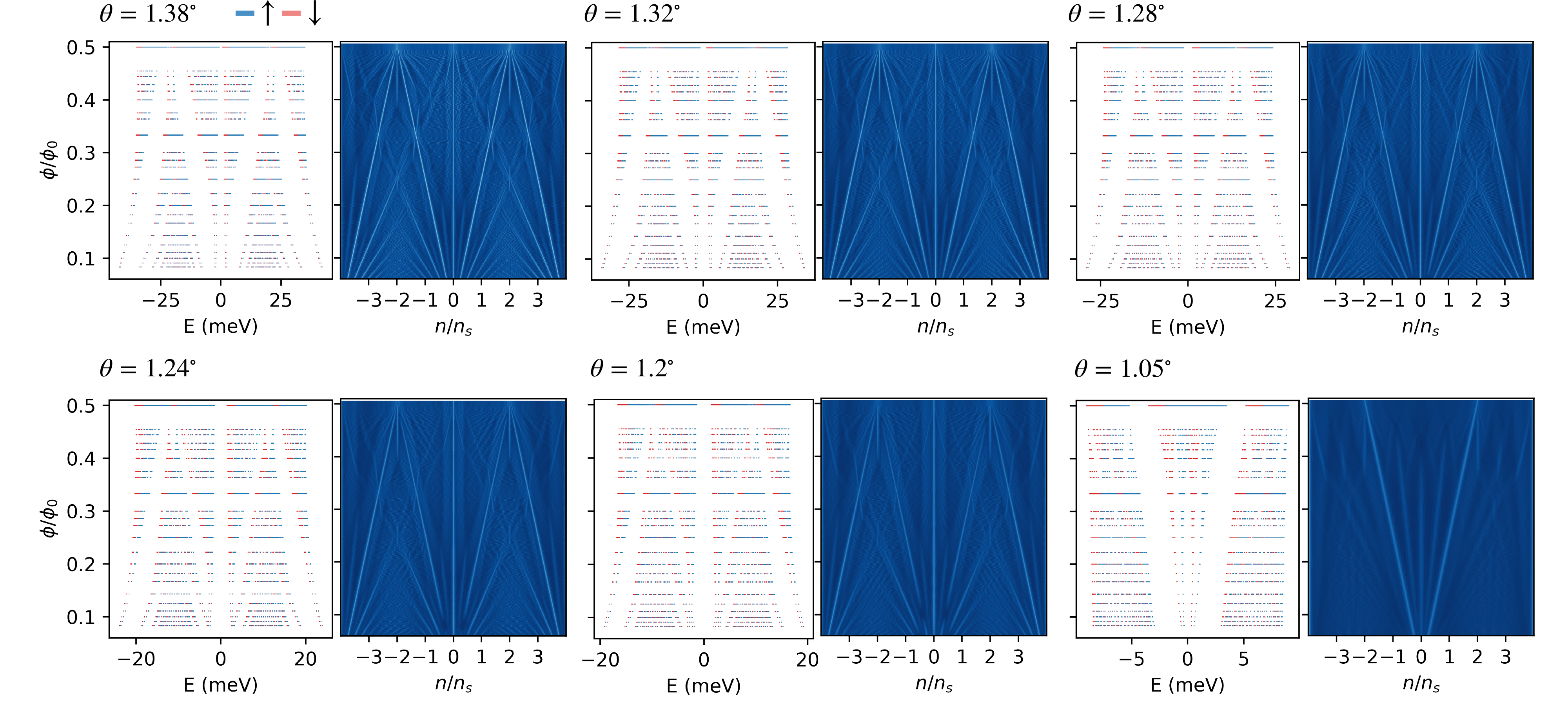}
\caption{\label{fig:strain_nonint} Non-interacting Hofstadter spectra and Wannier plots for various twist angles with heterostrain. In the Wannier plots, bright colors correspond to low density of states (gapped) and darker blue colors correponsd to high density of states (compressible). In obtaining the Wannier plots we used an energy broadening factor of $\gamma=0.1$meV. }
\end{figure}

\begin{figure}
\centering 
\includegraphics[width=0.8\linewidth]{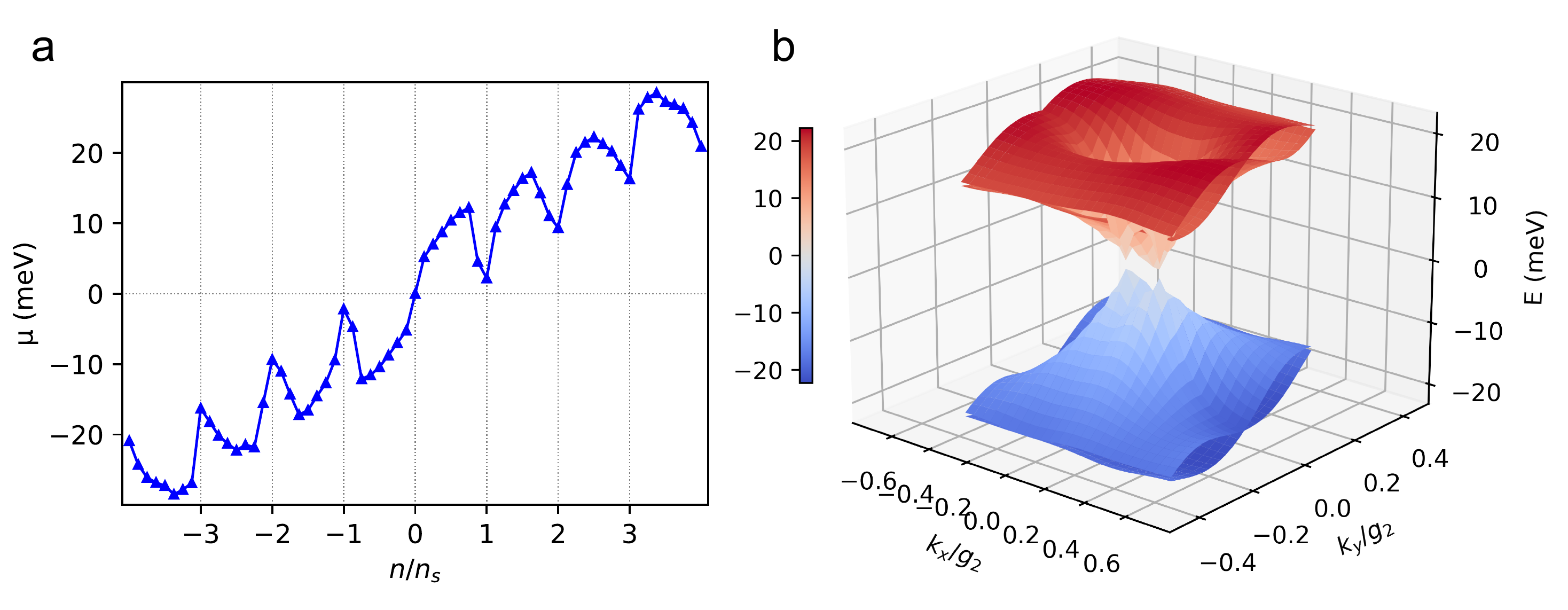}
\caption{\label{fig:strain_B0} (a) Chemical potential $\mu$ versus filling $n/n_s$ for $\theta=1.05^\circ$, zero magnetic field, and  with heterostrain. Results obtained for $16\times16$ mesh assuming moir\'e translation symmetry is preserved. (b) 3D plot of Hartree-Fock renormalized energy dispersions at $n/n_s=0$ for the same parameterization, showing semimetallic behavior. Dispersion obtained for $25\times25$ mesh.}
\end{figure}

\begin{figure}
\centering 
\includegraphics[width=0.6\linewidth]{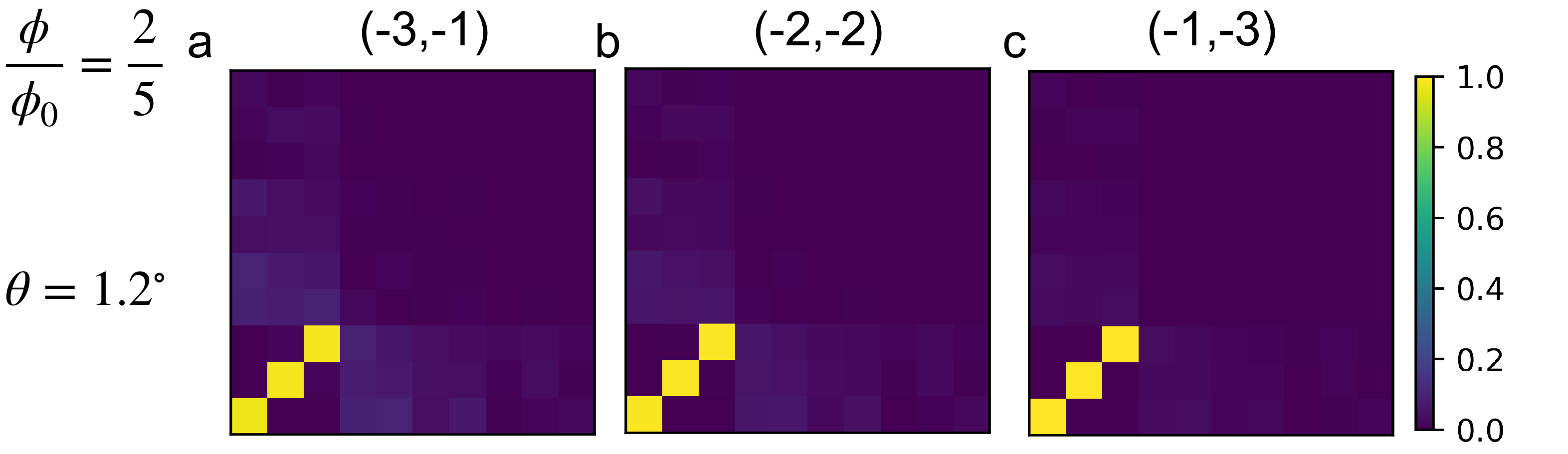}
\caption{\label{fig:strain_renormalized_chf} Absolute values of the density matrices at $\bk=\mathbf{0}$ of an occupied valley/spin flavor for the correlated Hofstadter ferromagnets (CHFs), expressed in the eigenbasis of $(0,-4)$. Results are obtained at flux ratio $\phi/\phi_0=2/5$ and in the presence of heterostrain. Here we show that the CHFs are flavor symmetry breaking states that predominantly occupy the lower Chern $-1$ group of renormalized magnetic subbands obtained at $(0,-4)$.}
\end{figure}

\begin{figure}
\centering 
\includegraphics[width=\linewidth]{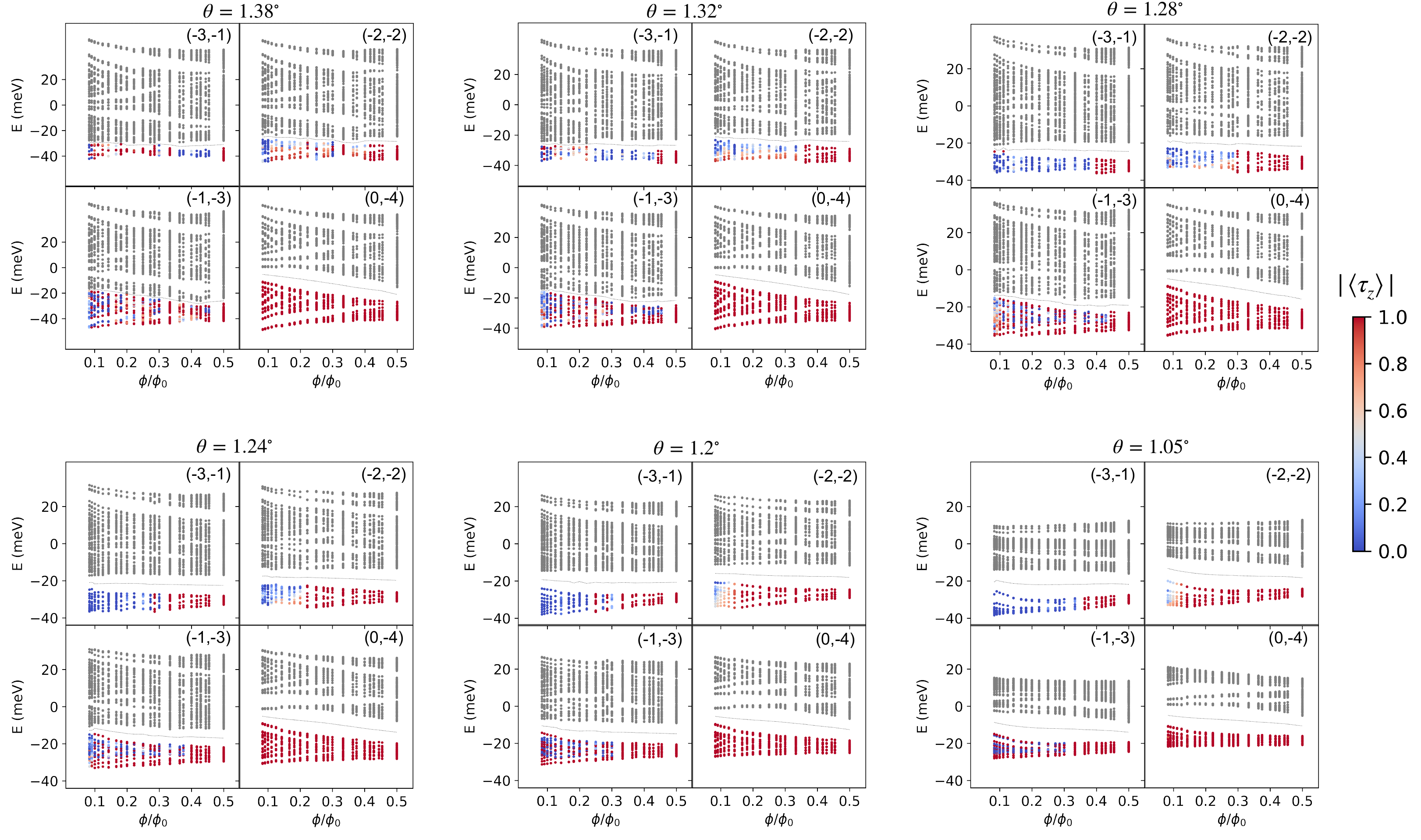}
\caption{\label{fig:strain_hf_energies} Hartree-Fock energy spectra along $(-3,-1)$, $(-2,-2)$, $(-1,-3)$, and $(0,-4)$, for all six twist angles studied. Single electronic states below the dashed lines are occupied, and colored by the absolute value of the state's valley polarization $|\langle \tau_z\rangle|$. $|\langle \tau_z\rangle|\rightarrow 1 (0)$ implies maximal valley polarization (intervalley mixing), in the same format as that of Fig.~3 of the main text. At higher $B$, CHFs which do not break valley/spin symmetries are energetically favorable. At lower $B$, CHFs transition into IKSs. Another transition from IKS into nearly compressible states (e.g., $1.32^\circ$ near $\phi/\phi_0=0.2$) is also observed at higher twist angles. At $1.38^\circ$, there could be QHFM states that are more energetically favorable than either IKS or CHF, e.g., $1.38^\circ$ at $\phi/\phi_0=1/3$ along $(-3,-1)$.}
\end{figure}

\begin{figure}
\centering 
\includegraphics[width=0.9\linewidth]{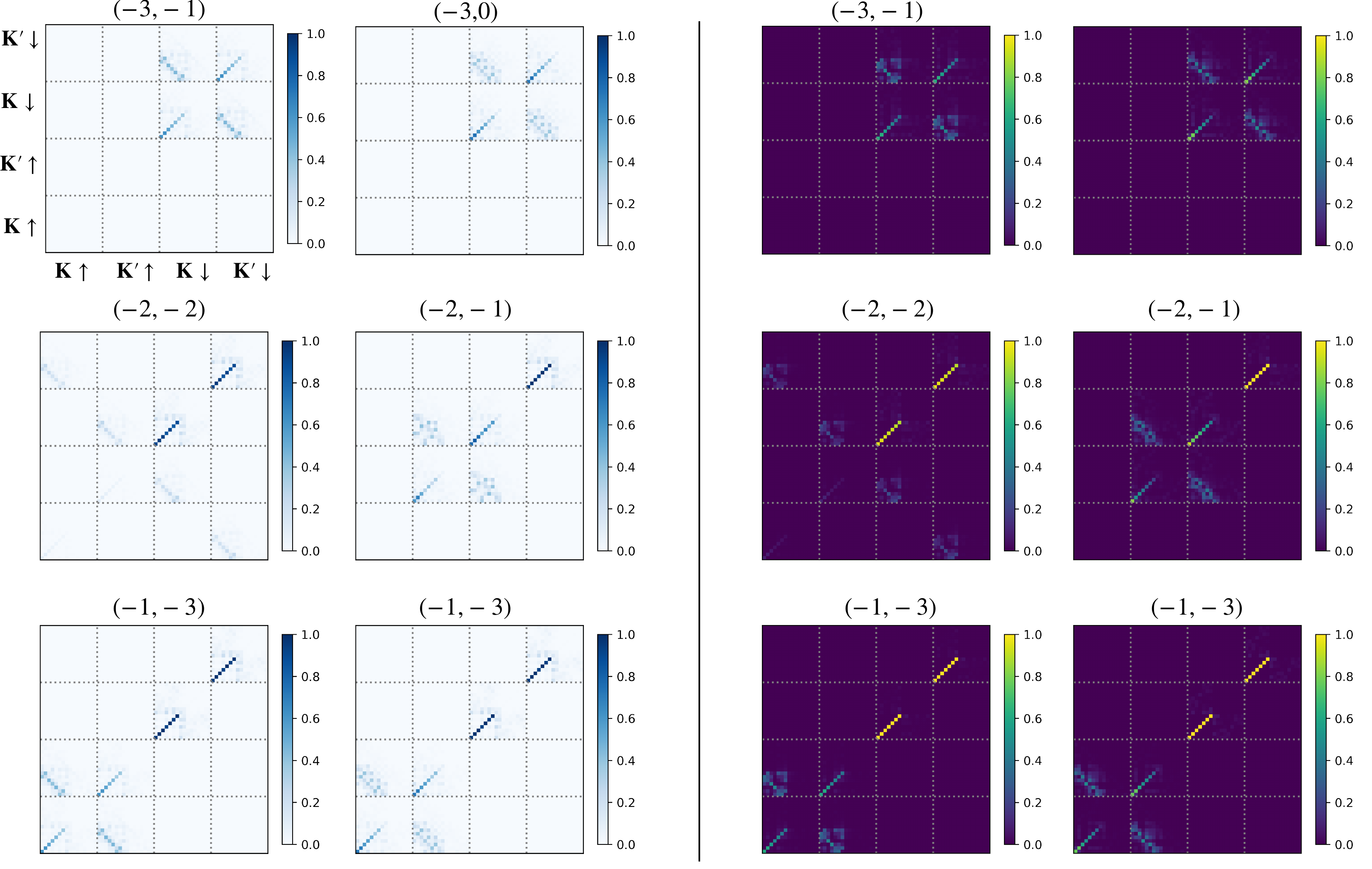}
\caption{\label{fig:strain_additionalCCIs} Left two columns: absolute values of the full density matrices at $\bk=\mathbf{0}$ for the intervalley coherent states at $(-3,-1)$, $(-2,-2)$ and $(-1,-3)$, as well as $(-3,0)$, $(-2,-1)$ and $(-1,-2)$. Results obtained for $\theta=1.05^\circ$, $\phi/\phi_0=1/8$, and with heterostrain. States along the two columns of the same row are both identified as IKS states, but do not share the same wavevector. Within our model parameters, we find along $(-3,-1)$, $(-2,-2)$ and $(-1,-3)$ the IKS wavevector to be $\frac{1}{8}\bg_2$,  and along $(-3,0)$, $(-2,-1)$ and $(-1,-2)$ to be $\frac{1}{2}\bg_1$. Right two columns are density matrices expressed in the eigenbasis of $(0,-4)$.}
\end{figure}

\begin{figure}
\centering 
\includegraphics[width=0.9\linewidth]{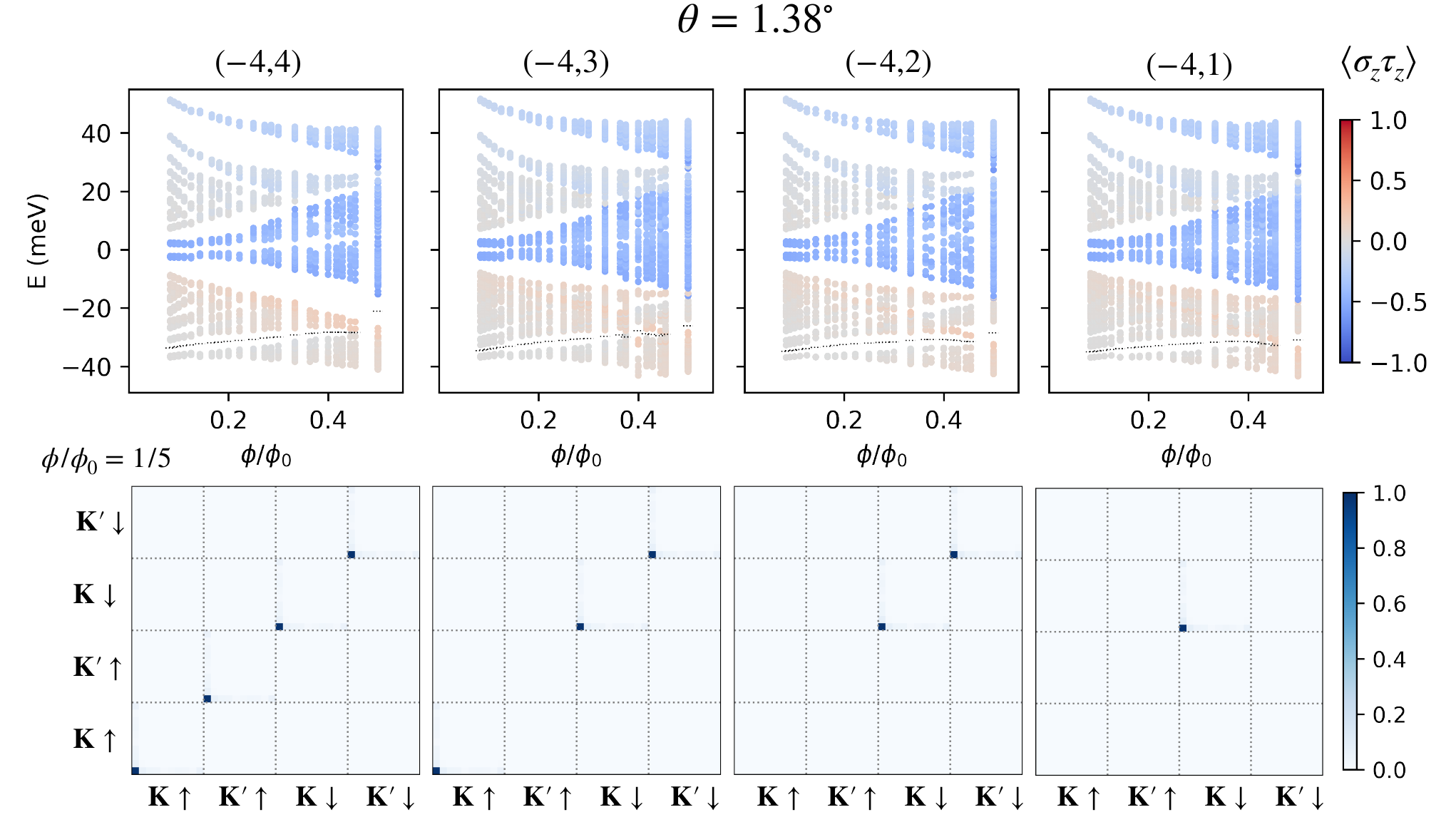}
\caption{\label{fig:strain_QHFM_edge} B-SCHF spectra (upper panel) and the absolute values of the respective density matrices at $\bk=\mathbf{0}$ (lower panel) for QHFM emanating from band bottom, described by Streda lines $(s,t)=(-4,4)$, $(-4,3)$, $(-4,2)$, and $(-4,1)$ (lower panels). Electronic states below the dashed lines are occupied. Results are obtained at twist angle $\theta=1.38^\circ$, and with heterostrain. The density matrices are obtained at $\phi/\phi_0=1/5$. They are predominantly valley/spin polarizations of the zeroth LL emanating from non-interacting band bottom, without intervalley coherence. The qualitative behaviors are the same for all twist angles studied. }
\end{figure}

\begin{figure}
\centering 
\includegraphics[width=\linewidth]{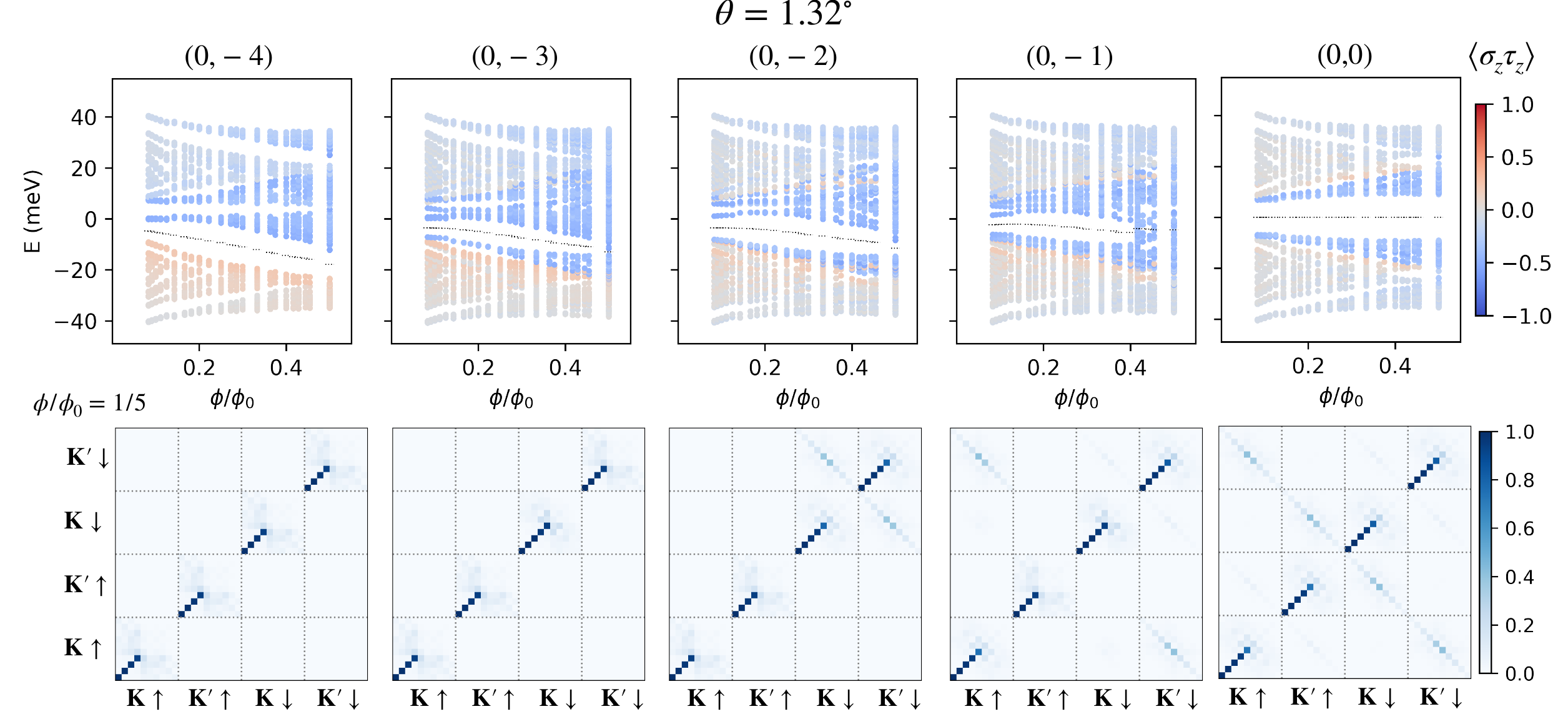}
\caption{\label{fig:strain_QHFM_cnp} B-SCHF spectra (upper panel) and the absolute values of the respective density matrices at $\bk=\mathbf{0}$ (lower panel) for QHFM emanating from the charge neutrality point, described by Streda lines $(s,t)=(0,-4)$, $(0,-3)$, $(0,-2)$, $(0,-1)$, and $(0,0)$(lower panels).  Electronic states below the dashed lines are occupied. Results obtained at twist angle $\theta=1.32^\circ$, and with heterostrain.  The density matrices are obtained at $\phi/\phi_0=1/5$. Note that QHFM from charge neutrality point are predominantly associated with the zeroth LLs of the Dirac points, and develop intevalley coherence, in contrast to QHFM from band bottom.  The qualitative behaviors are the same for all twist angles studied.}
\end{figure}

\begin{figure}
\centering 
\includegraphics[width=0.9\linewidth]{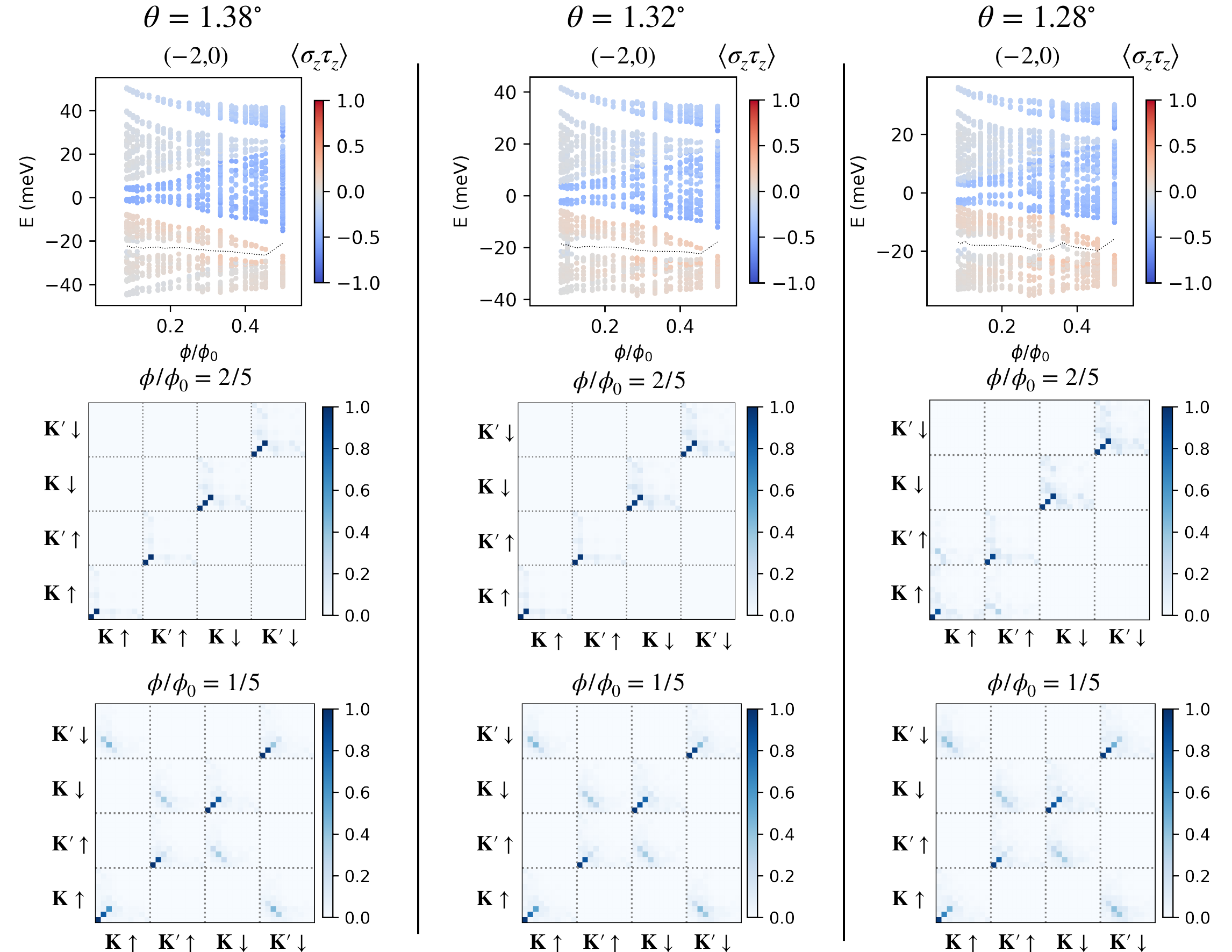}
\caption{\label{fig:strain_qsh} Gapped states along $(-2,0)$ at higher magnetic flux ratios for $1.38^\circ$, $1.32^\circ$ and $1.28^\circ$ (upper panel), as well as the absolute values of the representative density matrices at $\bk=\mathbf{0}$ at $\phi/\phi_0=p/q=2/5$ (middle panel) and $\phi/\phi_0=1/5$ (lower panel). In the upper panel, electronic states below the dashed lines are occupied. At higher $\phi/\phi_0$, these states are adiabatically connected to the non-interacting Quantum Spin Hall (QSH) insulators shown in the Wannier diagrams of Fig.~\ref{fig:strain_nonint}, where the magnetic subband group developed from the lowest LL for the up spin sector ($p$ diagonal squares at the lower left corner of the density matrix for a given valley/spin flavor), and the magnetic subband group below the zeroth LL near charge neutrality for the down spin sector ($q-p$ diagonal squares at the lower left corner of the density matrix for a given valley/spin flavor), are occupied. At $1.28^\circ$, some intervalley mixing is energetically favored than pure valley/spin polarization. At lower $\phi/\phi_0$, the QSHs evolve into IKS states as shown in the lower panel.}
\end{figure}

\begin{figure}
\centering 
\includegraphics[width=\linewidth]{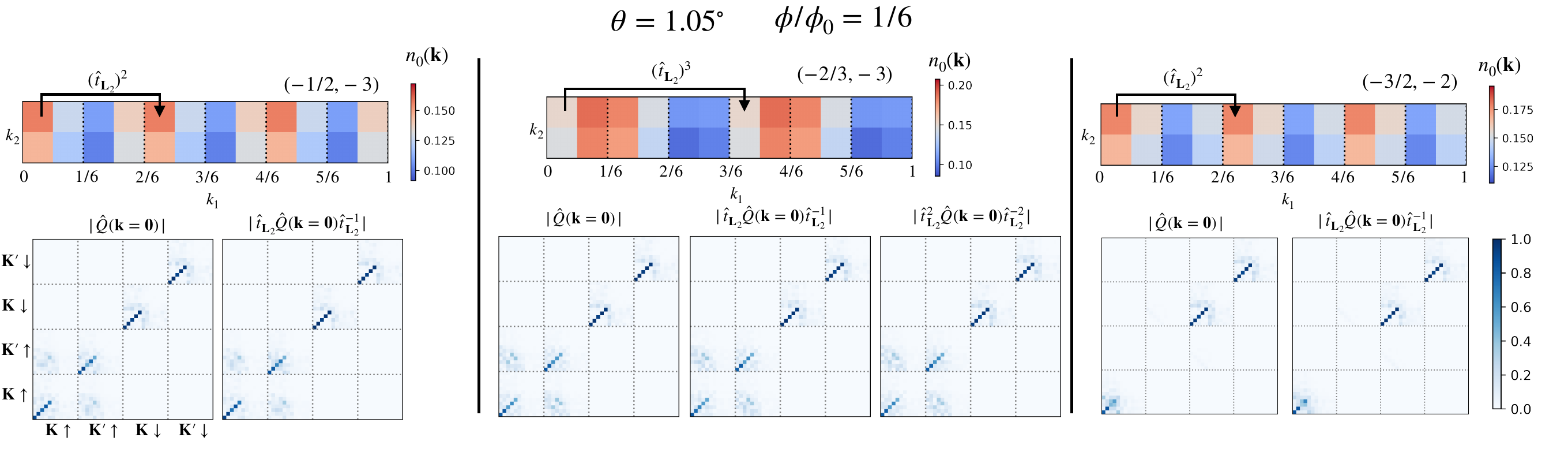}
\caption{\label{fig:strain_fractionalS} Examples of CCIs with fractional $s$. Results obtained at $1.05^\circ$ and $\phi/\phi_0=1/6$ with heterostrain. $n_0(\bk)$ is the occupation number of the zeroth LL emanating from the Dirac points of the non-interacting spectra, in the same notation as Fig.~7 of the main text.}
\end{figure}

\begin{figure}
\centering 
\includegraphics[width=\linewidth]{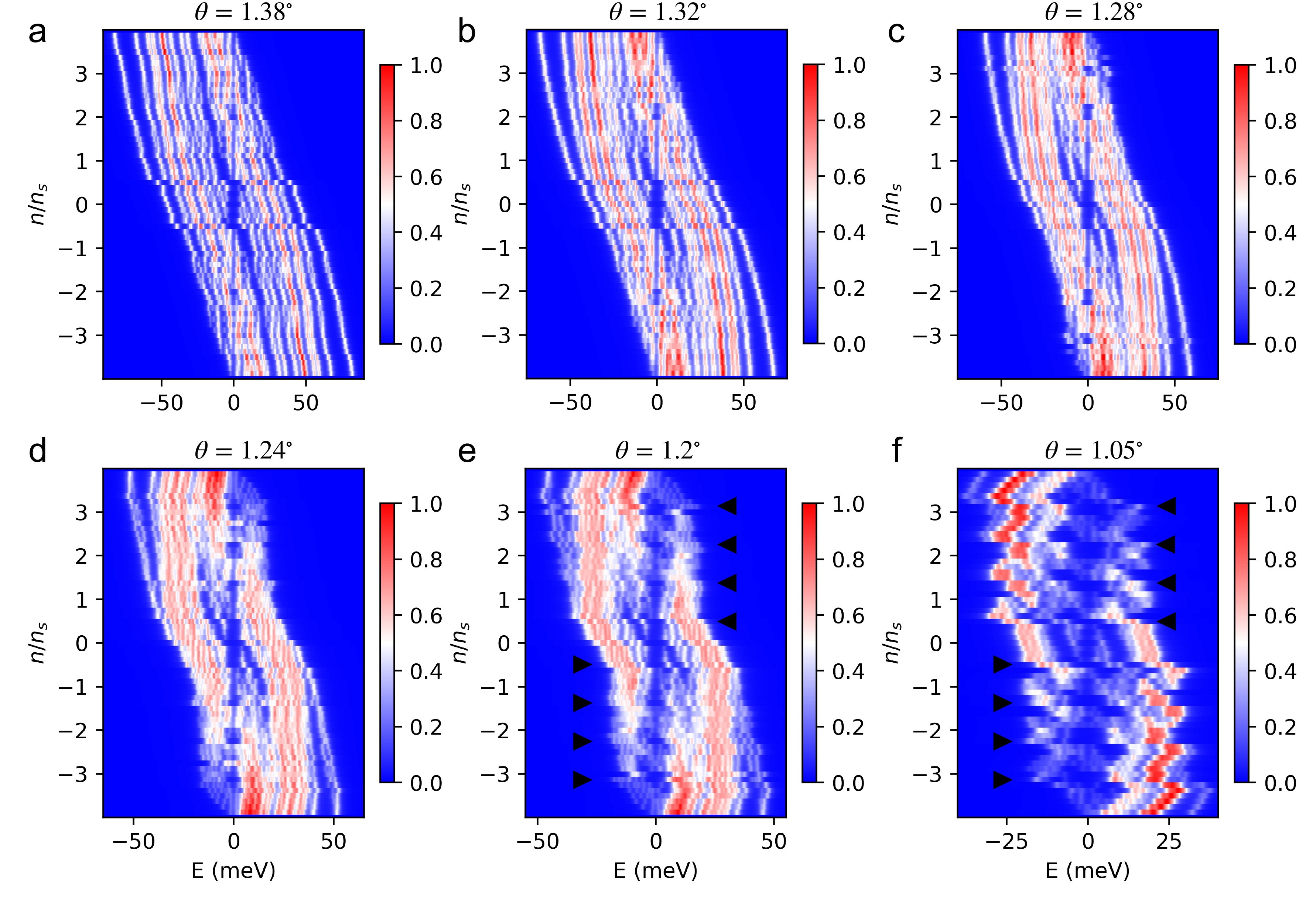}
\caption{\label{fig:ext_fig7} Calculated density of states $\mathcal{N}(E)\propto \frac{1}{\pi}\sum_{i} \frac{\gamma}{\gamma^2+(E-\epsilon_i)^2}$ at $\phi/\phi_0=1/8$ for various twist angles and with heterostrain. $\mathcal{N}(E)$ is normalized with respect to the maximum value. $E=0$ corresponds to the position of the chemical potential $\mu$ at a given filling $n/n_s$. Here we used an energy broadening factor of $\gamma=1\mathrm{meV}$. Dark arrows in (e) and (f) mark electron densities at $(s,t)=(0,\pm4),\pm(1,3),\pm(2,2),\pm(3,1)$. }
\end{figure}

\clearpage
\subsection{Extended results in the absence of heterostrain}
Here we present B-SCHF results obtained for various twist angles in the absence of heterostrain. 

\begin{figure}[h]
\centering 
\includegraphics[width=\linewidth]{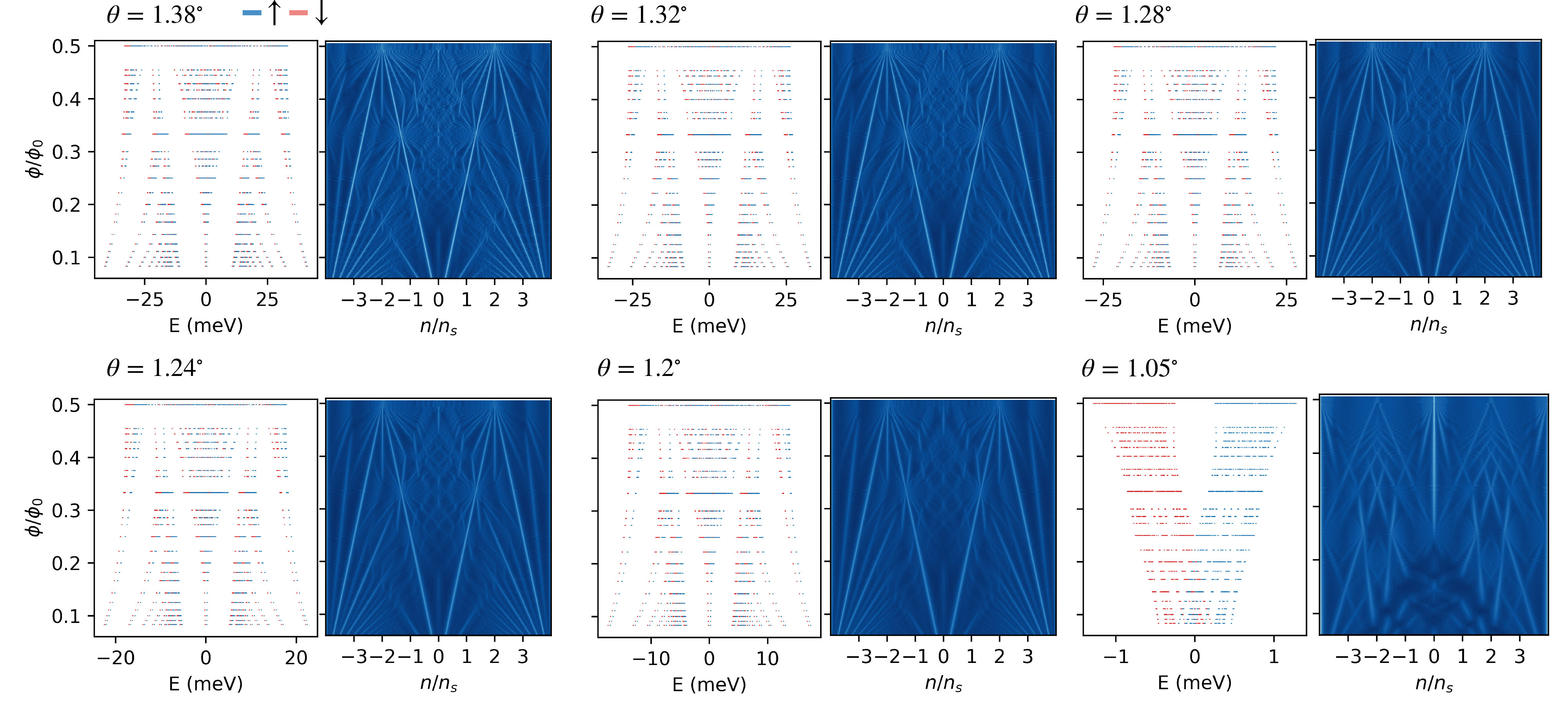}
\caption{\label{fig:nostrain_nonint} Non-interacting Hofstadter spectra and Wannier plots for various twist angles without heterostrain. In the Wannier plots, bright colors correspond to low density of states (gapped) and darker blue colors correponsd to high density of states (compressible). In obtaining the Wannier plots we used an energy broadening factor of $\gamma=0.1$meV, with the exception of $1.05^\circ$ where we used $\gamma=0.01$meV. }
\end{figure}


\begin{figure}
\centering 
\includegraphics[width=\linewidth]{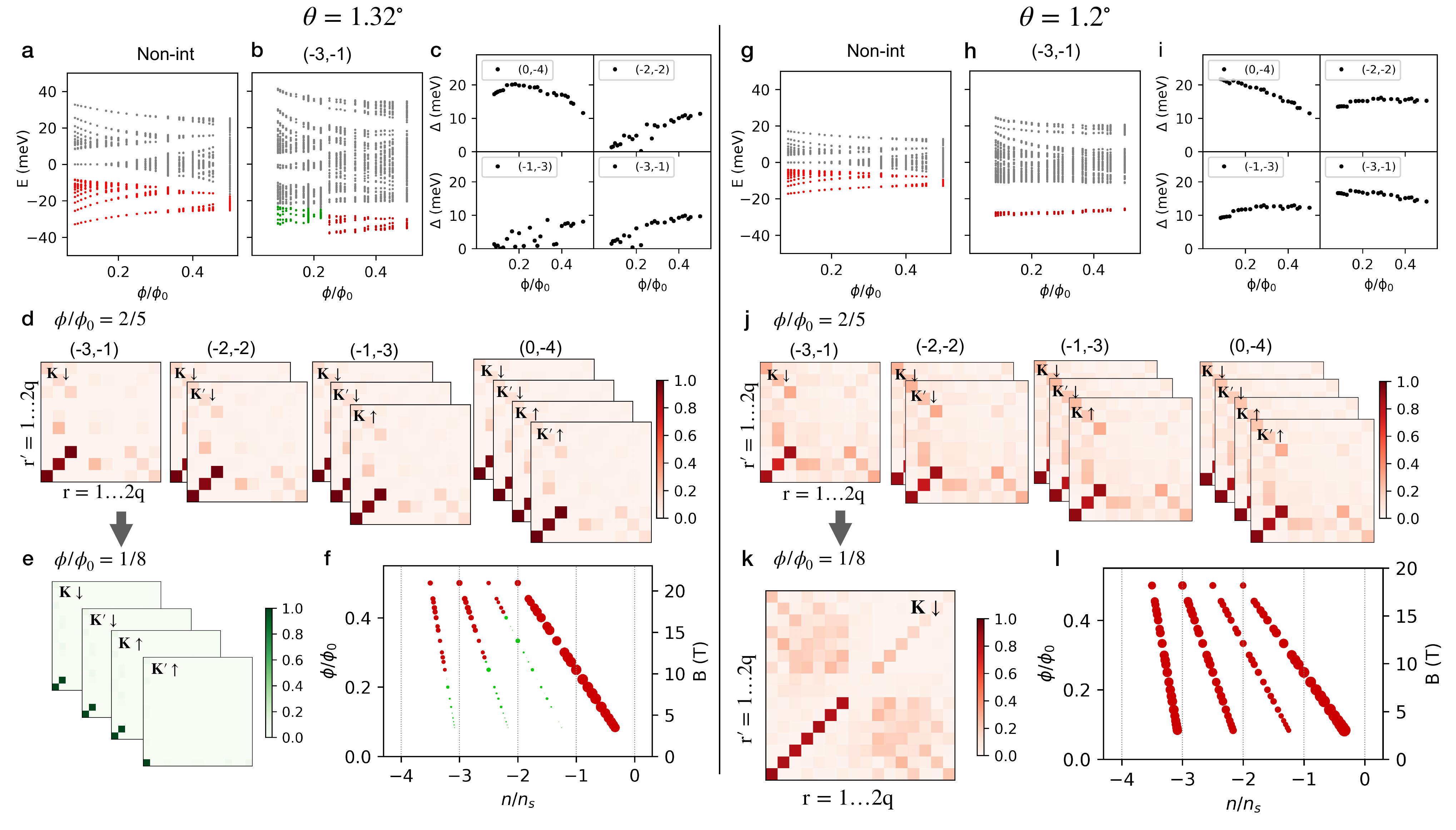}
\caption{\label{fig:nostrain_mainCCI}
Compilation of B-SCHF results at two representative twist angles, $\theta=1.32^\circ$ (a-f) and $\theta=1.2^\circ$ (g-l), in the absence of heterostrain. The non-interacting spectra are shown in (a) and (g). The renormalized spectra along $(s,t)=(-3,-1)$ are shown in (b) and (h). The occupied electronic states are colored in red and green at $1.32^\circ$, signifying the first order phase transition between CHFs (red) and nearly compressible states (green).  The occupied electronic states are colored in red at $1.2^\circ$, and identified as CHFs down to flux ratio $\phi/\phi_0=1/12$ studied here. (c) and (i) show the non-monotonic behaviors of single-particle excitation gap $\Delta$ along $(s,t)=(0,-4)$, $(-1,-3)$, $(-2,-2)$, and $(-3,-1)$ Density matrix analysis. (d) and (j) show the absolute value of the non-vanishing matrix elements of the density matrices at $\phi/\phi_0=2/5$ at $\bk=\mathbf{0}$. $(e)$ and $(j)$ are the absolute values of the representative density matrices at $\phi/\phi_0$ for a nearly compressible state (e) and an intervalley coherent state (k). (f) and (l) are replots of the single-particle excitation gap  $\Delta(n/n_s,\phi/\phi_0)$, colored in the same manner as (b) and (h) respectively. }
\end{figure}

\begin{figure}
\centering 
\includegraphics[width=0.9\linewidth]{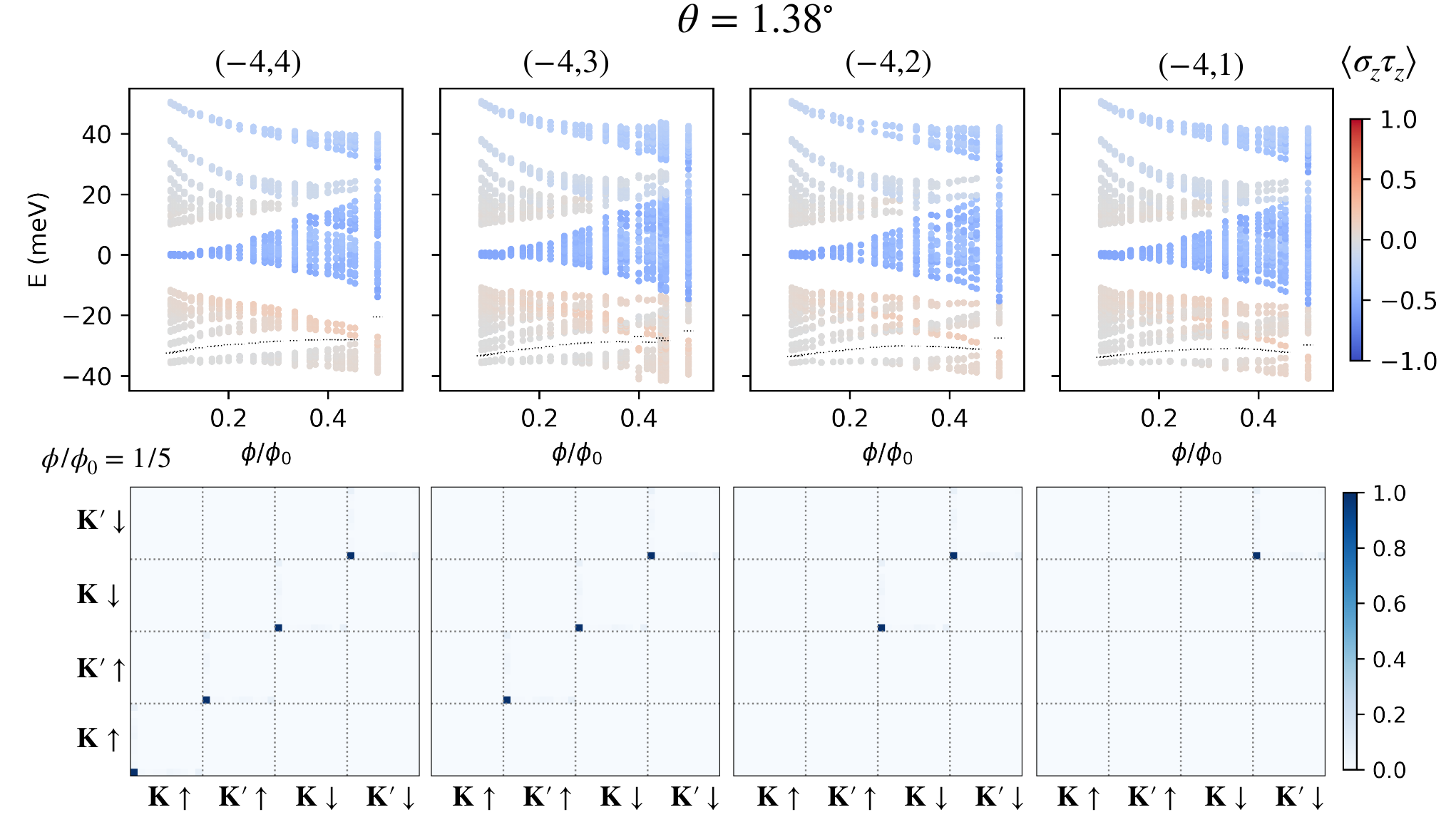}
\caption{\label{fig:nostrain_QHFM_edge} B-SCHF spectra (upper panel) and the absolute values of the respective density matrices at $\bk=\mathbf{0}$ (lower panel) for QHFM emanating from band bottom, described by Streda lines $(s,t)=(-4,4)$, $(-4,3)$, $(-4,2)$, and $(-4,1)$ (lower panels).  Electronic states below the dashed lines are occupied.  Results obtained at the higher twist angle $\theta=1.38^\circ$, and without heterostrain. The density matrices are obtained at $\phi/\phi_0=1/5$. They are predominantly valley/spin polarizations of the zeroth LL emanating from non-interacting band bottom, without intervalley coherence. The qualitative behaviors are the same for all twist angles.}
\end{figure}

\begin{figure}
\centering 
\includegraphics[width=\linewidth]{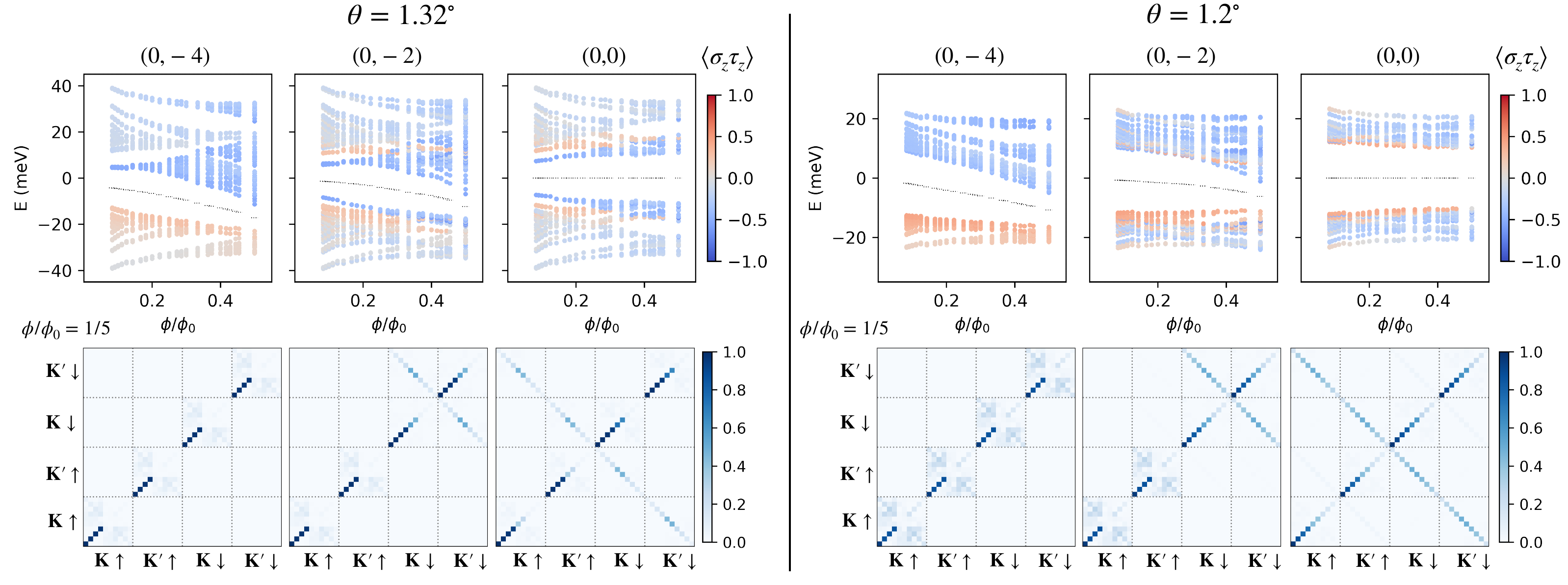}
\caption{\label{fig:nostrain_QHFM_cnp} B-SCHF spectra (upper panel) and the absolute values of the respective density matrices at $\bk=\mathbf{0}$ (lower panel) for QHFM emanating from the charge neutrality point, described by Streda lines $(s,t)=(0,-4)$, $(0,-2)$, and $(0,0)$ (lower panels).  Electronic states below the dashed lines are occupied.  Results obtained at twist angles $\theta=1.32^\circ$ and $1.2^\circ$ , and without heterostrain.  The density matrices are obtained at $\phi/\phi_0=1/5$. Note that QHFM from charge neutrality point are predominantly associated with the zeroth LLs of the Dirac points, and develop intervalley coherence, in contrast to QHFM from band bottom.  The qualitative behaviors are the same for all twist angles.}
\end{figure}

\begin{figure}
\centering 
\includegraphics[width=0.9\linewidth]{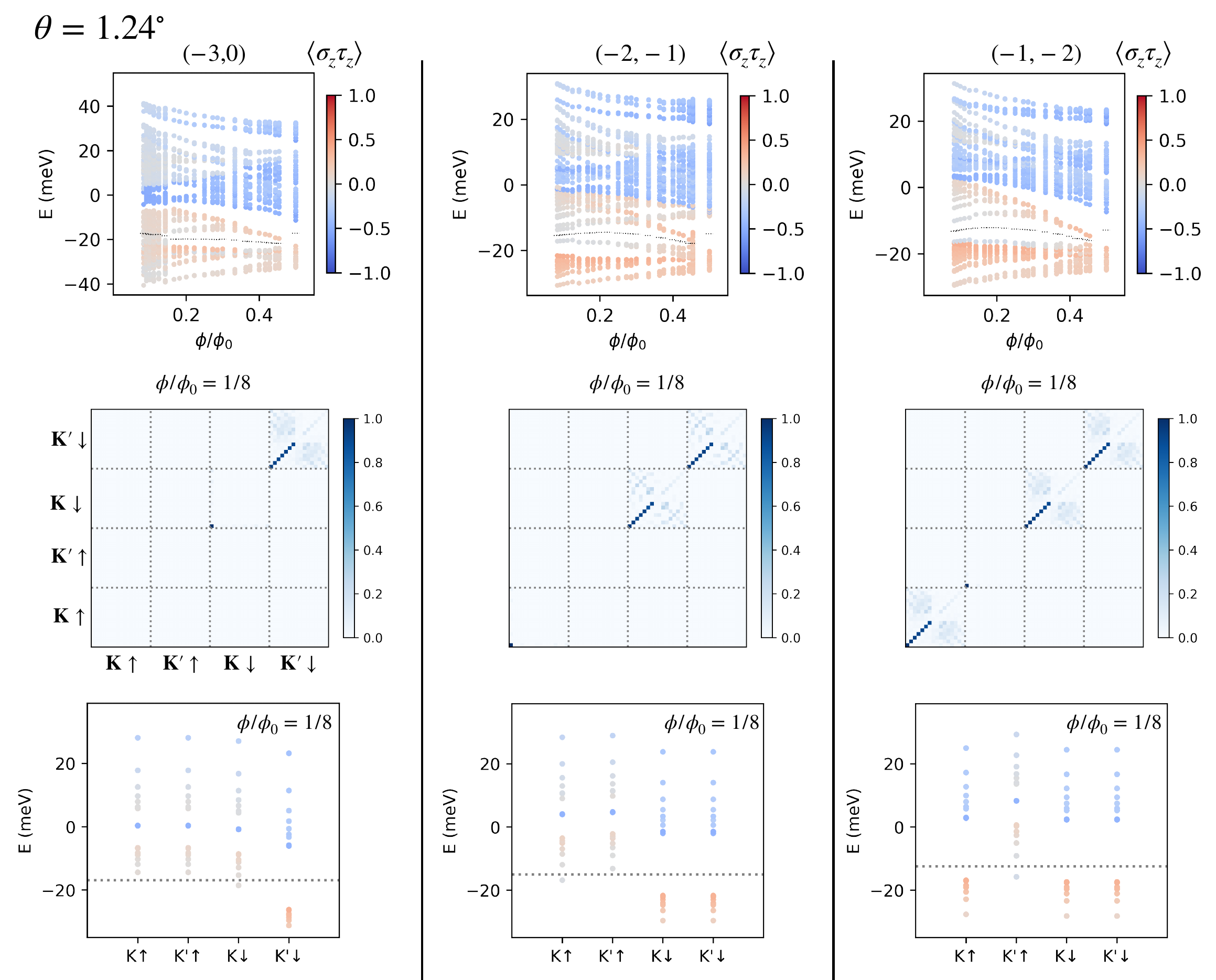}
\caption{\label{fig:nostrain_additionalCCIs} B-SCHF spectra (upper panel), representative density matrix $|\hat{Q}_{\eta s r,\eta's'r'}(\bk=\mathbf{0})|$ at $\phi/\phi_0=1/8$(middle panel) and flavor resolved spectra (lower panel), described by Streda lines $(s,t)=(-3,0)$, $(-2,-1)$, and $(-1,-2)$.  Electronic states below the dashed lines are occupied.  These are secondary CCIs with a $+1$ higher Chern number than the main CCIs along $(-3,-1)$ $(-2,-2)$ $(-1,-3)$. Results obtained at $\theta=1.24^\circ$, and without heterostrain. Unlike results with heterostrain (see Fig.~\ref{fig:strain_additionalCCIs}), here these states do not exhibit intervalley coherence. Rather, the middle and lower panels clearly demonstrate that the one extra Chern number compared to the main CCIs at $(-3,-1)$, $(-2,-2)$, and $(-1,-3)$ comes from populating the one zeroth LL from the band bottom.}
\end{figure}

\begin{figure}
\centering 
\includegraphics[width=\linewidth]{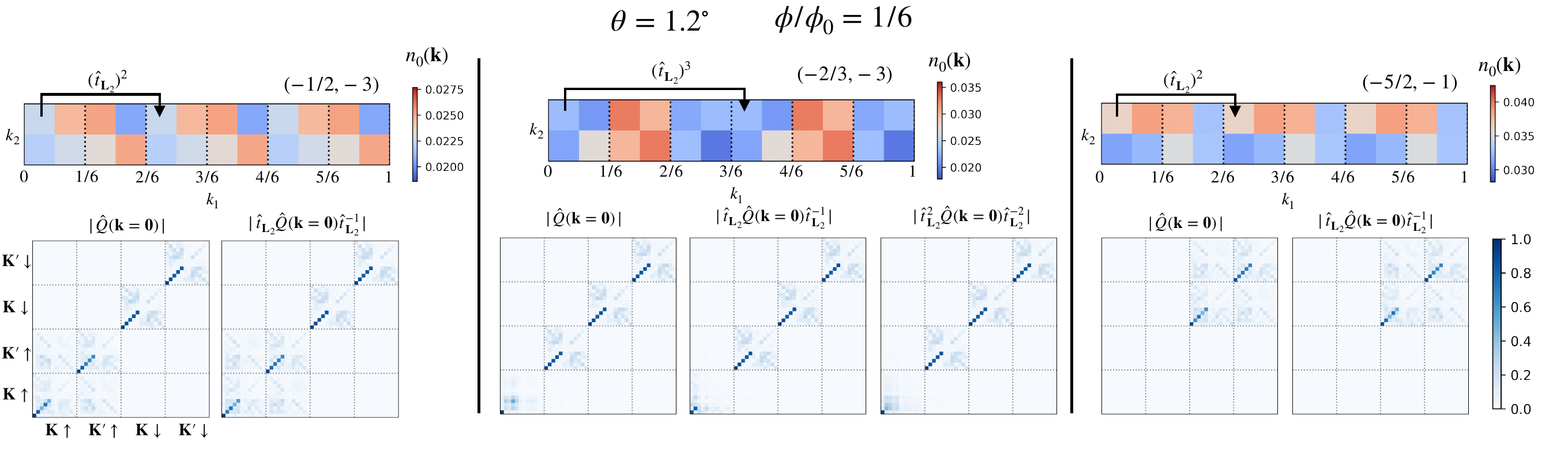}
\caption{\label{fig:nostrain_fractionalS} Examples of CCIs with fractional $s$.  Results are obtained at $1.2^\circ$, $\phi/\phi_0=1/5$, and without heterostrain. $n_0(\bk)$ is the occupation number of the zeroth LL emanating from the Dirac points of the non-interacting spectra, in the same notation as Fig.~7 of the main text.}
\end{figure}

\begin{figure}
\centering 
\includegraphics[width=0.9\linewidth]{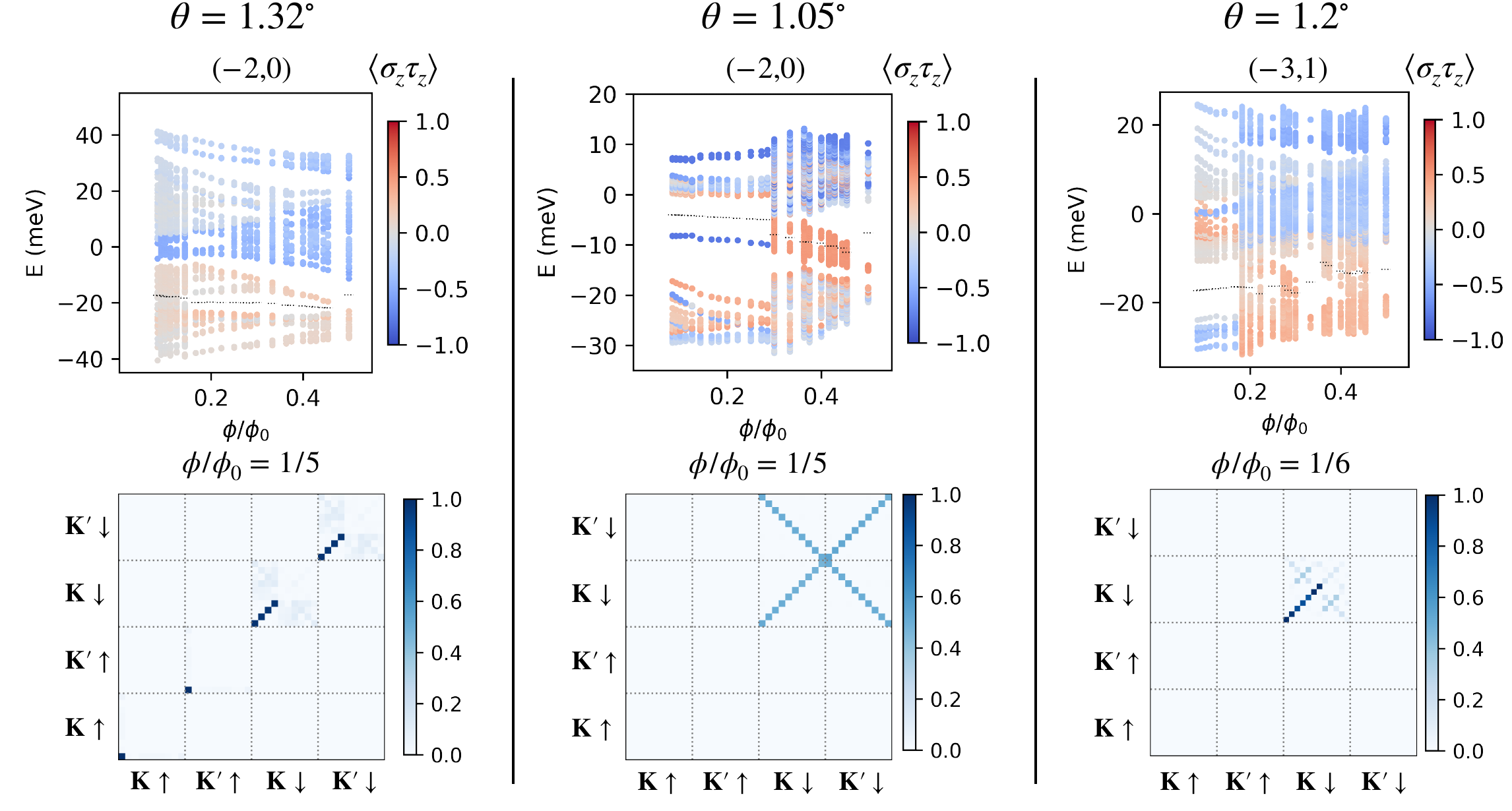}
\caption{\label{fig:nostrain_add_additionalCCIs} Additional correlated insulating states in magnetic phase diagram in Fig.~2 of the main text along $(-2,0)$ for $1.32^\circ$ and $1.05^\circ$, as well as the $(-3,1)$ gapped state for $1.2^\circ$.  Electronic states below the dashed lines are occupied.  }
\end{figure}

\begin{figure}
\centering 
\includegraphics[width=\linewidth]{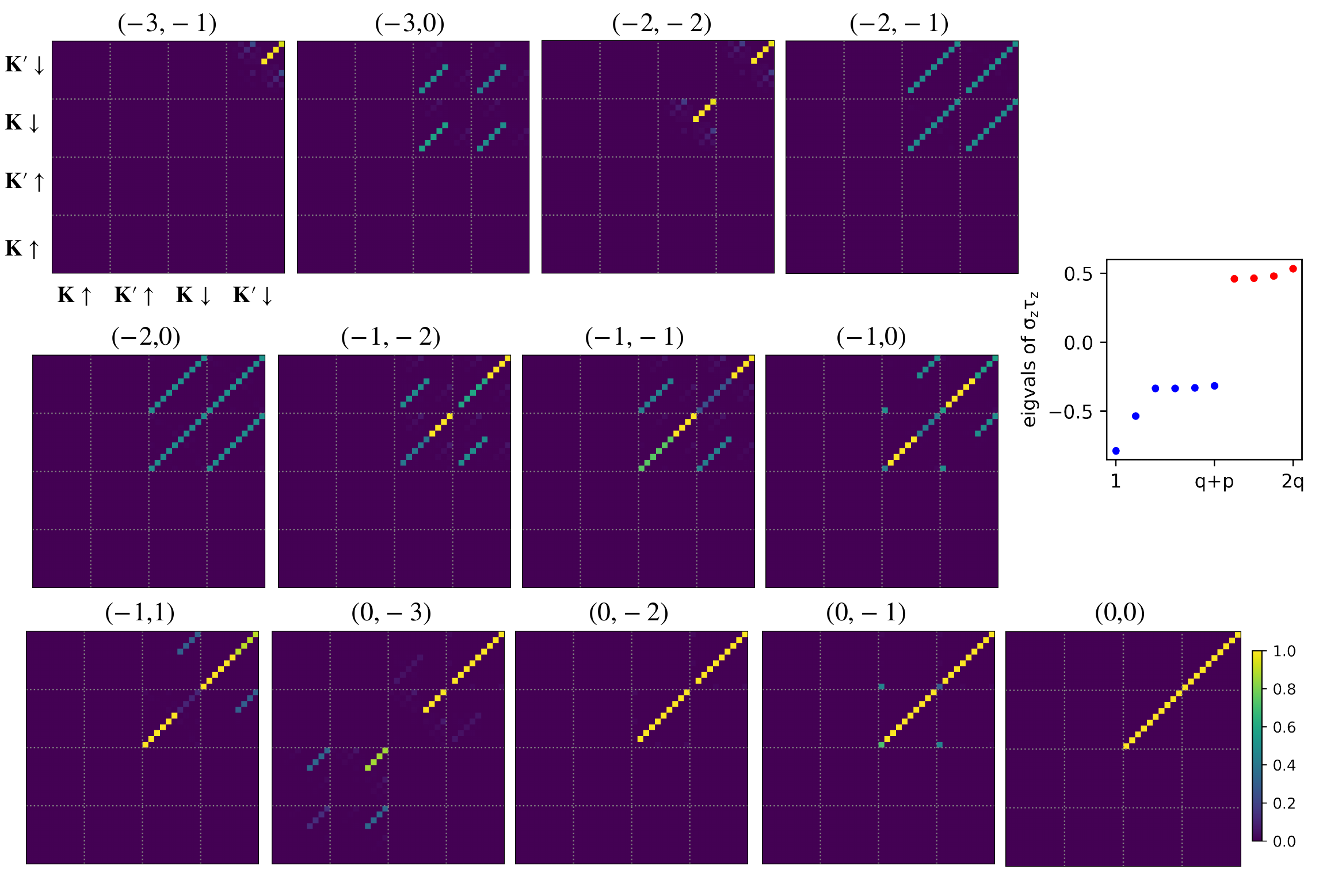}
\caption{\label{fig:nostrain_magic_angle} Absolute values of the full density matrices at $\bk=\mathbf{0}$ for all main gapped states at $1.05^\circ$ and $\phi/\phi_0=1/5$ in absence of heterostrain, expressed in the eigenbasis of the $\sigma_z\tau_z$ operator. At a general flux ratio $\phi/\phi_0=p/q$, the eigenstates of $\sigma_z\tau_z$ split into $q+p$ (blue) and $q-p$ (red) groups separated by a spectral gap, and are descendants of the zero field $+1$ and $-1$ Chern bands respectively. Near the magic angle, Coulomb interaction and magnetic field favor correlated gapped states that do not couple opposite Chern bands. It is important to note that  intervalley coherent states and valley/spin polarized states are extremely energetically competitive at the magic angle, and our converged states are often a mixture of these two kinds of states as illustrated in the density matrices.}
\end{figure}

\begin{figure}
\centering 
\includegraphics[width=\linewidth]{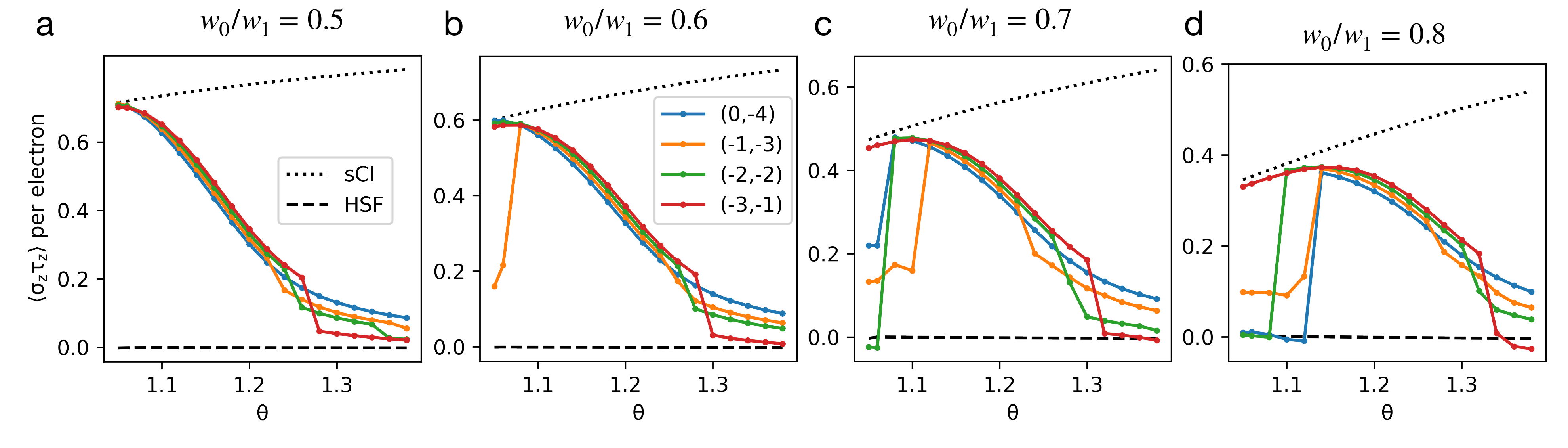}
\caption{\label{fig:nostrain_w0_w1} Averaged $\langle\sigma_z\tau_z \rangle$ for the occupied electronic states at $\phi/\phi_0=1/8$, twist angle $\theta=1.05^\circ$ in the absence of heterostrain, for different ratios of $w_0/w_1$. As $w_0/w_1$ is decreased, the correlated Hofstadter ferromagnets persist toward lower angles, before losing to other gapped states with intervalley coherence. The phase transition is first order, and marked by a collapse of $\langle\sigma_z\tau_z \rangle$.}
\end{figure}

\begin{figure}
\centering 
\includegraphics[width=0.8\linewidth]{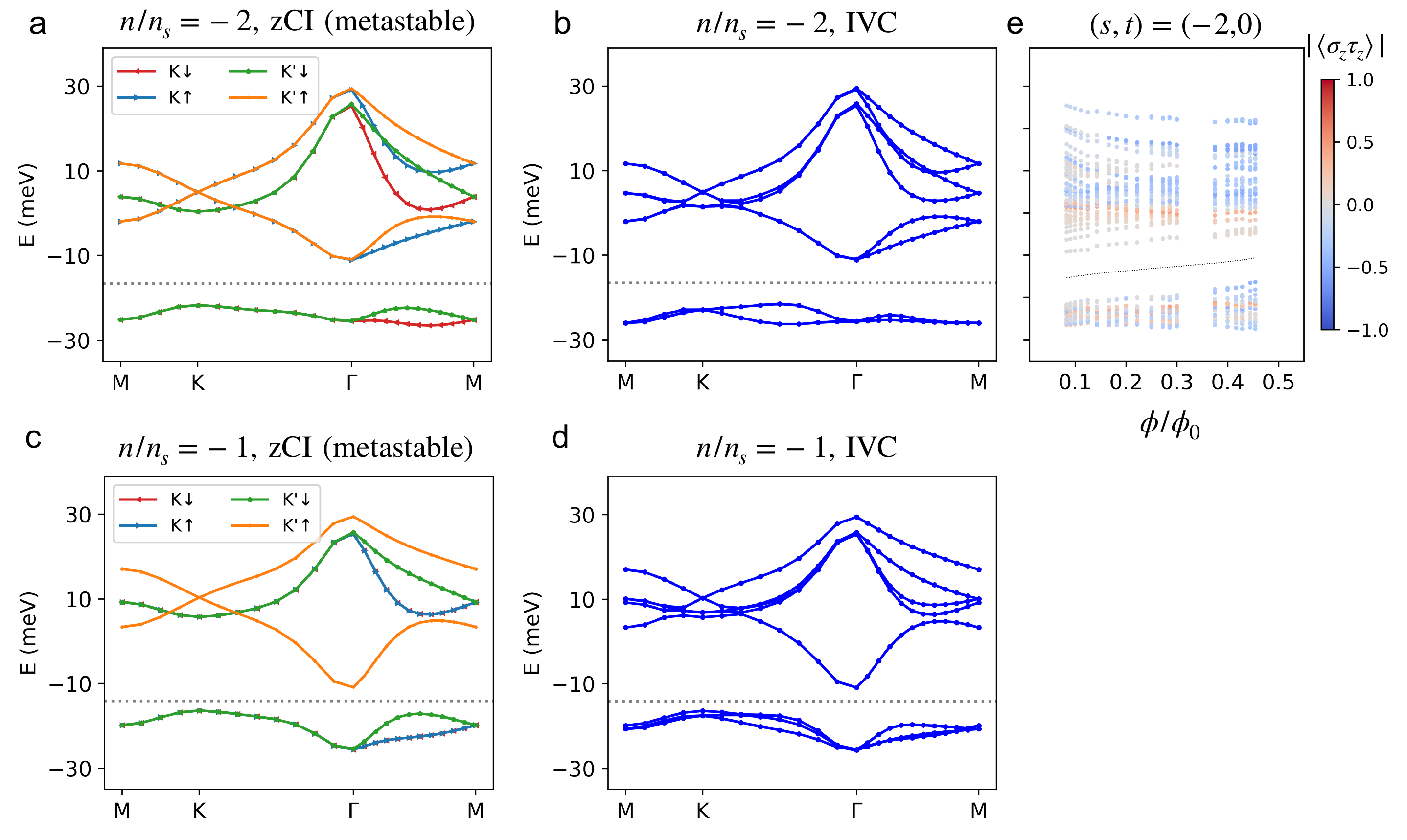}
\caption{\label{fig:figs20} $B=0$ Hartree-Fock study at $n/n_s=-2,-1$ at twist angle $1.20^\circ$, with no heterostrain. (a) At $n/n_s=-2$, Hartree-Fock spectra of a metastable zCI state with Chern number $-2$ , and (b) of the true IVC ground state. The Hartree-Fock energy difference between (a) and (b) is $\approx 0.52$meV per moir\'e unit cell. (c) At $n/n_s=-1$, Hartree-Fock spectra of a metastable zCI state with Chern number $-3$, and (d) of the true ground state. The Hartree-Fock energy difference between (c) and (d) is $\approx 0.56$meV per moir\'e unit cell. (e) At $n/n_s=-2$, the Hartree-Fock spectra along $(-2,0)$ smoothly extrapolates to the IVC ground state. At $n/n_s=-1$ and finite $B$, the true ground state loses to other gapped or nearly compressible states for all the magnetic flux ratios studied in this work. At low magnetic flux ratios we identify both $(-1,-1)$ and $(-1,-2)$ as populating the Landau quantized excitation spectra of the metastable zCI state in (c). }
\end{figure}


\end{widetext}

\end{document}